\documentclass[journal]{IEEEtran}

\newcommand{\href}[1]{}

\usepackage[colorlinks=true,pdfstartview=FitV,linkcolor=black,citecolor=black,urlcolor=blue,plainpages=false]{hyperref}

\usepackage[numbers,sort&compress]{natbib}

\usepackage[T1]{fontenc}
\usepackage[latin1]{inputenc}

\usepackage{amsmath}
\usepackage{amsthm}
\usepackage{amsfonts}
\usepackage{dsfont}
\usepackage{mathrsfs}
\usepackage{enumerate}
\usepackage{stfloats}

\usepackage[]{authblk}
\usepackage{times}
\usepackage{epsf}
\usepackage{psfrag}
\usepackage{epsfig}

\usepackage{amssymb}
\usepackage{amstext}
\usepackage{latexsym}
\usepackage{color}
\usepackage{ifthen}
\usepackage{multirow}

\usepackage{subfigure}

\newcommand{\Z}{{\mathds Z}}

\newcommand{\U}{{\mathcal U}}
\newcommand{\ep}{\epsilon}
\newcommand{\non}{\nonumber \\}
\newcommand{\C}{{\mathcal C}}
\newcommand{\D}{{\mathcal D}}

\newcommand{\I}{{\mathcal I}}
\newcommand{\N}{{\mathcal N}}
\newcommand{\IN}{{\mathbb N}}
\newcommand{\R}{{\mathbb R}}
\newcommand{\Os}{{\mathcal O}}

\newcommand{\dE}{D_\Sigma}
\newcommand{\q}[2]{Q_{s_{#1},d_{#2}}}
\newcommand{\p}[2]{P_{s_{#1},d_{#2}}}
\newcommand{\m}[2]{M_{s_{#1},d_{#2}}}
\newcommand{\z}[2]{Z_{s_{#1},d_{#2}}}

\newcommand{\MC}[3]{#1\leftrightarrow #2 \leftrightarrow #3}

\newcommand{\fMC}[4]{#1\leftrightarrow #2 \leftrightarrow #3 \leftrightarrow #4}

\newcommand{\defi}{\triangleq}

\newcommand{\goesto}{\rightarrow}

\newcommand{\concat}{\oplus}

\newcommand{\interf}{\hspace{1mm}{\stackrel{I}{\leadsto}}\hspace{1mm}}
\newcommand{\dinterf}{\hspace{1mm}{\stackrel{I}{\rightarrow}}\hspace{1mm}}

\newcommand{\st}{:}

\newcommand{\pr}{$^\prime$}

\newcounter{constcount}
\newcommand{\cons}{\stepcounter{constcount} K_{\arabic{constcount}}}
\newcommand{\curcons}{K_{\arabic{constcount}}}

\newcounter{numcount}
\newcommand{\eqnum}{\stackrel{(\roman{numcount})}{=}\stepcounter{numcount}}

\newcommand{\leqnum}{\stackrel{(\roman{numcount})}{\leq\;}\stepcounter{numcount}}

\newcommand{\cnt}{$(\roman{numcount})$\;\stepcounter{numcount}}

\newcommand{\rescnt}{\setcounter{numcount}{1}}

\newcounter{thmcnt}
  \let\Oldsection\section
  
\renewcommand{\section}{\stepcounter{thmcnt}\Oldsection}

\newtheorem{theorem}{Theorem}

\newtheorem{lemma}{Lemma}
\newtheorem{definition}{Definition}
\newtheorem{claim}{Claim}

\newenvironment{xlist}[1]{
  \begin{list}{}{
    \settowidth{\labelwidth}{#1}
    \setlength{\labelsep}{0.5cm}
    \setlength{\leftmargin}{\labelwidth}
    \addtolength{\leftmargin}{\labelsep}
    }
  }
{\end{list}}

\newenvironment{nproof}{\noindent \emph{Proof:}\,}{\hfill\IEEEQED}

\newcounter{examplecounter}
\newenvironment{example}
{
\stepcounter{examplecounter} {\vspace{4mm} \noindent \bf Example \arabic{examplecounter}.} \begin{it} \rm
}
{
\end{it} \vspace{2mm}}

\begin{document}

\title{Two-Unicast Wireless Networks:\\ Characterizing the Degrees-of-Freedom}

\author{Ilan Shomorony and A. Salman Avestimehr \thanks{The authors are with the School of Electrical and Computer Engineering, Cornell University, Ithaca, NY 14853 USA (e-mails: is256@cornell.edu, avestimehr@ece.cornell.edu). \\
\indent The research of A. S. Avestimehr and I. Shomorony was supported in part by the NSF CAREER award 0953117, U.S. Air Force Young Investigator Program award FA9550-11-1-0064, NSF TRUST Center, and NSF Grants CCF-1144000 and CCF-1161720. \\
\indent Manuscript received March 14, 2011; revised January 30, 2012. Date of current version August 5, 2012. Communicated by S. Jafar, Associate Editor for Communications. \\ 
\indent This paper was presented in part at the International Symposium on Information Theory 2011, St. Petersburg, Russia \cite{dof2unicastisit}}}


\maketitle

\begin{abstract}
We consider two-source two-destination (i.e., two-unicast) multi-hop wireless networks that have a layered structure with arbitrary connectivity. We show that, if the channel gains are chosen independently according to continuous distributions, then, with probability $1$, two-unicast layered Gaussian networks can only have $1$, $3/2$ or $2$ sum degrees-of-freedom (unless both source-destination pairs are disconnected, in which case no degrees-of-freedom can be achieved). 
We provide sufficient and necessary conditions for each case based on network connectivity and a new notion of source-destination paths with manageable interference. 
Our achievability scheme is based on forwarding the received signals at all nodes, except for a small fraction of them in at most two \emph{key layers}. Hence, we effectively create a ``condensed network'' that has at most four layers (including the sources layer and the destinations layer). We design the transmission strategies based on the structure of this condensed network.
 The converse results are obtained by developing information-theoretic inequalities that capture the structures of the network connectivity.
 Finally,
we extend this result and characterize the full degrees-of-freedom region of two-unicast layered wireless networks.
\end{abstract}

\begin{section}{Introduction}

Characterizing network capacity is one of the central problems in network information theory. While this problem is in general unsolved, there has been considerable success in two research fronts. The first one focuses on single-flow multi-hop networks, in which one source aims to send  the same message to one or more destinations, using multiple relay nodes. Since, in this scenario, all destination nodes are interested in the same message, there is effectively only one information stream in the network. Starting from the max-flow-min-cut theorem of Ford-Fulkerson \cite{FF56}, there has been significant progress on this problem.  For wireline networks, the maximum multicast flow was characterized in \cite{ACLY00}. In \cite{LinNetCod,KoetterMedard}, it was further shown that this maximum flow can be achieved using linear network codes.  

In \cite{ADTJ09}, the max-flow min-cut theorem was generalized for a class of linear deterministic networks with broadcast and interference. Inspired by this generalization, the multicast capacity of wireless networks was then characterized to within a gap that does not depend on the channel gains \cite{ADTJ09}, hence providing a constant-gap approximation of the capacity. Tighter capacity approximations were later derived in \cite{SuhasLatticeRelay,NoisyNetworkCoding}.

The second research direction focuses on multi-flow wireless networks with only one-hop between the sources and the destinations, i.e., the interference channel. While the capacity of  the interference channel remains unknown (except for special cases, such as \cite{KhandaniWIC,GerhardWIC,VenuWIC,SridharanKuserStrong,ElGamalCosta,SatoGInterference,CarleialVS}), there has been a variety of capacity approximations derived, such as constant-gap capacity approximations \cite{ETW,BreslerTse,BreslerTseEuro} and degrees-of-freedom characterizations \cite{CadambeJafar,Etkin,JafarShamai,MotahariKhandani,alignmentKhandani,JafarMIMOIC}.

However, once we go beyond single-hop, there is much less known about the capacity of multi-flow networks. Even in the simplest case with two sources and two destinations there are very few general results, such as \cite{HuMulticommodity}, where the maximum flow in two-unicast undirected wireline networks is characterized. 
For two-unicast directed wireline networks, \cite{Ness2unicast,ShenviDey,Cai2unicast} have provided graph-theoretic and cut-set based conditions under which rate $(1,1)$ can be achieved. In the wireless realm, constant-gap approximations of the capacity of specific two-hop networks (the ZZ and ZS networks) were obtained in \cite{MohajerZZ}. Furthermore, it was recently shown  that the network resulting from the concatenation of two or more \emph{fully connected} interference channels (the XX structure) admits the maximum of two degrees-of-freedom \cite{xx}. The achievability scheme relies on the notion of real interference alignment, which was introduced in \cite{MotahariKhandani}. 

In this paper, we consider two-unicast multi-hop wireless networks that have a layered structure with \emph{arbitrary connectivity}. We consider an AWGN channel model and assume that the channel gains (for each existing link) are independently drawn from a continuous distribution and remain fixed during the course of communication. Moreover, we assume that all channel gains are fully known at all nodes. Under these assumptions, we will show that, with probability 1 over the choice of the channel gains, two-unicast layered Gaussian networks can only have 1, $3/2$ or $2$ sum degrees-of-freedom (unless the source-destination pairs are disconnected, in which case we have $0$ degree-of-freedom). Furthermore, we will extend this result and show that there are only five possible degrees-of-freedom regions for two-unicast layered networks, and we will provide necessary and sufficient conditions for each case that are based only on properties of the network graph. 


\vspace{3mm}

\noindent \emph{\bf Paper outline and description of main contributions:}

\vspace{2mm}

In Section \ref{defns}, we provide some basic definitions and state our main result. Then, in Section \ref{overview}, we give a high-level description of the proof techniques and the intuition behind some of the arguments. We proceed to describing the networks in which only one degree-of-freedom can be achieved in Section \ref{onedeg}. More specifically, if we let $(s_1,d_1)$ and $(s_2,d_2)$ be the pairs of corresponding source and destination, we will show that the maximum achievable degrees-of-freedom is one if and only if we are in one of the following two cases: $(i)$ the network contains a node $v$ whose removal disconnects pairs $(s_1,d_1)$, $(s_2,d_2)$ and at least one of $(s_1,d_2)$ and $(s_2,d_1)$; or $(ii)$  the network contains an edge $(v_2,v_1)$ such that the removal of $v_1$ disconnects a destination from both sources and the removal of $v_2$ disconnects the non-corresponding source from both destinations. The conditions we present can be seen as a generalization of the graph-theoretic conditions given in \cite{ShenviDey} which characterize when a two-unicast wireline network does not support rate $(1,1)$. 

Then, in Section \ref{twodeg}, we consider the cases in which two degrees-of-freedom can be achieved. We will show that if our network graph contains a \emph{Butterfly} or a \emph{Grail} subgraph, then two degrees-of-freedom can be achieved. In order to describe the third class of networks which admit two degrees-of-freedom, we introduce the notion of \emph{manageable interference}. We will say that two disjoint source-destination paths have manageable interference if, intuitively, all the interference between them can be either avoided or neutralized. Once again, it is interesting to compare the description of the networks with two degrees-of-freedom to the graph-theoretic description of the wireline networks which support rate $(1,1)$. While in the wireline case it is possible to achieve rate $(1,1)$ in networks which contain a Butterfly, a Grail or two edge-disjoint paths from each source to its corresponding destination  (see \cite{ShenviDey,Cai2unicast}), in the wireless case, it is possible to achieve two degrees-of-freedom in networks which contain a Butterfly, a Grail or two vertex-disjoint paths from each source to its destination that have manageable interference.


In order to describe general achievability schemes that work for an arbitrary number of layers, we propose a new method which involves building a \emph{condensed} network, by identifying specific \emph{key layers} which will perform non-trivial relaying operations. All the nodes which do not belong to the key layers will be assumed to simply forward their received signals at all times. Therefore, an effective transfer matrix between any pair of consecutive key layers can be obtained and it can be used to define the edges and the channel gains of our condensed network. To achieve two degrees-of-freedom, we will consider two distinct relaying schemes for the nodes in the key layers. If our condensed network is a $2 \times 2 \times 2$ interference channel, then we will resort to the real interference alignment schemes provided in \cite{xx}. Otherwise, we will show that a linear coding scheme will suffice to achieve the sum degrees-of-freedom. Notice that, since we assume single antennas at all nodes, the cut-set bound tells us that we cannot hope to achieve more than two degrees-of-freedom, and this case requires no converse proof. 

In Section \ref{32deg}, we address all the networks which do not fall into the cases considered in Sections \ref{onedeg} and \ref{twodeg}. 
We will show that they all have $3/2$ degrees-of-freedom. Our achievability scheme is based on defining two distinct modes of operation for the network. During the first mode, specific nodes act as buffers, storing all the received signals in order to use them during the second mode of operation. Then, in the second mode, these stored signals can be either forwarded towards the destinations or used to neutralize the interference. This way, it is possible to achieve $3/2$ degrees-of-freedom by evenly dividing the amount of time the network operates in each mode. The converse result is obtained by finding information-theoretic inequalities which capture the fact that the interference, in this case, is not completely manageable. 

In Section \ref{ext}, we describe how the results regarding the sum degrees-of-freedom of two-unicast layered Gaussian networks can be extended to obtain the full degrees-of-freedom region. We show that there are only five possible degrees-of-freedom regions (assuming each source is connected to its destination) and we provide necessary and sufficient conditions for a network to have each of these regions. We do this by using the outer bound provided by the sum degrees-of-freedom and by describing achievability schemes for some specific extreme points in the degrees-of-freedom region, using real interference alignment. 

Finally, in Section \ref{concl}, we provide some concluding remarks. 

\end{section}

\begin{section}{Definitions and Main Results} \label{defns}

A \emph{multiple-unicast Gaussian network} $\N = (G,L)$ consists of a directed graph $G = (V,E)$, where $V$ is the vertex (or node) set and $E \subset V \times V$ is the edge set, and a set of source-destination pairs $L \subset V \times V$. We will focus on two-unicast (two-source two-destination) Gaussian networks, which means that $L = \{ (s_1,d_1),(s_2,d_2)\}$, for distinct vertices $s_1, s_2, d_1,d_2$. Moreover, we will assume that the network is \emph{layered}, meaning that the vertex set $V$ can be partitioned into $r$ subsets $V_1,V_2,...,V_r$ (called layers) in such a way that $E \subset \bigcup_{i=1}^{r-1} V_i \times V_{i+1}$, and $V_1 = \{s_1,s_2\}$, $V_r = \{d_1,d_2\}$.  For a vertex $v \in V_j$, we will let $\I(v)\defi \{u\in V_{j-1}\st (u,v) \in E\}$ (the input nodes) and $\Os(v)\defi\{u\in V_{j+1}\st (v,u) \in E\}$ (the output nodes). Furthermore, we will let $\ell (v)$ be the index corresponding to the layer containing $v$, i.e., $v \in V_{\ell(v)}$. Notice that the layers induce a natural ordering of the nodes. Thus, we may say, for example, that $v_a$ occurs before $v_b$ if $\ell(v_a) < \ell(v_b)$.

A real-valued channel gain $h_e$ is associated with each edge $e \in E$. Since we will often be referring to vertices by $v_i$, for $i \in \IN$, we will also use $h_{i,j}$ to represent the channel gain associated with edge $(v_i,v_j)$. We will assume that the channel coefficients $h_e$ are independently drawn from continuous distributions and are fixed during the course of communication. We also assume that all channel gains are fully known at all nodes. At time $m$, each node $v_i$ (with the exception of $d_1$ and $d_2$) transmits a real-valued signal $X_{v_i}[m]$ (or simply $X_i[m]$, when there is no ambiguity), which must satisfy an average power constraint $\frac1n \sum_{m=1}^n E\left[X_i^2[m]\right] \leq P$, $\forall \, v_i \in V$, for a communication session of duration $n$, where the expectation is taken with respect to any possible randomization involved. The signal received by node $v_j$ at time $m$ is given by
\[ Y_j[m] = \sum_{v_i \in \I(v_j)} h_{i,j} X_i[m] + N_j[m], \text{ for $m=1,2,...$ },\]
where $N_j[m]$ is the zero mean unit variance Gaussian discrete-time white noise process associated with node $v_j$. The transmitted signal from node $v_j$ (with the exception of $s_1$ and $s_2$) at time $m$ must be a (possibly randomized) function of its past received signals $Y_j[k]$, for $k = 1,...,m-1$.  Source $s_i$ picks a message $W_i$ that it wishes to communicate to $d_i$, and transmits signals $X_{s_i}[m]$, $m=1,...,n$, which are a function of $W_i$, for $i=1,2$. Each destination uses a decoder, which is a mapping $g_i: \R^n \rightarrow \{1,...,|W_i|\}$ from the $n$ received signals to the source message indices ($|W_i|$ is the number of messages that can be chosen). We say that rates $R_i \defi \frac{\log|W_i|}{n}$ for $i=1,2$ are achievable if the probability of error in the decoding of both messages by their corresponding destinations can be made arbitrarily close to 0 by choosing a sufficiently large $n$. The sum-capacity $C_\Sigma(P)$ is the supremum of the achievable sum-rates for power constraint $P$. 

\begin{definition} The sum degrees-of-freedom $\dE$ of a two-unicast Gaussian network is defined as
\[ \dE \defi \lim_{P\rightarrow \infty}\frac{C_\Sigma(P)}{\frac12 \log P}. \] 
\end{definition}
\emph{Remark:} $\dE$ will in general depend on $H = \{ h_e : e \in E\}$. However, we will show that with probability 1, $\dE$ only depends on the network graph $G$, and not on the values of $H$.


\vspace{2mm}

We now consider several definitions which will be used throughout the paper.

\begin{definition} A (directed) path between $v_1 \in V$ and $v_k \in V$ is an ordered set of nodes $\{v_1, v_2, ..., v_k\}$ such that $(v_i,v_{i+1}) \in E$ for $i=1,...,k-1$. We will commonly refer to a path between $v_1$ and  $v_k$ by $P_{v_1,v_k}$. We write $v_1 \leadsto v_k$, if there is a path between $v_1$ and $v_k$. Notice that for any node $v \in V$, $v \leadsto v$. 
\end{definition}

For simplicity, we will assume that any $v \in V$ belongs to at least one path $\p{i}j$ for $i \in \{1,2\}$ and $j\in\{1,2\}$. This is reasonable since a node that does not belong to any source-destination path does not alter the achievable rates in the network and can be removed. Moreover, we will always assume that $s_i \leadsto d_i$ for $i=1,2$, since $s_i \not\leadsto d_i$ implies that $R_i = 0$. 
In order to be able to ``cut and paste'' path segments we will also consider the following path operations. For a path $P_{v_a,v_b} = \{v_a,v_{a+1},...,v_b\}$, we will let $P_{v_a,v_b}[v_c,v_d] = \{v_c,v_{c+1},...,v_d\}$ if $a \leq c \leq d \leq b$. Moreover, if we have paths $P_{v_e,v_f}$ and $P_{v_f,v_g}$, we will let $P_{v_e,v_f} \concat P_{v_f,v_g}$ be the path which results from concatenating $P_{v_e,v_f}$ and $P_{v_f,v_g}$.

\begin{definition} Paths $P_{v_a,v_b}$ and $P_{v_c,v_d}$ are said to be \emph{disjoint} if $P_{v_a,v_b} \cap P_{v_c,v_d} = \emptyset$. (Notice that such paths are usually called vertex-disjoint, but here we will refer to them as simply disjoint)
\end{definition}

\begin{definition} For a subset of the vertices $S \subset V$, we say that $G[S]$ is the graph induced by $S$ on $G$, if $G[S]=(S,E_s)$, where $E_s=\{(v_i,v_j)\in E \st v_i,v_j \in S\}$.
\end{definition}

\begin{definition} 
We say that $\N' = (G',L')$ is a subnetwork of $\N = (G,L)$, if $G'= G[S]$, for some $S \subset V$ such that $L \subset S \times S$, and $L'= L$.
\end{definition}

For the next definitions, we assume we have two disjoint paths $\p11$ and $\p22$. Since we will often make statements which work for both $\p11$ and $\p22$, we will let $\bar i = 2$ if $i = 1$ and $\bar i = 1$ if $i = 2$.

\begin{definition} We will say that a node $v_a \notin \p{i}i$ causes interference on $\p{i}i$ and write $v_a \interf \p{i}i$, if we can find a node $v_b \in \p{i}i$ such that $(v_a,v_b) \in E$ and a path $P_{s_{\bar i},v_a}$ between $s_{\bar i}$ and $v_a$ such that $P_{s_{\bar i},v_a} \cap \p{i}i = \emptyset$, for $i=1,2$. Moreover, we will say the interference is direct, and write $v_a \dinterf \p{i}i$, if, in addition, $v_a \in \p{\bar i}{\bar i}$. Otherwise, we call the interference indirect. 
\end{definition}


Consider a subnetwork $(G[S],\{(s_1,d_1),(s_2,d_2)\})$ for some $S \supset (\p11 \cup \p22)$. We will define $n_{i}(G[S],\p i i) \defi |\{v \in S\st v\interf \p{i}i\}|$, for $i=1,2$. 
Notice that, in the definition of $n_i$, the path implied by $v \interf \p ii$ must exist in the subnetwork with graph $G[S]$. Moreover, we define $n_{i}^D(\p{\bar i}{\bar i},\p i i) \defi |\{v \in V\st v\dinterf \p{i}i \}|$.
When there is no ambiguity in the choice of our two disjoint paths $\p11$ and $\p22$, we will simplify the notation by using $n_{i}(G[S])$ and $n_{i}^D$.

\begin{definition} \label{manag}
Two disjoint paths $\p11$ and $\p22$ have manageable interference if we can find $S \subset V$ such that $\p11,\p22 \subset S$, $n_1(G[S]) \ne 1$ and $n_2(G[S]) \ne 1$. \end{definition}

The following example illustrates the definitions above.

\begin{example} \label{ex1} Consider the network depicted in Figure \ref{ex1fig}.
\begin{figure}[ht]
\begin{center}
\includegraphics[height=29mm]{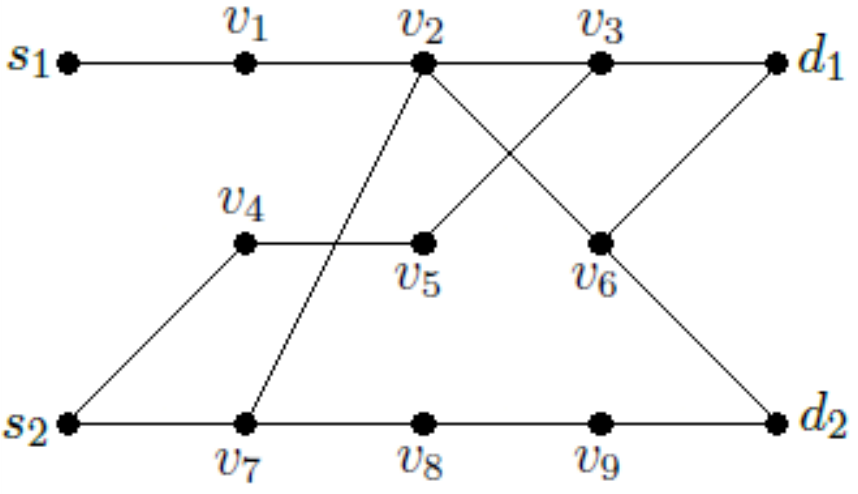}
\caption{Two-unicast layered network considered in Example \arabic{examplecounter}.} \label{ex1fig}
\end{center} 
\end{figure}
We have two disjoint paths from each source to its corresponding destination, given by $\p11 = \{ s_1,v_1,v_2,v_3,d_1 \}$ and $\p22 = \{ s_2, v_7, v_8, v_9, d_2 \}$. If we consider the entire network $\N = (G,L)$, then we have that $v_5 \interf \p11$ (since we have a path $P_{s_2,v_5} = \{s_2,v_4,v_5\}$ disjoint from $\p11$) and $v_7 \dinterf \p11$ (since we have a path $P_{s_2,v_7} = \{s_2,v_7\}$ disjoint from $\p11$ and $v_7 \in \p22$). Similarly, we have that $v_6 \interf \p22$. Thus, we conclude that $n_1^D(\p22,\p11) = 1$, $n_2^D(\p11,\p22) = 0$, $n_1(G,\p11) = 2$ and $n_2(G,\p22) = 1$. If instead we consider the subnetwork $\N = (G[S],L)$, where $S = \p11 \cup \p22$, we have $n_1(G[S],\p11) = n_1^D(\p22,\p11) = 1$ and $n_2(G[S],\p22) = n_2^D(\p11,\p22) = 0$. Finally, we consider the subnetwork $\N = (G[S'],L)$, where $S' = V \setminus \{v_6 \}$. Then we have $n_1(G[S'],\p11) = 2$ and $n_2(G[S'],\p22) = 0$, and we conclude that $\p11$ and $\p22$ have manageable interference. \end{example}

Now we state our main results.

\begin{figure*}[ht] 
     \centering
     \subfigure[]{
       \includegraphics[height=28mm]{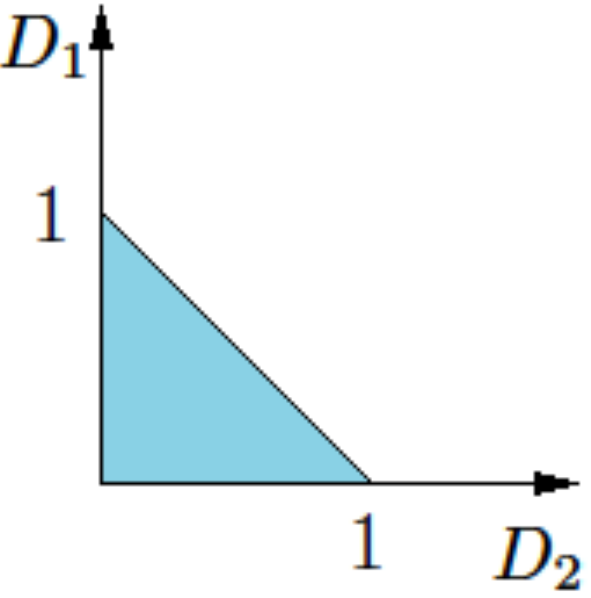} \label{reg1}} 
    \hspace{0mm}
    \subfigure[]{
       \includegraphics[height=28mm]{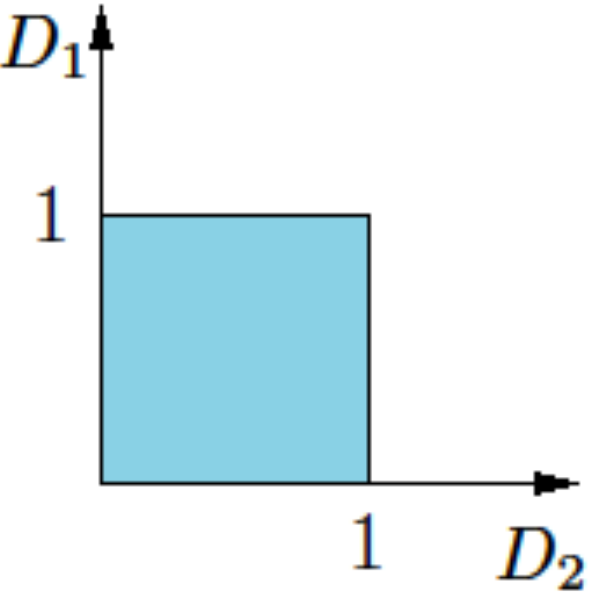} \label{reg2}}
    \hspace{0mm}
     \subfigure[]{
       \includegraphics[height=28mm]{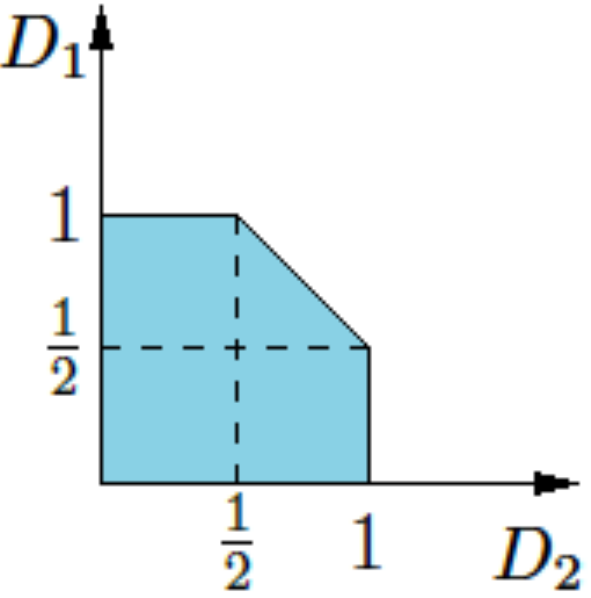} \label{reg32b}} 
    \hspace{0mm}
    \subfigure[]{
       \includegraphics[height=28mm]{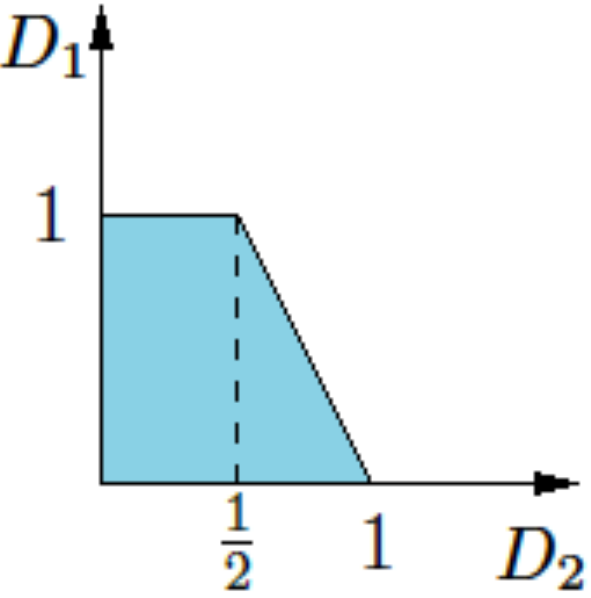} \label{reg32a}}
    \hspace{0mm}
    \subfigure[]{
       \includegraphics[height=28mm]{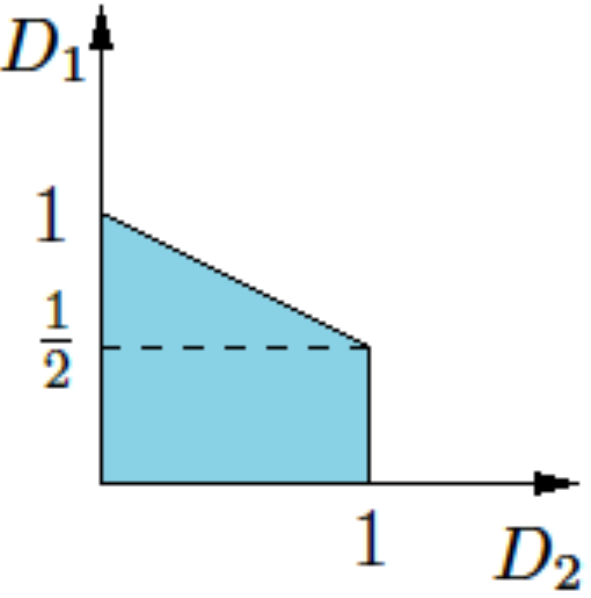} \label{reg32a_inv}}
     \caption{ Degrees-of-freedom region for networks in  case \ref{case1} (a);  case \ref{case2} (b); case \ref{case3} (c)  ; case \ref{case4} (d) ;  case \ref{case5} (e).} \label{regions}
\end{figure*}

\begin{theorem} \label{mainth}
For a two-unicast layered Gaussian network $\N = (G=(V,E), \{(s_1,d_1), (s_2,d_2)\})$, where the channel gains are independently drawn from continuous distributions, with probability 1, the sum degrees-of-freedom of $\N$, $\dE$, are given by

\vspace{2mm}

  \begin{xlist}{AB}
\item[A)\;] $\dE = 1$ if $\N$ contains a node $v$ whose removal disconnects $d_i$ from both sources and $s_{\bar i}$  from both destinations, for $i =1$ or $i=2$,
\item[A\pr)] $\dE = 1$ if $\N$ contains an edge $(v_2,v_1) \in E$ such that the removal of $v_1$ disconnects $d_i$ from both sources and the removal of $v_2$ disconnects $s_{\bar i}$  from both destinations, for $i =1$ or $i=2$,
\item[B)\;] $\dE = 2$ if $\N$ contains two disjoint paths $\p11$ and $\p22$ with manageable interference (see Definition \ref{manag}),
\item[B\pr)] $\dE = 2$ if $\N$ or any subnetwork does not contain two disjoint paths $\p11$ and $\p22$, but does not fall into case (A),
\item[C)\;] $\dE = \frac32$ in all other cases.
\end{xlist}

\end{theorem}

We also characterize the full degrees-of-freedom region of two-unicast layered Gaussian networks. We first define some basic notions.
 
\begin{definition}
The capacity region $\C(P)$ of a two-unicast Gaussian wireless network $\N$ with power constraint $P$ is the closure of the set of all pairs of achievable rates $(R_1,R_2)$.
\end{definition}

\begin{definition} \label{defn_region}
The degrees-of-freedom region of a two-unicast Gaussian network $\N$ 
is given by
\begin{align}
\vspace{2mm}
\D &= \left\{(D_1,D_2) \in \R^2_+ \st \forall \, w_1,w_2 \in \R_+, w_1 D_1 + w_2 D_2 \right. \nonumber \\ 
& \left. \leq \lim_{P \goesto \infty}\left( \sup_{(R_1,R_2) \in \C(P)}{\frac{w_1 R_1 + w_2 R_2}{\frac12 \log P}}\right) \right\}.
\end{align} 
\end{definition}
In order to simplify the characterization of the networks according to their degrees-of-freedom regions, we also consider the following definition.

\begin{definition} \label{semimanag}
Two disjoint paths $\p11$ and $\p22$ have $(s_i,d_i)$-manageable interference if we can find $S \subset V$ such that $\p11,\p22 \subset S$, $n_i(G[S]) \ne 1$, for $i=1$ or $2$. \end{definition}

For the degrees-of-freedom region of two-unicast Gaussian networks, we have the following result.

\begin{theorem} \label{extth}rofit
For a two-unicast layered Gaussian network $\N = (G=(V,E), \{(s_1,d_1), (s_2,d_2)\})$, where the channel gains are chosen according to independent continuous distributions, 
with probability 1, the degrees-of-freedom region $\D$ is given by

\vspace{2mm}

\begin{enumerate}[I.]
\item $\D = \left\{(D_1,D_2) \in \R^2_+  \st D_1 + D_2 \leq 1 \right\}$ if $\N$ falls in cases (A) or (A\pr) in Theorem \ref{mainth}, \label{case1}
\item $\D = \left\{(D_1,D_2) \in \R^2_+  \st D_1\leq 1, D_2 \leq 1 \right\}$ if $\N$ falls in cases (B) or (B\pr) in Theorem \ref{mainth}, \label{case2}
\item $\D = \left\{(D_1,D_2) \in \R^2_+  \st D_1\leq 1, D_2\leq 1, D_1+ D_2 \leq \frac32 \right\}$ if $\N$ is not in cases 
\ref{case1}, 
\ref{case2} and contains disjoint paths $\p11$ and $\p22$ whose interference is $(s_1,d_1)$ and $(s_2,d_2)$-manageable, \label{case3}
\item $\D = \left\{(D_1,D_2) \in \R^2_+  \st D_1\leq 1, D_1+ 2D_2 \leq 2 \right\}$ if $\N$ is not in cases 
\ref{case1}, 
\ref{case2} and \ref{case3} and contains paths $\q11$, $\z11$ and $\p22$, such that $\q11$ and $\p22$ are disjoint and have $(s_1,d_1)$-manageable interference, and $\z11$ and $\p22$ are disjoint and have $(s_2,d_2)$-manageable interference, \label{case4} 
\item $\D = \left\{(D_1,D_2) \in \R^2_+  \st D_2\leq 1,2 D_1+ D_2 \leq 2 \right\}$ if $\N$ is not in cases 
\ref{case1}, 
\ref{case2} and \ref{case3} and contains paths $\p11$, $\q22$ and $\z22$, such that $\q22$ and $\p11$ are disjoint and have $(s_1,d_1)$-manageable interference, and $\z22$ and $\p11$ are disjoint and have $(s_2,d_2)$-manageable interference. \label{case5} 
\end{enumerate}
Moreover, any two-unicast layered Gaussian network $\N$ falls into one of the cases described above. These five regions are depicted in Figure \ref{regions}.
\end{theorem}

\end{section}

\begin{section}{Proof overview} \label{overview}

Even though Theorem \ref{mainth} can be seen as a simple consequence of Theorem \ref{extth}, we will first prove Theorem \ref{mainth}. Theorem \ref{extth} will then follow as an extension of it.
We will consider cases (A), (A\pr), (B), (B\pr) and (C) sequentially. The intuition behind (A) is as follows. Let $W_1$ be the message from $s_1$ and $W_2$ be the message from $s_2$. If the removal of $v$ disconnects $d_1$ from both sources, then by knowing the received signal at $v$ we should be able to decode $W_1$. Then, since $v$ also disconnects $d_2$ from $s_2$, loosely speaking, all the information about $W_2$ goes through $v$. Therefore, $v$ can use the knowledge about $W_1$ to remove any interference due to signals about $W_1$, thus being able to decode $W_2$ as well. Since a single node can decode both messages, we have that $\dE \leq 1$, and it follows that $\dE = 1$, since 1 degree-of-freedom is trivially achievable from the fact that $s_1 \leadsto d_1$ and $s_2 \leadsto d_2$. 
\begin{figure}[ht]
\begin{center}
\includegraphics[height=26mm]{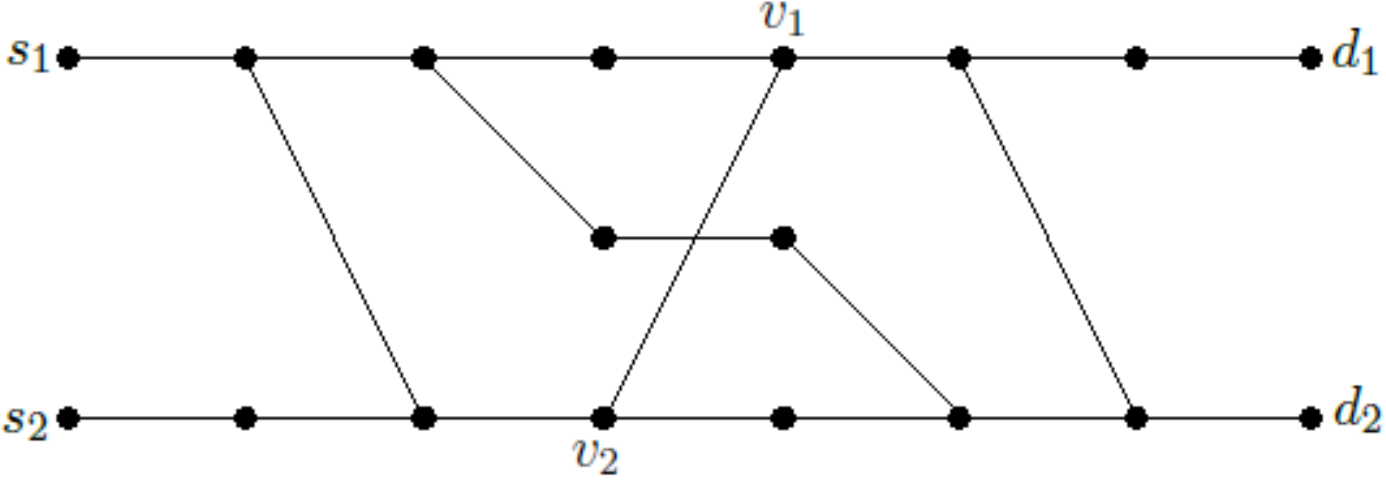}
\caption{An example of a network in case (A\pr).} \label{app}
\end{center} 
\end{figure}
The intuition behind (A\pr) is similar. If the removal of $v_1$ disconnects $d_1$ from both sources, then by knowing the received signal at $v_1$ we should be able to decode $W_1$. Since the removal of $v_2$ disconnects $s_2$ from both terminals, all the information regarding $W_2$ goes through $v_2$. This means that all the information received at $v_1$ which does not come from $v_2$ is about $W_1$ and, thus, by knowing the received signal at $v_1$, one can remove the part regarding $W_1$ and obtain the part of the transmitted signal at $v_2$ regarding $W_2$. But this implies that from $v_1$ we should be able to decode both $W_1$ and $W_2$, which implies $\dE \leq 1$. An example of a network that would fall in (A\pr) is shown in Figure \ref{app}.
\begin{figure*}[bp]
     \centering
     \subfigure[\,]{
       \includegraphics[height=23mm]{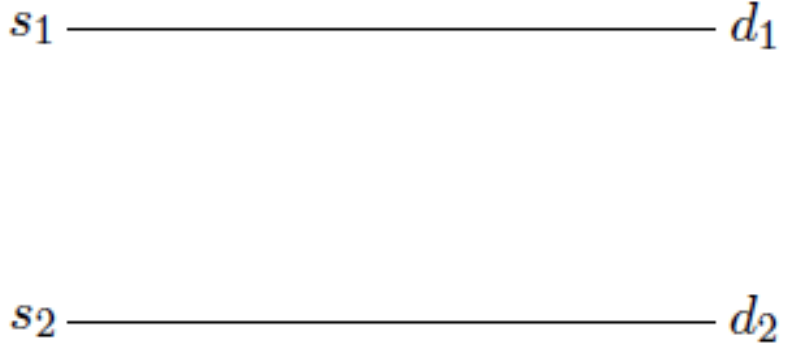} }
    \hspace{1mm}
    \subfigure[\,]{
       \includegraphics[height=23mm]{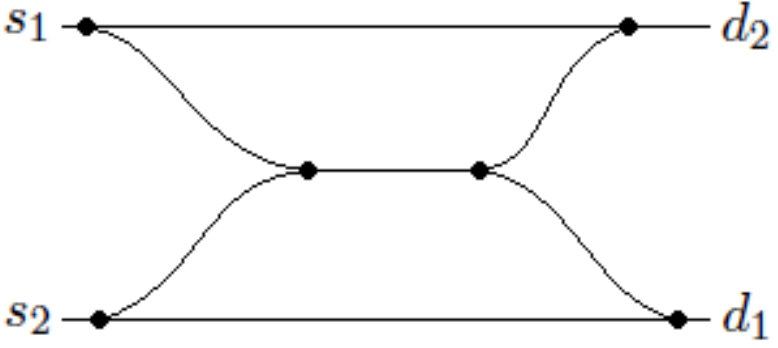}}
    \hspace{1mm}
    \subfigure[\,]{
       \includegraphics[height=23mm]{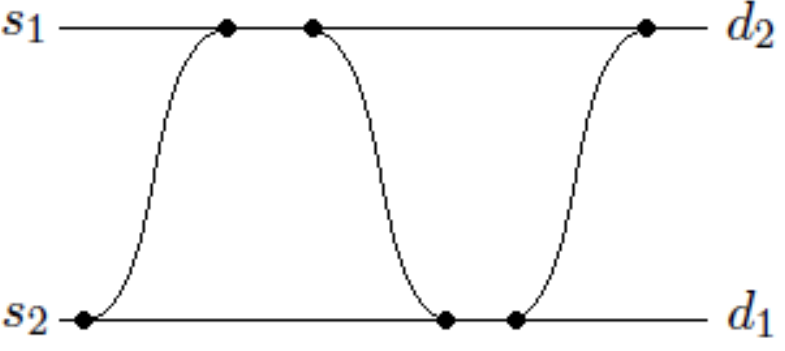}}
     \caption{Three categories of networks which are not in case (A). Notice that not all nodes are explicitly shown for the sake of generality. Each line represents a path, not necessarily a link, with any number of nodes. \label{nets1}}
\end{figure*}

To prove (B) and (B\pr), we will provide several achievability schemes for 2 degrees-of-freedom. For networks in (B), i.e., networks which contain two disjoint paths with manageable interference, depending on the network topology, we will either consider simple amplify-and-forward schemes or schemes based on real interference alignment, as described in \cite{xx}. If the network is in (B\pr), we will first restrict ourselves to the subnetwork which satisfies the description in (B\pr). Then, we will use a result from the double unicast problem for wireline networks to claim that the subnetwork must contain one of the three structures shown in Figure \ref{nets1}. But since we are assuming that the subnetwork has no two disjoint paths, we must have either the structure in Figure \ref{nets1}b or the structure in Figure \ref{nets1}c. We provide an amplify-and-forward achievability scheme in each case.


For case (C), we only need to consider networks which have two disjoint paths $\p11$ and $\p22$, but do not have two disjoint paths with manageable interference. This is because all networks which do not contain two disjoint paths $\p11$ and $\p22$ must fall into (A) or (B\pr). 
Moreover, any network that has two disjoint paths with manageable inteference will fall into (B). We will identify two main classes of networks in (C), depicted in Figure \ref{casesc}, and for each of these classes we will first provide an achievability scheme, based on two separate modes of operation for the network, which achieves $\frac32$ degrees-of-freedom. Then, we will show that the non-existence of two disjoint paths with manageable interference implies that either the network falls into (B\pr) or $\dE \leq \frac32$.
\begin{figure*}
     \centering
       \includegraphics[height=23mm]{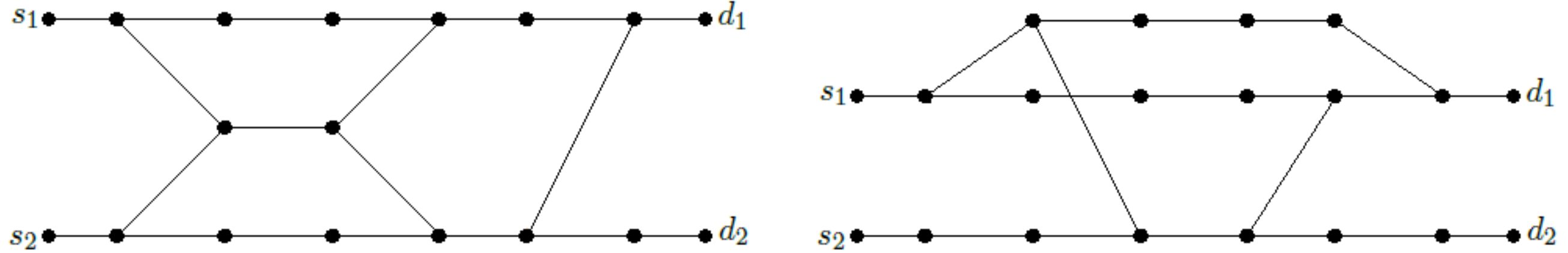} 
     \caption{Examples of the two classes of networks in case (C).} \label{casesc}
\end{figure*}

We will then build upon the result from Theorem \ref{mainth} to obtain Theorem \ref{extth}. For networks in cases (A), (A\pr), (B) and (B\pr), we will notice that the degrees-of-freedom region can be readily obtained from the sum degrees-of-freedom. For networks in case (C), we use the fact that the they must contain two disjoint paths that do not have manageable interference to infer properties about the network connectivity. Then, we combinine the outer bound provided by the sum degrees-of-freedom with achievability schemes for the extreme points to characterize the degrees-of-freedom region. Some of the extreme points will require the use of real interference alignment schemes.

\end{section}
\begin{section}{Networks with only one degree-of-freedom} 
\label{onedeg}

In this section, we will provide converse results for networks that fall in cases (A) and (A\pr). 
For the converse proofs, necessary for (A), (A\pr) and (C), we will derive information inequalities which allow us to bound the achievable sum-rates, and thus the degrees-of-freedom. We start by considering (A), and we assume WLOG that we have a node $v$ whose removal disconnects $d_1$ from both sources and $s_2$ from both destinations. We assume that the communication session lasts $n$ time steps, and for a node $v_j \in V$, we let $X_j^n$, $Y_j^n$ and $N_j^n$ be length $n$ vectors whose entries are, respectively, the transmitted signals $X_j[1], ..., X_j[n]$, the received signals $Y_j[1],...,Y_j[n]$ and the noise terms $N_j[1],...,N_j[n]$. For a set of nodes $S$, we will define $X_S$ to be the set of all $X_i$'s, for $v_i \in S$. Then, if we have $X_S^n$, we have a set of length $n$ vectors. We let $W_1$ and $W_2$ be independent random variables corresponding to uniform choices over the messages on sources $s_1$ and $s_2$ respectively. Then we have \rescnt
\begin{align}
n R_1 & = H(W_1) = I(W_1;Y_{d_1}^n) + H(W_1|Y_{d_1}^n) \nonumber \\
& \leqnum I(W_1;Y_{d_1}^n)+n\epsilon_n \leqnum I(X_{s_1}^n;Y_v^n)  + n\epsilon_n \label{aeq1}
\end{align} \rescnt
where \cnt follows from Fano's inequality, where $\epsilon_n \goesto 0$ as $n \goesto \infty$; and \cnt follows because the removal of $v$ disconnects $d_1$ from both sources; thus we have $\fMC{W_1}{ X_{s_1}^n}{Y_v^n}{Y_{d_1}^n}$. For $R_2$, we have \rescnt
\begin{align}
n R_2 & = H(W_2) = I(W_2;Y_{d_2}^n) + H(W_2|Y_{d_2}^n) \nonumber \\ 
& \leq I(W_2;Y_{d_2}^n)+n\epsilon_n \leqnum I( X_{s_2}^n;Y_v^n,X_{s_1}^n) + n\epsilon_n \nonumber \\ 
&\leqnum I( X_{s_2}^n;Y_v^n|X_{s_1}^n)  + n\epsilon_n \label{aeq2}
\end{align} \rescnt
where \cnt follows because the removal of $v$ disconnects $d_2$ from $s_2$, and, as a consequence, the removal of $v$ and $s_1$ disconnects $d_2$ from both sources, and we have $\fMC{W_2}{X_{s_2}^n}{(Y_v^n,X_{s_1}^n)}{Y_{d_2}^n}$; and \cnt follows since $X_{s_1}^n$ is independent of $X_{s_2}^n$. Now, by adding inequalities (\ref{aeq1}) and (\ref{aeq2}), we obtain
\rescnt
\begin{align} 
n (R_1+ R_2) & = I(X_{s_1}^n;Y_v^n) + I( X_{s_2}^n;Y_v^n|X_{s_1}^n) + n\epsilon_n \non 
& = I(X_{s_1}^n, X_{s_2}^n;Y_v^n) + n\epsilon_n \nonumber \\
& = h(Y_v^n) - h(Y_v^n|X_{s_1}^n, X_{s_2}^n)  + n\epsilon_n  \non 
& \leq h(Y_v^n) - h(Y_v^n|X_{s_1}^n, X_{s_2}^n,X_{\I(v)}^n)  + n\epsilon_n \nonumber \\
& \leq  h(Y_v^n) - h(N_v^n)  + n\epsilon_n \non 
& \leq \frac n 2 \log\left( 1+ \left(\textstyle{\sum_{u \in \I(v)}} |h_{u,v}|\right)^2 P \right) \non 
& \quad \quad \quad  - \frac n 2 \log (2 \pi e)  + n\epsilon_n \nonumber \\
& = \frac n 2 \log\left( \frac{1+ \left(\textstyle{\sum_{u \in \I(v)}} |h_{u,v}|\right)^2 P}{2 \pi e} \right)  \non
& \quad \quad \quad + n\epsilon_n \leq \frac n 2 \log (\beta P)  + n\epsilon_n \label{aeq3},
\end{align}
where $\beta$ is a constant which does not depend on $P$, for $P$ sufficiently large. Therefore we conclude that
\[ \dE \leq \lim_{P \goesto \infty}\lim_{n \goesto \infty}\frac{\log( \beta P) + 2 \epsilon_n }{\log P} = 1. \]

In order to simplify the converse proofs for (A\pr) and (C), we will consider a decomposition of the additive Gaussian noise $N_j$ associated with each node $v_j$. More specifically, if $m = |\I(v_j)|$, we break the noise at node $v_j$ into $m$ independent noise components, each with variance $1/m$. Then we associate each of these components with one of the incoming edges, and we can define, for $v_i \in \I(v_j)$,
\[ \tilde{X}_{i,j} \defi h_{i,j}X_i+N_{i,j},\]
where $N_{i,j}$ is the noise term associated with the edge $(v_i,v_j)$. Clearly, we have $N_j = \sum_{i:v_i\in\I(v_j)} N_{i,j}$, and $N_j$ has unit variance. Notice that we can now write, for a node $v_j$, $Y_j = \sum_{i:v_i\in\I(v_j)}\tilde{X}_{i,j}$. Moreover, we will define
\[ \tilde{X}_i \defi \{\tilde{X}_{i,j} : j \text{ s.t. } v_j \in \Os(v_i) \}.\]
As before, we let $\tilde X_S$ be the set of all $\tilde X_i$'s, for $v_i \in S$, and $\tilde X_i^n$ be a length $n$ vector with all the $\tilde X_i[m]$'s, for $m=1,...,n$.

In order to find upper bounds to the rates, we will often be interested in showing that certain conditional mutual information terms can be upper bounded by a constant. In particular, if we have a Z structure across two layers in the network, such as the one shown in Figure \ref{z}a, we would like to say that $I(X_c^n;\tilde X_c^n|Y_b^n,\tilde X_a^n)$ can be upper bounded by a constant that does not depend on $P$. 
\begin{figure}[ht]
     \centering
     \subfigure[]{
       \includegraphics[height=24mm]{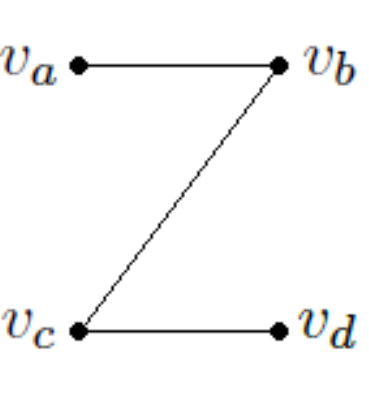} }
    \hspace{5mm}
    \subfigure[]{
       \includegraphics[height=24mm]{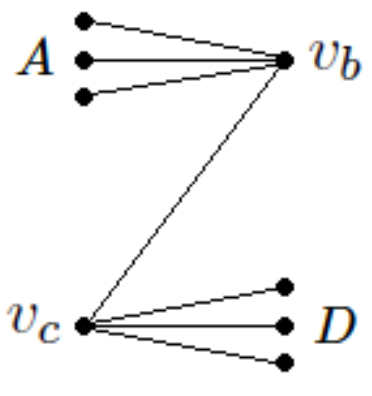}}
     \caption{The Z structure.} \label{z}
\end{figure}
Intuitively, the reason is that, given $\tilde X_a^n$ and $Y_b^n$, one can subtract $\tilde X_{a,b}^n$ from $Y_b^n$ and obtain $\tilde X_{c,b}^n$. This means that ``almost all'' information in $\tilde X_c^n$ can be deduced from $(Y_b^n,\tilde X_a^n)$, and thus the conditional mutual information cannot be very large. This reasoning is formalized in the following lemma, where we generalize the Z structure to one where $|\I(v_b)|\geq 2$ and $|\Os(v_c)|\geq 2$, as shown in Figure \ref{z}b. Moreover, we generalize this notion to the case where the mutual information may be conditioned on other signals as well, provided that these signals do not contain information about $N_{c,d}^n$, for some $v_d \in D$. The proof can be found in Appendix \ref{prooflemmaconstant}.

\begin{lemma} \label{const} 
Suppose we have nodes $v_b$ and $v_c$ such that $(v_c,v_b) \in E$, and let $A = \I(v_b) \setminus \{v_c\}$ and $D = \Os(v_c) \setminus \{v_b\}$. Suppose, in addition, that we have a set of nodes $S$ such that, if $u \in \Os(v_c)$ and $w \in S$, we have $u \not\leadsto w$, and a set of nodes $T$ with the property that, if $u \in D$ and $w \in T$, then $u \not\leadsto w$. Then, we have
\[ I(X_S^n;\tilde{X}_c^n|Y_b^n,\tilde X_A^n,X_T^n) \leq n K,\]
where $K$ is a constant that is only a function of the channel gains and the network graph $G$. \end{lemma}

\noindent \emph{Remarks:}  If, in the statement of Lemma \ref{const}, we condition the mutual information on $\tilde X_T^n$ instead of $X_T^n$ the same result holds. 
Also, if instead of conditioning on $\tilde X_A^n$ and $Y_b^n$ we condition on $\tilde X_{c,b}^n$, the same result holds, since, in the proof, we use $\tilde X_A^n$ and $Y_b^n$ to construct $\tilde X_{c,b}^n$. We will consider these cases to be covered by Lemma \ref{const} as well.

\vspace{2mm}

We can now proceed to the proof of case (A') in Theorem \ref{mainth}. We assume WLOG that we have an edge $(v_2,v_1) \in E$ such that the removal of $v_1$ disconnects $d_1$ from both sources and the removal of $v_2$ disconnects $s_2$ from both destinations. We let $A \defi \{ v \in V \st s_2 \not\leadsto v \}$, and we notice that $\I(v_1) \setminus \{v_2\} \subset A$, since, otherwise, we would have a node $v_a \in \I(v_1) \setminus \{v_2\}$ such that $s_2 \leadsto v_a$, and this would contradict the fact that the removal of $v_2$ disconnects $s_2$ from $d_1$. 
Moreover, $v_2 \notin A$, because all paths from $s_2$ to $d_2$ contain $v_2$ and we must have at least one such path. 
Thus we have \rescnt
\begin{align}
nR_1 & \leq I(W_1;Y_{d_1}^n)+n\epsilon_n \leqnum I(\tilde X_A^n;Y_1^n)+n\epsilon_n \non 
& = I(\tilde X_A^n,X_2^n;Y_1^n) - I(X_2^n;Y_1^n|\tilde X_A^n) +n\epsilon_n \nonumber \\
& \leqnum \frac n 2\log P + nK_1 - I(X_2^n;Y_1^n|\tilde X_A^n) +n\epsilon_n, \label{l2eq1}
\end{align} \rescnt
where 
\cnt follows because $v_1$ disconnects $d_1$ from both sources and $s_1 \in A$, thus we have $\fMC{W_1}{\tilde X_A^n}{Y_1^n}{Y_{d_1}^n}$; and \cnt follows because $\I(v_1) \setminus \{v_2\} \subset A$ and $v_2 \notin A$, hence we can upper-bound $I(\tilde X_A^n, X_2^n;Y_1^n)$ as 
\begin{align} 
I(\tilde X_A^n,X_2^n;Y_1^n) & = h(Y_1^n) - h(Y_1^n|\tilde X_A^n, X_2^n) \non 
& = h(Y_1^n) - h(N_{2,1}^n) \nonumber \\
&\leq \frac n 2 \log\left( \frac{1+ \left(\textstyle{\sum_{u \in \I(v_1)}} |h_{u,v_1}|\right)^2 P}{2 \pi e / |\I(v_1)|} \right) \nonumber \\
& \leq \frac n 2 \log (\gamma P) \leq  \frac n2 \log P + n\cons,
\label{deriv}
\end{align} where $\gamma$ and $\curcons$ are constants which are independent of $P$, for sufficiently large $P$. 

Next we notice that, since the removal of $v_2$ disconnects $d_2$ from $s_2$ and the removal of $A$ disconnects $d_2$ from $s_1$, the removal of $v_2$ and $A$ disconnects $d_2$ from both sources. Thus we have \rescnt
\begin{align}
n R_2 & \leq I(W_2;Y_{d_2}^n)+n\epsilon_n \leqnum I(W_2;\tilde X_2^n,\tilde X_A^n) +n\epsilon_n \non 
& \eqnum I(W_2;\tilde X_2^n|\tilde X_A^n) +n\epsilon_n \non 
& \leqnum I(X_2^n;\tilde X_2^n|\tilde X_A^n) +n\epsilon_n \nonumber \\
&\leq I(X_2^n;\tilde X_2^n,Y_1^n|\tilde X_A^n) +n\epsilon_n \non 
& = I(X_2^n;Y_1^n|\tilde X_A^n)+I(X_2^n;\tilde X_2^n|\tilde X_A^n,Y_1^n)+n\epsilon_n \nonumber \\
& \leqnum I(X_2^n;Y_1^n|\tilde X_A^n)+n \cons+n\epsilon_n, \label{l2eq2}
\end{align} \rescnt
where \cnt follows from the fact that the removal of $v_2$ and $A$ disconnects $d_2$ from both sources, which implies $\MC{W_2}{(\tilde X_2^n,\tilde X_A^n)}{Y_{d_2}^n}$; \cnt follows from the fact that $W_2$ is independent of $\tilde X_A^n$; \cnt follows from the fact that, given $\tilde X_A^n$, we have $\MC{W_2}{X_2^n}{\tilde X_2^n}$; \cnt follows from the application of Lemma \ref{const} to $I(X_2^n;\tilde X_2^n|\tilde X_A^n,Y_1^n)$, since $\I(v_1)\setminus\{v_2\} \subset A$. Finally, by adding (\ref{l2eq1}) and (\ref{l2eq2}) we obtain
\begin{align*}
n(R_1+R_2) \leq \frac n 2 \log P + n(K_1+K_2) + n\epsilon_n, \end{align*} 
and we conclude that $\dE \leq 1$. Since one degree-of-freedom is trivially achievable, we have $\dE = 1$ for both (A) and (A\pr). 

\end{section}

\begin{section}{Networks with two degrees-of-freedom} \label{twodeg}

In this section, we will provide achievability schemes for the networks which fall into cases (B) and (B\pr).
In order to describe these schemes we will proceed as follows. We will first identify the \emph{key layers}, whose nodes will be responsible for performing non-trivial relaying operations. All the nodes which do not belong to the key layers will simply forward their received signal. This will allow us to build a \emph{condensed} version of the network. The condensed network only contains the nodes in the key layers, $V_1$ and $V_r$. The edges and respective channel gains are determined according to the effective transfer matrices between two consecutive layers of the condensed network, which are obtained by assuming that all intermediate nodes which are not in the key layers, $V_1$ or $V_r$ are simply forwarding their received signals. An example is shown in Figure \ref{cond1}. 

\begin{figure}[ht]
     \centering
     \subfigure[]{
       \includegraphics[height=30mm]{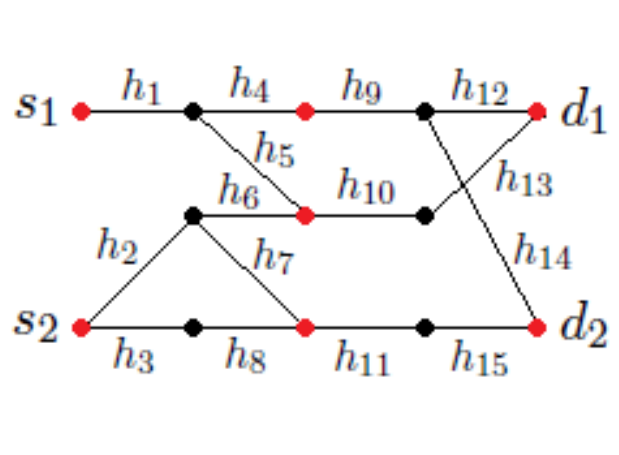} }
    \subfigure[]{
       \includegraphics[height=30mm]{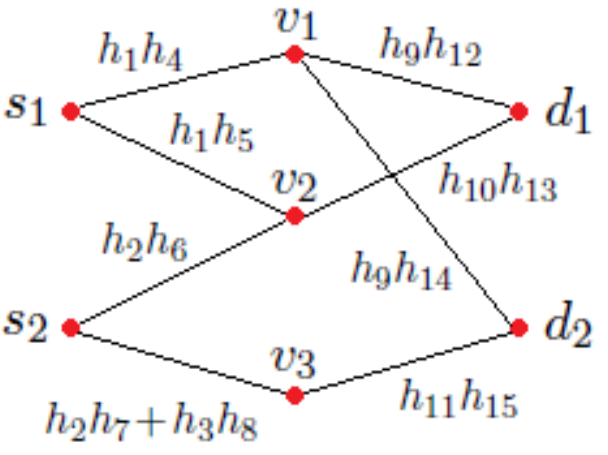}}
	\caption{A 5-layer network (a) and its 3-layer condensed version (b)}
	\label{cond1}
\end{figure}

We will refer to the effective channel gains of the edges in the condensed network by $\hat h (v,u)$, where $v$ is the starting node and $u$ is the ending node. For example, in Figure \ref{cond1}, we have $\hat h(s_2,v_3) = h_2 h_7 + h_3 h_8$ and $\hat h(v_2,d_2) = 0$. 
Notice that, in the condensed network, the effective additive noises at the nodes are not necessarily independent and identically distributed. However, they are still drawn from continuous distributions, which will be sufficient for us.

The condensed networks will be useful since we will conclude that entire classes of layered networks will possess essentially the same condensed network, and therefore we may describe a single achievability scheme for all the networks in that class.
We will describe achievability schemes for $\dE = 2$ in essentially two ways, according to the structure of the condensed network. If the resulting condensed network is a $2 \times 2 \times 2$ interference channel, then we will use the scheme described in \cite{xx} to achieve $\dE = 2$. Otherwise, we will describe a simple amplify-and-forward scheme that guarantees that the end-to-end transfer matrix for the condensed network (and thus for the original network as well) is of the form
\[ 
\begin{bmatrix}
  \beta_1 & 0 \\
  0 & \beta_2 \\
\end{bmatrix},
\]
for $\beta_1,\beta_2 \ne 0$. Thus we have $Y_{d_i} = \beta_i X_{s_i} + N_{d_i}^\text{eff}$, for $i=1,2$, where $ N_{d_i}^\text{eff}$ is the effective additive noise at $d_i$. Since the scaling factors used at the key layers and the noise variances are functions of the channel gains only (and not the power $P$ of the signals transmitted by the sources), we have essentially two parallel point-to-point AWGN channels. In order to make sure that the output power constraint is satisfied at all nodes, we will restrict the sources to using power $\alpha P$, for some $\alpha \in (0,1)$. It is not difficult to see that, for $P$ sufficiently large, $\alpha$ can be chosen independent of $P$. The effective additive noises at the destinations will be linear combinations of the individual Gaussian noises at each node, where the coefficients are functions of the channel gains $h_e$. Therefore, $\sigma_i^2$, the variance of the additive Gaussian noise at destination $d_i$, is not a function of $P$, and each source-destination pair $(s_i,d_i)$, for $i =1,2$, can use Gaussian random codes to achieve rate
\[ R_i = \frac12 \log\left(1+\frac{\alpha \beta_i^2 P}{\sigma_i^2}\right),\]
and, therefore, one degree-of-freedom. We conclude that we achieve $\dE = 2$.


First, we will consider (B), in which case we have two disjoint paths with manageable interference.

\begin{subsection}{Two disjoint paths with manageable interference} 

We let $\p11$ and $\p22$ be our two disjoint paths such that we have $S \subset V$ containing $\p11$ and $\p22$ and satisfying $n_1(G[S])\ne 1$ and $n_2(G[S])\ne 1$. In general, we will assume that $S$ is chosen to be minimal, and all the nodes in $V \setminus S$ are removed from the network.
If we have $n_1(G[S]) = 0$ and $n_2(G[S]) = 0$, then achieving $\dE = 2$ is trivial: we have two disjoint paths $\p11$ and $\p22$ with no interference whatsoever. For networks where $n_i(G[S]) \geq 2$, for $i =1$ or $i=2$, we will define $v_p^i$ to be the first node on $\p{i}i$ whose removal disconnects $d_i$ from $s_{\bar{i}}$. Notice that $V_{\ell(v_p^i)-1}$ is the layer containing $\I(v_p^i)$. This layer will be used as one of the key layers. Intuitively, this is the last layer where we can choose the scaling used at the nodes so that the interference on $\p{i}i$ is canceled. 
 If $n_i(G[S]) \geq 2$ and $n_{\bar i}(G[S]) =0$, for $i=1$ or $i=2$,  our condensed network will be a two-hop network formed by layers $V_1, V_{\ell(v_p^i)-1}$ and $V_r$. If $n_1(G[S]) \geq 2$ and $n_2(G[S]) \geq 2$, our condensed network will be a three-hop network formed by layers $V_1, V_{\ell(v_p^1)-1}, V_{\ell(v_p^2)-1}$ and $V_r$ (unless $\ell(v_p^1)=\ell(v_p^2)$, in which case the condensed network will be a two-hop network).
We will need the following technical lemma about $v_p^i$, whose proof can be found in the Appendix.

\begin{lemma} \label{lem:paths}
Assume $n_i(G[S]) \geq 2$, for $i =1$ or $i=2$, and let $v_p^i$ be defined as above. Then, there exist two paths $P_{s_1,v_p^i}$ and $P_{s_2,v_p^i}$ such that $P_{s_1,v_p^i} \cap P_{s_2,v_p^i} = \{v_p^i\}$. 
\end{lemma}

The importance of Lemma \ref{lem:paths} is that it guarantees that the transfer matrix between $(s_1,s_2)$ and two nodes in $\I(v_p^i)$ will be invertible with probability 1. This will be further explained later but, intuitively, it is necessary to give the nodes in $\I(v_p^i)$ freedom to cancel the interference from $s_{\bar i}$ on $\p{i}i$. A second useful property about $v_p^i$ is now stated in the form of another Lemma.

\begin{lemma} \label{lem:two}
Assume $n_i(G[S]) \geq 2$, for $i =1$ or $i=2$, and let $v_p^i$ be defined as above. Then, there are (at least) two nodes $v_1,v_2 \in \I(v_p^i)$ such that $s_{\bar i} \leadsto v_1$ and  $s_{\bar i} \leadsto v_2$.
\end{lemma} 
\begin{nproof}
Since $n_i(G[S]) \geq 2$, we have that $s_{\bar i} \leadsto d_i$. Thus, since the removal of $v_p^i$ disconnects $s_{\bar i}$ from $d_i$, we must have at least one node $v_1 \in \I(v_p^i)$ such that $s_{\bar i} \leadsto v_1$. If we suppose by contradiction that $v_1$ is the only such node, then we have that $v_1$ disconnects $s_{\bar i}$ from $d_i$. If $v_1 \in \p{i}i$ we contradict our choice of $v_p^i$. If $v_1 \notin \p{i}i$, then we contradict the fact that $n_i(G[S]) \geq 2$.
\end{nproof}

\vspace{2mm}

The importance of the property in Lemma \ref{lem:two} is that it guarantees that, with probability 1, at least two nodes in $\I(v_p^i)$ will have in their received signal a component which corresponds to the transmitted signal from $s_{\bar i}$. Intuitively, this means that, we can cancel the interference from $s_{\bar i}$ on $\p{i}i$, while still allowing the signal from $s_{\bar i}$ to reach $d_{\bar i}$.
We now consider the case in which we have $n_1(G[S]) \geq 2$ and $n_2(G[S]) = 0$.

\vspace{2mm}


\begin{subsubsection}{$n_1(G[S]) \geq 2$, $n_2(G[S]) = 0$} 
\label{allup}

Notice that in this case only $v_p^1$ is defined. Thus, we will consider the condensed network formed by layers $V_1, V_{\ell(v_p^1)-1}$ and $V_r$, with $m = |V_{\ell(v_p^1)-1}|$. Our condensed network should look like the network in Figure \ref{condallup}.
\begin{figure}[ht]
	\centering
		\includegraphics[height=32mm]{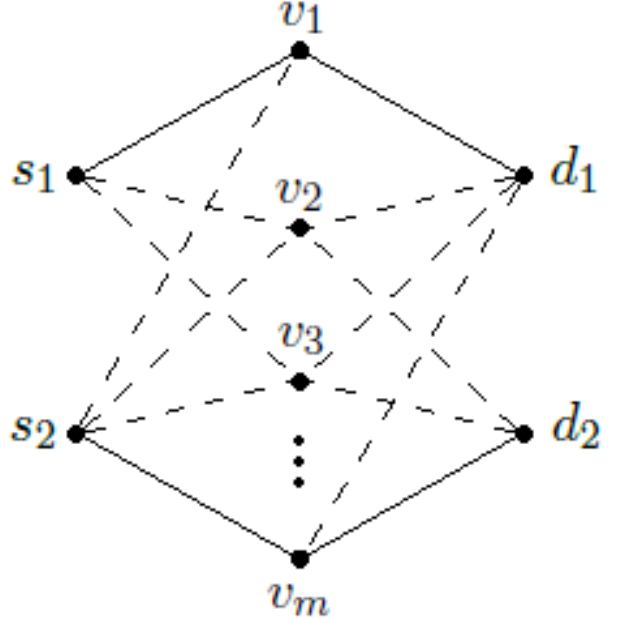}
	\caption{Illustration of a condensed network with $n_1(G[S]) \geq 2$ and $n_2(G[S]) = 0$.}
	\label{condallup}
\end{figure}
The solid lines correspond to edges that must exist in the condensed network, due to the existence of two disjoint paths $\p11$ and $\p22$. The dashed lines correspond to edges that may or may not exist. To each of the nodes $v_i$, $i = 1,...,m$ in the intermediate layer, we associate a variable $x_i$ which will be the scaling factor used by node $v_i$. Our task is to show that the end-to-end transfer matrix, given by
\begin{align}
  T &= \begin{bmatrix}
  \hat h(v_1,d_1) & \hat h(v_2,d_1) & \cdots & \hat h(v_m,d_1) \\
  \hat h(v_1,d_2) & \hat h(v_2,d_2) & \cdots & \hat h(v_m,d_2) \\
\end{bmatrix} \non
& \begin{bmatrix}
  x_1 & 0 & \cdots & 0 \\
  0 & x_2 & \cdots & 0 \\
  \vdots & & \ddots & \vdots \\
  0 & 0 & \cdots & x_m \\
\end{bmatrix}
\begin{bmatrix}
  \hat h(s_1,v_1) & \hat h(s_2,v_1) \\
  \hat h(s_1,v_2) & \hat h(s_2,v_2) \\
  \vdots & \vdots \\
  \hat h(s_1,v_m) & \hat h(s_2,v_m) \\
\end{bmatrix} \nonumber \\
& = \begin{bmatrix}
  T_{1,1} & T_{1,2} \\
  T_{2,1} & T_{2,2} \\
\end{bmatrix},
\label{transfer1}
\end{align}
where $T_{j,k} = \sum_{i=1}^m \hat h(s_j,v_i) \hat h(v_i,d_k) x_i$,
can be made diagonal with non-zero diagonal entries by an appropriate choice of $x_1,...,x_m$. Since, in this case, $n_2(G[S]) = 0$, there is no path from $s_1$ to $d_2$, and therefore we must have $\hat h(s_1,v_i) \hat h(v_i,d_2) = 0$ for $i = 1,...,m$ and $T_{2,1}$ 
is always 0. From the use of Lemma \ref{lem:paths}, we know that for two nodes $v_a, v_b \in \I(v_p^1) \subset V_{\ell(v_p^1)-1}$ with associated variables $x_a$ and $x_b$, we must have two disjoint paths $P_{s_1,v_a}$ and $P_{s_2,v_b}$. From Lemma \ref{lem:two}, we know that there is a node $v_c \in \I(v_p^1) \subset V_{\ell(v_p^1)-1}$, such that $s_2 \leadsto v_c$ and $c \ne m$. We now claim that if the matrices 
\begin{align*} & M_1 = \begin{bmatrix}
  \hat h(s_1,v_a)\hat h(v_a,d_1) & \hat h(s_1,v_b)\hat h(v_b,d_1) \\
  \hat h(s_2,v_a)\hat h(v_a,d_1) & \hat h(s_2,v_b)\hat h(v_b,d_1) \\
\end{bmatrix} \text{ and } \\
& M_2 = \begin{bmatrix}
  \hat h(s_2,v_c) \hat h(v_c,d_1) & \hat h(s_2,v_m) \hat h(v_m,d_1) \\
  \hat h(s_2,v_c) \hat h(v_c,d_2) & \hat h(s_2,v_m) \hat h(v_m,d_2) \\
\end{bmatrix} 
\end{align*}
are both full-rank, then we can choose $x_1, ..., x_m$ so that $T$ is diagonal with non-zero diagonal entries. To see this, we first consider ${\bf x'} = [x_1' \; ... \; x_m']$, where $x_j' = 0$ for $j \ne a,b$, and $[x_a'\; x_b']^T = M_1^{-1} [1 \; 0]^T$. This choice of scaling factors guarantees that $T_{1,1} = 1$ and $T_{1,2} = 0$. If $T_{2,2} \ne 0$ we are done. Otherwise, if $T_{2,2} = 0$, we let ${\bf x''} = [x_1'' \; ... \; x_m'']$, where $x_j'' = 0$ for $j \ne c,m$ and $[x_c''\;x_m'']^T = M_2^{-1} [0 \; 1]^T$. This choice guarantees that $T_{1,2} = 0$ and $T_{2,2} = 1$. If we have $T_{1,1} \ne 0$, we are done. Otherwise, we set ${\bf x'''} = {\bf x'} + {\bf x''}$. By linearity, this choice will guarantee that $T$ is the identity matrix. 

Next we show that, with probability 1, $M_1$ and $M_2$ (which are just functions of the channel gains in the original network) are full-rank. First we consider the transfer matrix between $(s_1,s_2)$ and $(v_a,v_b)$, given by 
\[ Z_1 = \begin{bmatrix}
  \hat h(s_1,v_a) & \hat h(s_2,v_a) \\
  \hat h(s_1,v_b) & \hat h(s_2,v_b) \\
\end{bmatrix}. \]
The determinant of $Z_1$ can be seen as a polynomial where the variables are the channel gains $h_e$ from the original network. All we need to show is that this polynomial is not identically zero. Then, since the $h_e$'s are drawn independently from continuous distributions, $\det Z_1$ will be non-zero with probability 1. To see that this polynomial is not identically zero we notice that the existence of two disjoint paths $P_{s_1,v_a}$ and $P_{s_2,v_b}$ guarantees that, if we set $h_e = 1$ if $e$ connects two consecutive vertices of $P_{s_1,v_a}$ or $P_{s_2,v_b}$ and $h_e = 0$ otherwise, $Z_1$ will be the identity matrix. Therefore, $Z_1$ will be invertible, and thus $\det Z_1$ cannot be identically zero. Now, we  notice that
\begin{align*} 
\det M_1 &= \begin{vmatrix}
  \hat h(s_1,v_a)\hat h(v_a,d_1) & \hat h(s_1,v_b)\hat h(v_b,d_1) \\
  \hat h(s_2,v_a)\hat h(v_a,d_1) & \hat h(s_2,v_b)\hat h(v_b,d_1) \\
\end{vmatrix} \\ & =
\hat h(v_a,d_1)\hat h(v_b,d_1) \begin{vmatrix}
  \hat h(s_1,v_a) & \hat h(s_1,v_b) \\
  \hat h(s_2,v_a) & \hat h(s_2,v_b) \\
\end{vmatrix}  \\
&= \hat h(v_a,d_1)\hat h(v_b,d_1) \det Z_1.
\end{align*}
Since $v_a \leadsto d_1$ and $v_b \leadsto d_1$, we have that $\hat h(v_a,d_1)\hat h(v_b,d_1)$ is also a non-identically zero polynomial in the $h_e$'s, and therefore $M_1$ is invertible with probability 1. To show that $M_2$ is invertible with probability 1, we will follow very similar steps. We notice that the transfer matrix between $(v_c,v_m)$ and $(d_1,d_2)$ is given by
\[ Z_2 = \begin{bmatrix}
  \hat h(v_c,d_1) & \hat h(v_m,d_1) \\
  \hat h(v_c,d_2) & \hat h(v_m,d_2) \\
\end{bmatrix}. \]
Since $v_c \in \I(v_p^1)$ and $v_p^1 \in \p11$, we clearly have two disjoint paths $P_{v_c,d_1} = (v_c,v_p^1) \concat \p11[v_p^1,d_1]$ and $P_{v_m,d_2}= \p22[v_m,d_2]$. This implies that $\det Z_2$ is non-identically zero, and therefore non-zero with probability 1. Then, we notice that
\begin{align*} 
\det M_2 &= \begin{vmatrix}
  \hat h(s_2,v_c) \hat h(v_c,d_1) & \hat h(s_2,v_m) \hat h(v_m,d_1) \\
  \hat h(s_2,v_c) \hat h(v_c,d_2) & \hat h(s_2,v_m) \hat h(v_m,d_2) \\
\end{vmatrix} \\ &= 
\hat h(s_2,v_c) \hat h(s_2,v_m) \begin{vmatrix}
  \hat h(v_c,d_1) & \hat h(v_m,d_1) \\
  \hat h(v_c,d_2) & \hat h(v_m,d_2) \\
\end{vmatrix}  \\
&= \hat h(s_2,v_c) \hat h(s_2,v_m) \det Z_2,
\end{align*}
and, since $s_2 \leadsto v_c$, $s_2 \leadsto v_m$, we have that $\hat h(s_2,v_c) \hat h(s_2,v_m)$ is a non-identically zero polynomial in the $h_e$'s and therefore so is $\det M_2$. This proves that $M_2$ is full-rank with probability 1, and thus we conclude the proof when $n_1(G[S]) \geq 2$, $n_2(G[S]) = 0$. The case where $n_1(G[S]) = 0$, $n_2(G[S]) \geq 2$ follows in the exact same way.


\end{subsubsection}


Next, we consider the cases in which $n_1(G[S]) \geq 2$ and $n_2(G[S]) \geq 2$. We will use  $V_{\ell(v_p^2)-1}$ and $V_{\ell(v_p^1)-1}$ as our key layers. We can assume WLOG that $\ell(v_p^2)\leq \ell(v_p^1)$. We consider the case where $\ell(v_p^2) < \ell(v_p^1)$ and the case where $\ell(v_p^2) = \ell(v_p^1)$ separately.

\vspace{2mm}

\begin{subsubsection}{$n_1(G[S]) \geq 2$, $n_2(G[S]) \geq 2$ and $\ell(v_p^2) < \ell(v_p^1)$}
\label{downup}

We let $m = |V_{\ell(v_p^1)-1}|$ and $n = |V_{\ell(v_p^2)-1}|$. Our condensed network will be of the form shown in Figure \ref{conddownup}a.
\begin{figure}[ht]
     \centering
     \subfigure[]{
       \includegraphics[height=33mm]{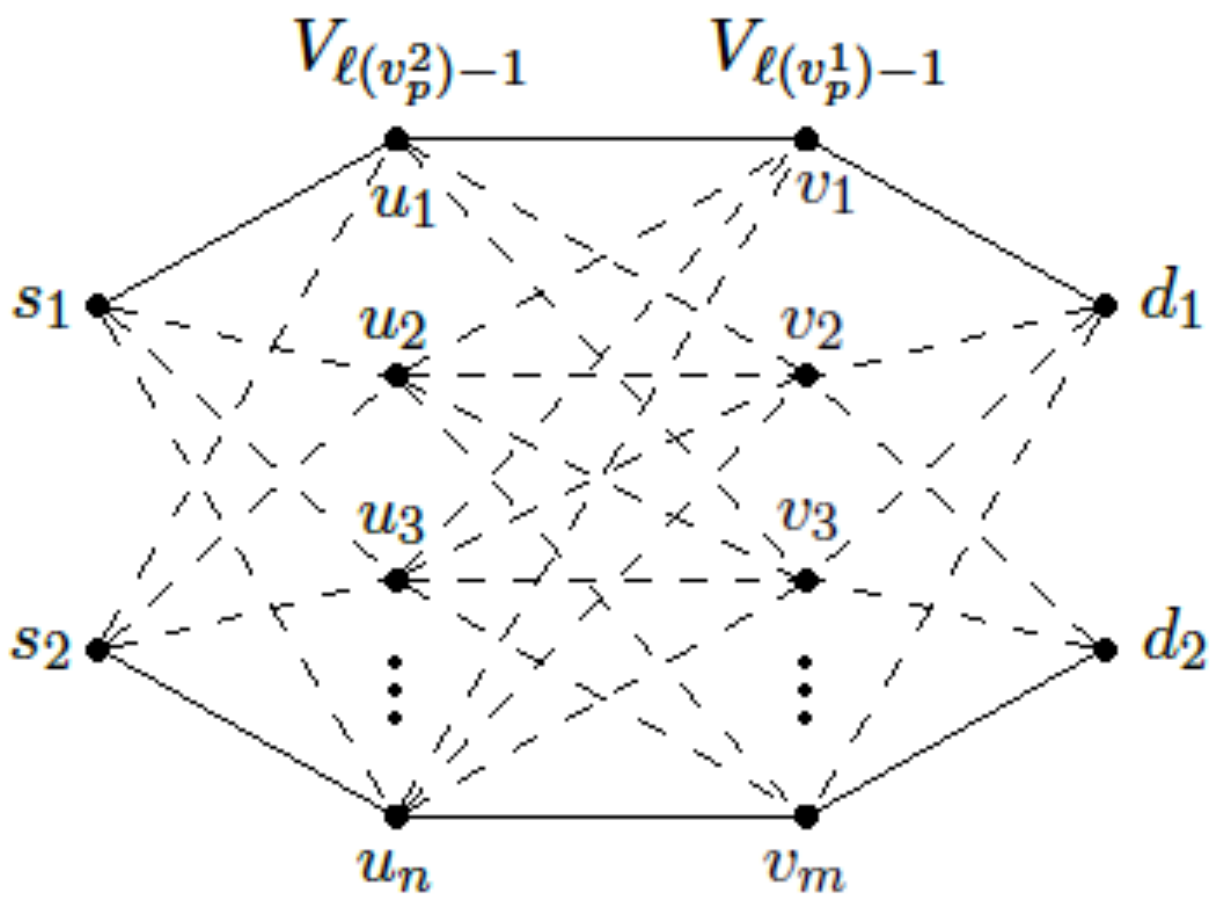} } 
    \hspace{2mm}
    \subfigure[]{
       \includegraphics[height=32mm]{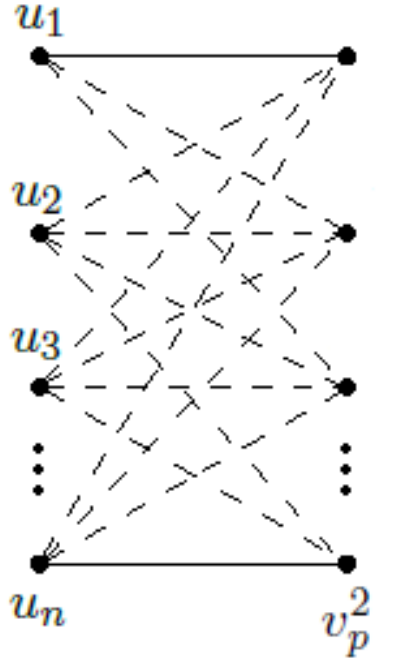}}
	\caption{(a) Illustration of the condensed network in the case where $n_1(G[S]) \geq 2$, $n_2(G[S]) \geq 2$ and $\ell(v_p^2) < \ell(v_p^1)$; (b) Illustration of the connectivity between layers $V_{\ell(v_p^2)-1}$ and $V_{\ell(v_p^2)}$ in the original network.}
	\label{conddownup}
\end{figure}
Once again, the solid lines correspond to edges that must exist in the condensed network, due to the existence of two disjoint paths $\p11$ and $\p22$, and the dashed lines correspond to edges that may or may not exist. We name the nodes in $V_{\ell(v_p^1)-1}$, $v_1, v_2,..., v_m$ and the nodes in $V_{\ell(v_p^2)-1}$, $u_1, u_2, ..., u_n$. Moreover, to each of the nodes $v_i$, $i = 1,...,m$, we associate a variable $x_i$ which will be the scaling factor used by node $v_i$, and to each of the nodes $u_i$, $i = 1,...,n$ we associate a variable $y_i$ which will be the scaling factor used by node $u_i$. 

We will again show that, with probability 1, there is a choice of $x_1,..., x_m$ and $y_1,...,y_n$ such that the effective end-to-end transfer matrix is diagonal with non-zero diagonal entries. This time, however, we will proceed in two steps. First we will show that, with probability 1, we can choose $y_1, ..., y_n$ such that, for some $v_a,v_b \in \I(v_p^1)$, the transfer matrix between $(s_1,s_2)$ and $(v_a, v_b)$ is invertible and the transfer matrix between $(s_1,s_2)$ and $v_m$ is of the form $[0 \; \beta]$ for $\beta \ne 0$. Then, by ``supressing'' the key layer $V_{\ell(v_p^2)-1}$, we will essentially be in the case we described in \ref{allup}, and thus we can choose $x_1,...,x_m$ so that the end-to-end transfer matrix is as desired.

In order to describe how we choose $y_1,...,y_n$ we must first consider the connectivity between the nodes in $V_{\ell(v_p^2)-1}$ and its consecutive layer, $V_{\ell(v_p^2)}$, in the original network. This layer transition can be depicted as in Figure \ref{conddownup}b.
We will now show that, with probability 1, it is possible to choose $y_1, ..., y_n$ all non-zero, such that the transfer matrix $F$ between $(s_1,s_2)$ and $v_p^2$ is of the form $[0 \; \alpha]$ for $\alpha \ne 0$. We first notice that $F$ is given by
\[  \begin{bmatrix}
  \sum_{i=1}^n \hat h(s_1,u_i) h_{(u_i,v_p^2)} y_i & \sum_{i=1}^n \hat h(s_2,u_i) h_{(u_i,v_p^2)} y_i \\
\end{bmatrix}.
\]

From Lemma \ref{lem:two}, we know that there are at least two nodes $u_c,u_d \in \I(v_p^2)$ such that $s_1 \leadsto u_c$ and $s_1 \leadsto u_d$. This implies that $\hat h(s_1,u_c) h_{(u_c,v_p^2)}$ and $\hat h(s_1,u_d) h_{(u_d,v_p^2)}$, if viewed as polynomials on the channel gains, are not identically zero. Thus, with probability 1, they will be non-zero, and $\sum_{i=1}^n \hat h(s_1,u_i) h_{(u_i,v_p^2)} y_i$ will have non-zero coefficients in front of $y_c$ and $y_d$. This means that we can choose ${\bf y'} = (y_1',...,y_n')$, with $y_1',...,y_n'$ all non-zero, so that $F_1 = \sum_{i=1}^n \hat h(s_1,u_i) h_{(u_i,v_p^2)} y_i' = 0$. If we have $F_2 = \sum_{i=1}^n \hat h(s_2,u_i) h_{(u_i,v_p^2)} y_i' \ne 0$, then we are done. Otherwise, if $F_2 = 0$, we proceed as follows. From Lemma \ref{lem:paths}, we know that we can choose $u_a, u_b \in \I(v_p^2) \subset V_{\ell(v_p^2)-1}$ so that we have two disjoint paths $P_{s_1,u_a}$ and $P_{s_2,u_b}$. Therefore, the transfer matrix between $(s_1,s_2)$ and $(u_a,u_b)$, given by 
\[ K =  \begin{bmatrix}
  \hat h(s_1,u_a) & \hat h(s_1,u_b) \\
  \hat h(s_2,u_a) & \hat h(s_2,u_b) \\
\end{bmatrix}, \]
is full-rank with probability 1. This also implies that the matrix
\[ M =  \begin{bmatrix}
  \hat h(s_1,u_a)h_{(u_a,v_p^2)} & \hat h(s_1,u_b)h_{(u_b,v_p^2)} \\
  \hat h(s_2,u_a)h_{(u_a,v_p^2)} & \hat h(s_2,u_b)h_{(u_b,v_p^2)} \\
\end{bmatrix} \]
is full-rank with probability 1, because we have $\det M = h_{(u_a,v_p^2)} h_{(u_b,v_p^2)} \det K$, and, since $u_a, u_b \in \I(v_p^2)$, we have that $h_{(u_a,v_p^2)} h_{(u_b,v_p^2)}$ is non-zero with probability 1. The matrix $M$ allows us to build ${\bf y''} = (y_1'',...,y_n'')$ by setting $y_i''=0$, for $i \ne a,b$, and $[y_a''\; y_b'']^T = M^{-1} [ 0 \; 1]^T$. 
This choice guarantees that $F = [0 \; 1]$ as desired, but we do not have $y_1'',...,y_n''$ all non-zero. However, it is easy to see that if we set ${\bf y'''} = {\bf y''} + \alpha {\bf y'}$, for some $\alpha \ne 0$, we will have $y_1''',...,y_n'''$ all non-zero and $F = [0 \; \alpha]$. 

We conclude that we can choose $y_1,...,y_n$ all non-zero and have $F = [0 \; \alpha]$ with $\alpha \ne 0$. Moreover, since there exists a path from $v_p^2$ to $v_m$, and there exists no path from $s_1$ to $v_m$ which does not contain $v_p^2$, we conclude that, with probability 1, our choice of $y_1,...,y_m$ will make the transfer matrix from $(s_1,s_2)$ to $v_m$ be of the form $[0 \; \beta]$ for $\beta \ne 0$.

Next, we would like to prove that, with this choice of $y_1,...,y_m$, there exist nodes $v_a, v_b \in \I(v_p^1)$, such that the transfer matrix between $(s_1,s_2)$ and $(v_a, v_b)$ is full-rank. First, we notice that, from Lemma \ref{lem:paths}, there exist two nodes $v_e, v_f \in \I(v_p^1)$, such that we have two disjoint paths $P_{s_1,v_e}$ and $P_{s_2,v_f}$. However, we cannot proceed as before to conclude that the transfer matrix between $(s_1,s_2)$ and $(v_e,v_f)$ is full-rank with probability 1, because our variables $y_1,...,y_m$ were not chosen independently from the channel gains. Nonetheless, if we let $\tilde H$ be the set of all 
$h_{(u_j,v_p^2)}$ for $j=1,...,n$ and all the channel gains that appear in $\hat h(s_i,u_j)$, for $i=1,2$ and  $j=1,...,n$, we notice that our choice of $y_1,...,y_n$ only depends on $\tilde H$. Therefore, we assume that all the channel gains in $\tilde H$ are drawn according to their distributions, and are from now on viewed as constants. Then, we can also fix $y_1,...,y_n$, following the steps described previously, and view them as constants.

First, we assume that neither $P_{s_1,v_e}$ nor $P_{s_2,v_f}$ contain $v_p^2$. In this case we will show that we can set $v_a = v_e$ and $v_b = v_f$. The determinant of the transfer matrix between $(s_1,s_2)$ and $(v_e,v_f)$ can be seen as a polynomial where the variables are the channel gains which are not in $\tilde H$. Notice that all the channel gains not in $\tilde H$ are still independent (since the choice of $y_1,...,y_m$ was made independent of them) and have a continuous distribution. Thus, we will show that, with probability 1 over the choice of the channel gains in $\tilde H$, there exists a choice of the channel gains which are not in $\tilde H$, such that the transfer matrix between $(s_1,s_2)$ and $(v_e,v_f)$ is invertible. Therefore, the determinant of the transfer matrix between $(s_1,s_2)$ and $(v_e,v_f)$ is not identically zero, and will be non-zero with probability 1 over the choice of the channel gains not in $\tilde H$. 

Since $P_{s_1,v_e}$ and $P_{s_2,v_f}$ are disjoint, there are distinct nodes $u_e$ and $u_f$ in $V_{\ell(v_p^2)-1}$, such that $u_e \in P_{s_1,v_e}$ and $u_f \in P_{s_1,v_f}$. For any $h_e \notin \tilde H$, we will set $h_e = 1$ if $e$ connects two consecutive vertices of $P_{s_1,v_e}$ or $P_{s_2,v_f}$ and $h_e = 0$ otherwise. Therefore, the transfer matrix between $(u_e,u_f)$ and $(v_e,v_f)$ is the identity matrix. Thus, we have that the transfer matrix between $(s_1,s_2)$ and $(v_e,v_f)$ is given by
\begin{equation}
\begin{bmatrix}
  1 & 0 \\
  0 & 1 \\
\end{bmatrix}
\begin{bmatrix}
  y_e & 0 \\
  0 & y_f \\
\end{bmatrix} 
\begin{bmatrix}
  \hat h(s_1,u_e) & \hat h(s_2,u_e) \\
  \hat h(s_1,u_f) & \hat h(s_2,u_f) \\
\end{bmatrix}. \label{trasfef} 
\end{equation}
The existence of disjoint paths $P_{s_1,v_e}$ and $P_{s_2,v_f}$ implies the existence of disjoint paths $P_{s_1,u_e}=P_{s_1,v_e}[s_1,u_e]$ and $P_{s_2,u_f}=P_{s_2,v_f}[s_2,u_f]$. Therefore, with probability 1 over the choice of the channel gains in $\tilde H$ (since they were drawn independently first, according to their continuous distributions), 
\[ \begin{bmatrix}
  \hat h(s_1,u_e) & \hat h(s_2,u_e) \\
  \hat h(s_1,u_f) & \hat h(s_2,u_f) \\
\end{bmatrix} \]
is full-rank. Therefore, since we chose $y_e$ and $y_f$ to be non-zero, the transfer matrix in (\ref{trasfef}) must be full-rank, which implies that the transfer matrix between $(s_1,s_2)$ and $(v_e,v_f)$ is full-rank with probability 1 if $y_1,...,y_n$ are chosen as described above.

Now, we consider the situations in which either $P_{s_1,v_e}$ or $P_{s_2,v_f}$ contains $v_p^2$. We will show that, in any case, for some $v_g,v_h \in \I(v_p^1)$, we can find either 
\begin{enumerate}[i. ]
\item two other disjoint paths $P_{s_1,v_g}$ and $P_{s_2,v_h}$ not containing $v_p^2$, or 
\item two disjoint paths $P_{s_1,v_g}$ and $P_{v_p^2,v_h}$. 
\end{enumerate}

If we suppose $v_p^2 \in P_{s_2,v_f}$, then we are clearly in case ii, by setting $g = e$ and $h = f$, and setting $P_{v_p^2,v_h} = P_{s_2,v_f}[v_p^2,v_f]$. Thus, we suppose that $v_p^2 \in P_{s_1,v_e}$. If we let $w_f$ be the node from $P_{s_2,v_f}$ in the layer containing $v_p^2$, we have two disjoint paths $P_{s_1,v_p^2} = P_{s_1,v_e}[s_1,v_p^2]$ and $P_{s_2,w_f}=P_{s_2,v_f}[s_2,w_f]$. We also let $w_1$ be the node from $\p11$ in $V_{\ell(v_p^2)}$. Then we let $v_l$ be the last common node between $P_{s_2,w_f}$ and $\p11 \cup \p22$. If $v_l \in \p11$ (Figure \ref{zigzag}a), we must have disjoint paths $P_{s_1,w_f}$ and $P_{s_2,v_p^2}$. This implies that we have a path $P_{s_1,v_f} = P_{s_1,w_f} \concat P_{s_2,v_f}[w_f,v_f]$ and a path $P_{v_p^2,v_e} = P_{s_1,v_e}[v_p^2,v_e]$ which are disjoint, and we are in case ii. Note that this case also includes $w_f = w_1$. If, instead, $v_l \in \p22$ (Figure \ref{zigzag}b), we must have disjoint paths $P_{s_1,w_1}$ and $P_{s_2,w_f}$.
\begin{figure}[ht]
     \centering
     \subfigure[$v_l \in \p11$]{
       \includegraphics[height=23mm]{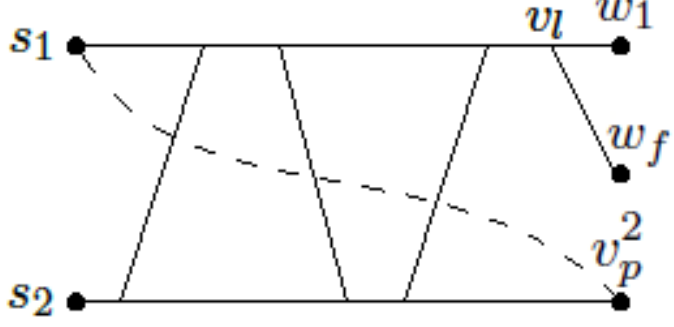} } 
    \hspace{1mm}
    \subfigure[$v_l \in \p11$]{
       \includegraphics[height=23mm]{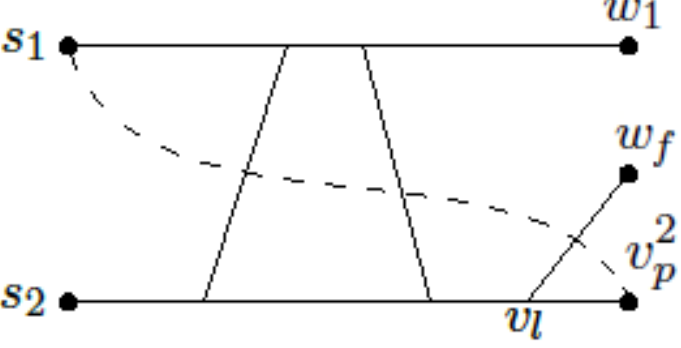}}
     \caption{Illustration of the two possible cases for $v_l$ (the last common node between $P_{s_2,u_f}$ and $\p11 \cup \p22$).} \label{zigzag}
\end{figure}
We also clearly have two disjoint paths $P_{v_p^2,v_e} = P_{s_1,v_e}[v_p^2,v_e]$ and $P_{w_f,v_f} = P_{s_2,v_f}[w_f,v_f]$. 
Thus, we let $v_r$ be the first common node between $\p11$ and $P_{v_p^2,v_e} \cup P_{w_f,v_f}$. If $v_r \in P_{w_f,v_f}$ (Figure \ref{mess}a), then we have two disjoint paths $P_{s_1,v_f} = \p11[s_1,v_r]\concat P_{w_f,v_f}[v_r,v_f]$ and $P_{v_p^2,v_e}$. Therefore, we are in case ii. If $v_r \in P_{v_p^2,v_e}$ (Figure \ref{mess}b), then we have two disjoint paths $P_{w_1,v_e} = P_{s_1,d_1}[w_1,v_r] \concat P_{v_p^2,v_e}[v_r,v_e]$ and $P_{w_f,v_f}$. Therefore, we can 
build two disjoint paths $P_{s_1,v_e}' = P_{s_1,w_1} \concat P_{w_1,v_e}$ and $P_{s_2,v_f}' = P_{s_2,w_f} \concat P_{w_f,v_f}$ not containing $v_p^2$, and we are in case i. Finally, if $v_r$ does not exist, we clearly have the disjoint paths $P_{s_1,d_1}[s_1,v_1]$ and $P_{v_p^2,v_e}$, and, since $v_1 \in \I(v_p^1)$, we are in case ii. 
\begin{figure}[ht]
     \centering
     \subfigure[$v_l \in \p11$]{
       \includegraphics[height=23mm]{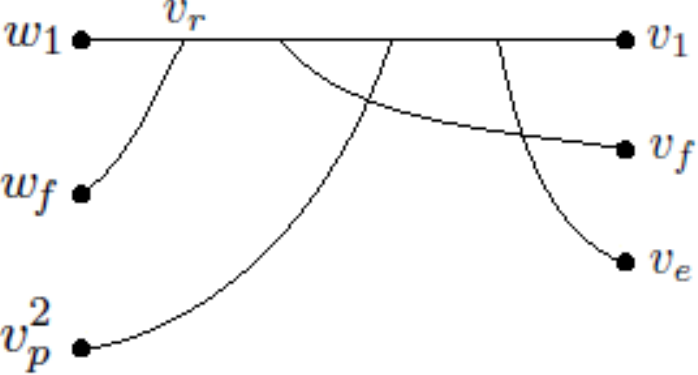} } 
    \hspace{1mm}
    \subfigure[$v_l \in \p11$]{
       \includegraphics[height=23mm]{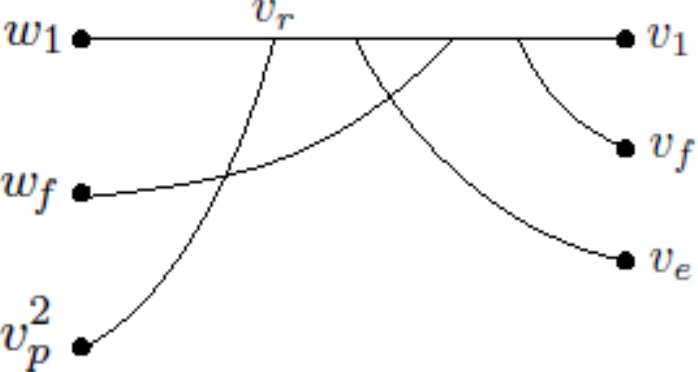}}
     \caption{Illustration of the two possible cases for $v_r$ (the first common node between $\p11$ and $P_{v_p^2,v_e} \cup P_{w_f,v_f}$.} \label{mess}
\end{figure}

Since case i was already taken care of, we only need to consider case ii. We will show that, if we have two disjoint paths $P_{s_1,v_g}$ and $P_{v_p^2,v_h}$, and if we choose $y_1,...,y_n$ as described previously, then, for some $v_g,v_h \in \I(v_p^1)$, the transfer matrix between $(s_1,s_2)$ and $(v_g,v_h)$ will be full-rank with probability 1 over the choice of the channel gains not in $\tilde H$. We will look at the determinant of the transfer matrix between $(s_1,s_2)$ and $(v_g,v_h)$ as a polynomial on the channel gains not in $\tilde H$, since the channel gains in $\tilde H$ and the scaling factors $y_1,...,y_n$ have already been fixed. Then we can show that this determinant is not identically zero by showing that for a specific choice of the channel gains not in $\tilde H$, the transfer matrix between $(s_1,s_2)$ and $(v_g,v_h)$ is full-rank. For any $h_e$ not in $\tilde H$, we will choose $h_e = 1$ if $e$ is connecting two consecutive vertices of $P_{s_1,v_g}$ or $P_{v_p^2,v_h}$, and $h_e = 0$ otherwise. This means that the transfer matrix between $(u_g,v_p^2)$ and $(v_g,v_h)$ is the identity matrix. Then, if we let $u_g$ be the node from $P_{s_1,v_g}$ in layer $V_{\ell(v_p^2)-1}$, the transfer matrix between $(s_1,s_2)$ and $(v_g,v_h)$ is given by
\begin{align}
& \begin{bmatrix}
  1 & 0 \\
  0 & 1 \\
\end{bmatrix}
\begin{bmatrix}
  y_g & 0 \\
  0 & 1 \\
\end{bmatrix} \non
& \begin{bmatrix}
  \hat h(s_1,u_g) & \hat h(s_2,u_g) \\
	\sum_{i=1}^n \hat h(s_1,u_i) h_{u_i,v_p^2} y_i & \sum_{i=1}^n \hat h(s_2,u_i) h_{u_i,v_p^2} y_i
\end{bmatrix} \nonumber \\
& \quad \quad \quad = 
\begin{bmatrix}
  y_g \hat h(s_1,u_g) & y_g \hat h(s_2,u_g) \\
	0 & \alpha
\end{bmatrix},
 \label{trasfgh} 
\end{align}
where we used the fact that our choice of $y_1, ..., y_n$ guarantees that the transfer matrix between $(s_1,s_2)$ and $v_p^2$ is $[0 \; \alpha]$ for some $\alpha \ne 0$. Now, since there exists a path from $s_1$ to $u_g$, $\hat h(s_1,u_g)$ is non-zero with probability 1 over the choice of the channel gains in $\tilde H$. Therefore, since $y_g$ was chosen to be non-zero, the transfer matrix in (\ref{trasfgh}) is upper-triangular (with non-zero diagonal entries) and thus full-rank. 

Therefore we proved that we can find $v_a,v_b \in \I(v_p^1)$ so that the transfer matrix between $(s_1,s_2)$ and $(v_a,v_b)$ is full-rank with probability 1, after the choice of the scaling factors $y_1,..., y_m$. Next, we consider supressing the layer $V_{\ell(v_p^2)-1}$ from the condensed network by incorporating our choice of $y_1, ..., y_n$ into the terms $\hat h(s_i,v_j)$ for $i = 1,2$ and $j=1,...,m$. We will show that the resulting condensed network is equivalent to the one considered in \ref{allup}. As in \ref{allup}, the end-to-end transfer matrix can now be written as
\begin{align}
\begin{bmatrix}
  \sum_{i=1}^m \hat h(s_1,v_i) \hat h(v_i,d_1) x_i & \sum_{i=1}^m \hat h(s_2,v_i) \hat h(v_i,d_1) x_i \\
  \noalign{\medskip}
  \sum_{i=1}^m \hat h(s_1,v_i) \hat h(v_i,d_2) x_i & \sum_{i=1}^m \hat h(s_2,v_i) \hat h(v_i,d_2) x_i \\
\end{bmatrix}.  \label{transfer2}
\end{align}

As we noted before, the transfer matrix between $(s_1,s_2)$ and $v_m$ is of the form $[0 \; \beta]$ for some $\beta \ne 0$. This implies that $\hat h(s_1,v_m) = 0$ and $\hat h(s_2,v_m) = \beta \ne 0$. Moreover, since $v_p^2$ disconnects $d_2$ from $s_1$, we conclude that $\hat h(s_1,v_i)\hat h(v_i,d_2) = 0$ for $i=2,...,m$. Otherwise, this would either imply the existence of a path between $s_1$ and $d_2$ not containing $v_p^2$ or contradict the fact that the transfer matrix between $(s_1,s_2)$ and $v_p^2$ is of the form $[0 \; \alpha]$. Thus, we conclude that $\sum_{i=1}^m \hat h(s_1,v_i) \hat h(v_i,d_2) x_i = 0$. As shown in \ref{allup}, if we can find $v_a,v_b$ and $v_c$, $c \ne m$, such that the matrices
\begin{align*}
&M_1 = \begin{bmatrix}
  \hat h(s_1,v_a)\hat h(v_a,d_1) & \hat h(s_1,v_b)\hat h(v_b,d_1) \\
  \hat h(s_2,v_a)\hat h(v_a,d_1) & \hat h(s_2,v_b)\hat h(v_b,d_1) \\
\end{bmatrix} \text{ and } \\
& M_2 = \begin{bmatrix}
  \hat h(s_2,v_c) \hat h(v_c,d_1) & \hat h(s_2,v_m) \hat h(v_m,d_1) \\
  \hat h(s_2,v_c) \hat h(v_c,d_2) & \hat h(s_2,v_m) \hat h(v_m,d_2) \\
\end{bmatrix} 
\end{align*}
are both full-rank, then it is possible to choose $x_1,...,x_m$ so that the end-to-end transfer matrix in (\ref{transfer2}) is diagonal with non-zero diagonal entries. We will choose $v_a$ and $v_b$ to be the two nodes in $\I(v_p^1)$ for which the transfer matrix from $(s_1,s_2)$ to $(v_a,v_b)$
\[ \begin{bmatrix}
  \hat h(s_1,v_a) & \hat h(s_2,v_a) \\
  \hat h(s_1,v_b) & \hat h(s_2,v_b) \\
\end{bmatrix} \]
 is full-rank with probability 1. Then, we will notice that 
\begin{align*} 
\det M_1 &= \begin{vmatrix}
  \hat h(s_1,v_a)\hat h(v_a,d_1) & \hat h(s_1,v_b)\hat h(v_b,d_1) \\
  \hat h(s_2,v_a)\hat h(v_a,d_1) & \hat h(s_2,v_b)\hat h(v_b,d_1) \\
\end{vmatrix} \\ & = 
\hat h(v_a,d_1)\hat h(v_b,d_1) \begin{vmatrix}
  \hat h(s_1,v_a) & \hat h(s_1,v_b) \\
  \hat h(s_2,v_a) & \hat h(s_2,v_b) \\
\end{vmatrix}. 
\end{align*}
Since $v_a,v_b \in \I(v_p^1)$, we have that $v_a \leadsto d_1$ and $v_b \leadsto d_1$, and $\hat h(v_a,d_1)\hat h(v_b,d_1)$ is non-zero with probability 1. Therefore, $M_1$ is invertible with probability 1. 

As we did in \ref{allup}, we use Lemma \ref{lem:two} to guarantee that we can choose $v_c \in \I(v_p^1)$ such that $s_2 \leadsto v_c$ and $c \ne m$. Then, we notice that the transfer matrix between $(v_c,v_m)$ and $(d_1,d_2)$ is given by
\begin{equation} \begin{bmatrix}
  \hat h(v_c,d_1) & \hat h(v_m,d_1) \\
  \hat h(v_c,d_2) & \hat h(v_m,d_2) \\
\end{bmatrix}. \label{trans3} \end{equation}
Since $v_c \in \I(v_p^1)$ and $v_p^1 \in \p11$, we clearly have two disjoint paths $P_{v_c,d_1}$ and $P_{v_m,d_2}$. This implies that the transfer matrix in (\ref{trans3}) is invertible with probability 1. Then, we notice that
\begin{align*} 
\det M_2 &= \begin{vmatrix}
  \hat h(s_2,v_c) \hat h(v_c,d_1) & \hat h(s_2,v_m) \hat h(v_m,d_1) \\
  \hat h(s_2,v_c) \hat h(v_c,d_2) & \hat h(s_2,v_m) \hat h(v_m,d_2) \\
\end{vmatrix} \\ &= 
\hat h(s_2,v_c) \hat h(s_2,v_m) \begin{vmatrix}
  \hat h(v_c,d_1) & \hat h(v_m,d_1) \\
  \hat h(v_c,d_2) & \hat h(v_m,d_2) \\
\end{vmatrix}.  
\end{align*}
As we noticed before, our choice of $y_1,...,y_n$ guarantees that $\hat h(s_2,v_m) = \beta \ne 0$. Since $s_2 \leadsto v_c$, there must be at least one path $P_{s_2,v_c}$. If $P_{s_2,v_c}$ does not contain $v_p^2$, then the fact that we chose $y_1,...,y_m$ to be non-zero guarantees that $\hat h(s_2,v_c)$ is non-zero with probability 1. If $P_{s_2,v_c}$ contains $v_p^2$, then the fact that the transfer matrix between $(s_1,s_2)$ and $v_p^2$ is $[0 \; \alpha]$ for $\alpha \ne 0$ guarantees that $\hat h(s_2,v_c)$ is non-zero with probability 1. Either way, we conclude that $M_2$ is invertible with probability 1. This concludes the proof when $n_1(G[S]) \geq 2$, $n_2(G[S]) \geq 2$ and $\ell(v_p^1) > \ell(v_p^2)$.

\end{subsubsection}

\vspace{3mm}

Next, we consider the situations in which $\ell(v_p^1) = \ell(v_p^2)$. In this case, our condensed network will only contain three layers, $V_1$, $V_{\ell(v_p^1)-1} = V_{\ell(v_p^2)-1}$ and $V_r$. We will use two different approaches, depending on the size of $V_{\ell(v_p^1)-1}$.

\begin{subsubsection}{$n_1(G[S]) \geq 2$, $n_2(G[S]) \geq 2$, $\ell(v_p^2) = \ell(v_p^1)$ and $|V_{\ell(v_p^1)-1}| = 2$}
\label{onelayertwo}

Our condensed network should look like the network in Figure \ref{condonelayertwo}.
\begin{figure}[ht]
	\centering
		\includegraphics[height=23mm]{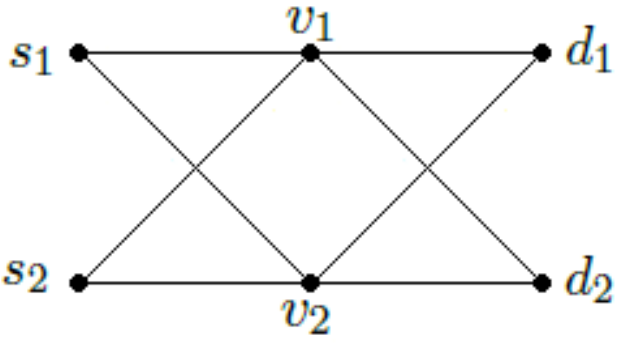}
	\caption{Illustration of the condensed network for the case where $n_1(G[S]) \geq 2$, $n_2(G[S]) \geq 2$, $\ell(v_p^2) = \ell(v_p^1)$ and $|V_{\ell(v_p^1)-1}| = 2$.}
	\label{condonelayertwo}
\end{figure}
The nodes in $V_{\ell(v_p^1)-1}$ are named according to Figure \ref{condonelayertwo}. We notice that all the edges in the condensed network must in fact exist. This can be justified as follows. Lemma \ref{lem:paths} guarantees that $|\I(v_p^1)| \geq 2$ and $|\I(v_p^2)| \geq 2$. Thus we must have $\I(v_p^1)=\I(v_p^2)=\{v_1,v_2\}$, which justifies the existence of edges $(v_i,d_j)$ for $i \in \{1,2\}$ and $j \in \{1,2\}$. Moreover, from Lemma \ref{lem:two}, we have that there must be two distinct nodes $v_a,v_b$ in $\I(v_p^1)$ such that $s_2 \leadsto v_a$ and $s_2 \leadsto v_b$. This justifies the existence of $(s_2,v_1)$ and $(s_2,v_2)$. Similarly, we can apply Lemma \ref{lem:two} to $v_p^2$ to justify the existence of $(s_1,v_2)$ and $(s_1,v_1)$. 

The edge structure of the condensed network guarantees that, with probability 1, the transfer matrix between $(s_1,s_2)$ and $(v_1,v_2)$ and the transfer matrix between $(v_1,v_2)$ and $(d_1,d_2)$, given respectively by 
\[
\begin{bmatrix}
  \hat h(s_1,v_1) & \hat h(s_2,v_1)  \\
  \hat h(s_1,v_2) & \hat h(s_2,v_2)  \\
\end{bmatrix} \text{ and }
\begin{bmatrix}
  \hat h(v_1,d_1) & \hat h(v_2,d_1)  \\
  \hat h(v_1,d_2) & \hat h(v_2,d_2)  \\
\end{bmatrix},
\]
have only non-zero entries. Furthermore, from our previous discussions, we know that the existence of disjoint paths $P_{s_1,v_1}=\p11[s_1,v_1]$ and $P_{s_2,v_2}=\p22[s_2,v_2]$ guarantees that the transfer matrix between $(s_1,s_2)$ and $(v_1,v_2)$ is full-rank with probability 1. Similarly, the existence of disjoint paths $P_{v_1,d_1}$ and $P_{v_2,d_2}$ guarantees that the transfer matrix between $(v_1,v_2)$ and $(d_1,d_2)$ is full-rank with probability 1. Therefore, we essentially have the $2 \times 2 \times 2$ interference channel described in \cite{xx}. The only difference is that additive noises at $v_1, v_2, d_1$ and $d_2$ are not independent and Gaussian. However, they still have a variance which does not depend on the power $P$ (only on the channel gains), and thus the same scheme described in \cite{xx} will achieve $\dE = 2$.

\end{subsubsection}

\begin{subsubsection}{$n_1(G[S]) \geq 2$, $n_2(G[S]) \geq 2$, $\ell(v_p^2) = \ell(v_p^1)$ and $|V_{\ell(v_p^1)-1}| \geq 3$}
\label{onelayer}

In this case, our condensed network is shown in Figure \ref{condonelayer}.
\begin{figure}[ht]
	\centering
		\includegraphics[height=33mm]{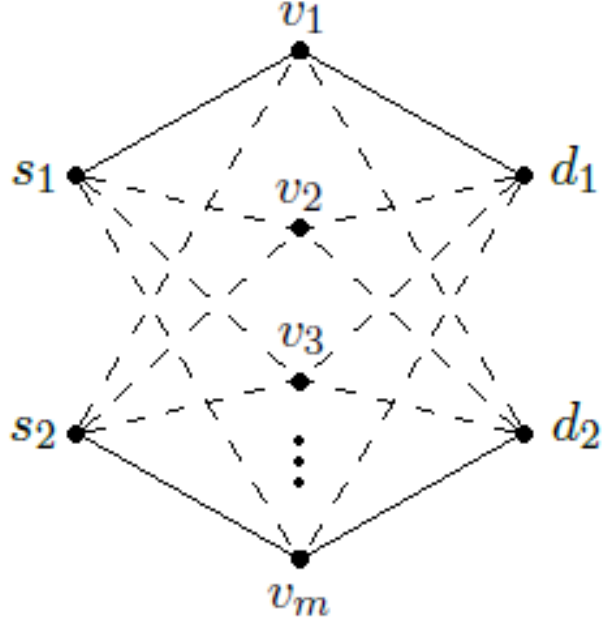}
	\caption{Illustration of the condensed network for the case where $n_1(G[S]) \geq 2$, $n_2(G[S]) \geq 2$, $\ell(v_p^2) = \ell(v_p^1)$ and $|V_{\ell(v_p^1)-1}| \geq 3$.}
	\label{condonelayer}
\end{figure}
Once again we let $v_1,...,v_m$ be the nodes in $V_{\ell(v_p^1)-1} = V_{\ell(v_p^2)-1}$, and to each of the nodes $v_i$, $i = 1,...,m$, we associate a variable $x_i$ which will be the scaling factor used by node $v_i$. We will show that the end-to-end transfer matrix, given by
\begin{align}
\begin{bmatrix}
  \sum_{i=1}^m \hat h(s_1,v_i) \hat h(v_i,d_1) x_i & \sum_{i=1}^m \hat h(s_2,v_i) \hat h(v_i,d_1) x_i \\
  \noalign{\medskip}
  \sum_{i=1}^m \hat h(s_1,v_i) \hat h(v_i,d_2) x_i & \sum_{i=1}^m \hat h(s_2,v_i) \hat h(v_i,d_2) x_i \\
\end{bmatrix},  \label{transfer3}
\end{align}
can be made diagonal with non-zero diagonal entries by an appropriate choice of $x_1,...,x_m$. First, we notice that we can assume that, in the original network, any layer $V_i$ for $i \geq \ell(v_p^1)$ only contains two nodes. This is because any node in such layer $V_i$ which is not in $\p11$ nor $\p22$ can be removed since it may not contribute to $n_1(G[S])$ nor $n_2(G[S])$ (or that would contradict the fact that $v_p^1$ disconnects $d_1$ from $s_2$ and $v_p^2$ disconnects $d_2$ from $s_1$). Therefore, the edge configuration between $V_{\ell(v_p^1)-1}$ and $\{d_1,d_2\}$ in the condensed network is the same as the edge configuration between $V_{\ell(v_p^1)-1}$ and $V_{\ell(v_p^1)}$ in the original network. It is then easy to see that each $\hat h(v_i,d_j)$, for $i=1,...,m$ and $j=1,2$, when seen as a polynomial in the channel gains, is composed of a single product of variables $h_e$, one of which is not shared by any other $\hat h$. 


Next we claim that if we can find two sets of nodes $\{v_a, v_b, v_c\} \subset V_{\ell(v_p^1)-1}$ and $\{v_d,v_e,v_f\} \subset V_{\ell(v_p^1)-1}$, such that the matrices
\begin{align} 
&M_1 = \left[ \begin{array}{cc}
  \hat h(s_1,v_a)\hat h(v_a,d_1) & \hat h(s_1,v_b)\hat h(v_b,d_1) \\ 
  \hat h(s_2,v_a)\hat h(v_a,d_1) & \hat h(s_2,v_b)\hat h(v_b,d_1) \\ 
  \hat h(s_1,v_a)\hat h(v_a,d_2) & \hat h(s_1,v_b)\hat h(v_b,d_2) \\ 
\end{array} \right. \non
& \quad \quad \quad\quad \quad \left. \begin{array}{c}
  \hat h(s_1,v_c)\hat h(v_c,d_1) \\
  \hat h(s_2,v_c)\hat h(v_c,d_1) \\
  \hat h(s_1,v_b)\hat h(v_b,d_2) \\ 
\end{array} \right]
\text{ and } \nonumber \\
& M_2 = \left[\begin{array}{cc}
  \hat h(s_2,v_d)\hat h(v_d,d_2) & \hat h(s_2,v_e)\hat h(v_e,d_2) \\ 
  \hat h(s_1,v_d)\hat h(v_d,d_2) & \hat h(s_1,v_e)\hat h(v_e,d_2) \\ 
  \hat h(s_2,v_d)\hat h(v_d,d_1) & \hat h(s_2,v_e)\hat h(v_e,d_1) \\ 
\end{array} \right. \non
& \quad \quad \quad\quad \quad \left. \begin{array}{c}
  \hat h(s_2,v_f)\hat h(v_f,d_2) \\
  \hat h(s_1,v_f)\hat h(v_f,d_2) \\
  \hat h(s_2,v_f)\hat h(v_f,d_1) \\
\end{array} \right] \nonumber
\end{align}
are full-rank, then we can choose $x_1, ..., x_m$ such that the transfer matrix in (\ref{transfer3}) is diagonal with non-zero diagonal entries. To see this, suppose $M_1$ and $M_2$ are full-rank. Then we can set ${\bf x'} = [x_1' \; ... \; x_m']$, where $x_j' = 0$ for $j \ne a,b,c$, and $[x_a'\; x_b' \; x_c' ]^T = M_1^{-1} [1 \; 0 \; 0]^T$. This guarantees that the transfer matrix in (\ref{transfer3}) is of the form 
\[ 
\begin{bmatrix}
  1 & 0 \\
  0 & \gamma \\
\end{bmatrix}.
\]
If $\gamma \ne 0$, we achieve our goal with ${\bf x'}$. If $\gamma = 0$, we set ${\bf x''} = [x_1'' \; ... \; x_m'']$, where $x_j'' = 0$ for $j \ne d,e,f$, and $[x_d'\; x_e' \; x_f' ]^T = M_2^{-1} [1 \; 0 \; 0]^T$. This guarantees that the transfer matrix in (\ref{transfer3}) is of the form 
\[ 
\begin{bmatrix}
  \delta & 0 \\
  0 & 1 \\
\end{bmatrix}.
\]
If $\delta \ne 0$, we achieve our goal with ${\bf x''}$. If $\delta = 0$, then we let ${\bf x'''} = {\bf x'} + {\bf x''}$, and the transfer matrix in (\ref{transfer3}) becomes the identity matrix. 

Next, we show that we can either find $\{v_a, v_b, v_c\}$ and $\{v_d,v_e,v_f\}$ as described above, or we can remove nodes from $V_{\ell(v_p^1)-1}$ so that we have a $2 \times 2 \times 2$ interference channel (case \ref{onelayertwo}). We start by applying Lemma \ref{lem:paths} to $v_p^1$. Then we can find $u_a, u_b \in \I(v_p^1)$ so that there are two disjoint paths $P_{s_1,u_a}$ and $P_{s_2,u_b}$. Then, from Lemma \ref{lem:two} applied to $v_p^2$, we know that there exist nodes $u_c, u_d \in \I(v_p^2)$ such that $s_1 \leadsto u_c$ and $s_1 \leadsto u_d$. Suppose $\{u_a,u_b\} \ne \{u_c,u_d\}$. Then we can assume WLOG that $u_c \ne u_a,u_b$.  We choose $\{v_a,v_b,v_c\}=\{u_a,u_b,u_c\}$ and we have 
{\small
\begin{align*} 
&\det M_1 = \\ &\begin{vmatrix}
  \hat h(s_1,u_a)\hat h(u_a,d_1) & \hat h(s_1,u_b)\hat h(u_b,d_1) & \hat h(s_1,u_c)\hat h(u_c,d_1) \\
  \hat h(s_2,u_a)\hat h(u_a,d_1) & \hat h(s_2,u_b)\hat h(u_b,d_1) & \hat h(s_2,u_c)\hat h(u_c,d_1) \\
  \hat h(s_1,u_a)\hat h(u_a,d_2) & \hat h(s_1,u_b)\hat h(u_b,d_2) & \hat h(s_1,u_c)\hat h(u_c,d_2) \\
\end{vmatrix} \\
& = \hat h(s_1,u_c)\hat h(u_c,d_1)
\begin{vmatrix}
  \hat h(s_2,u_a)\hat h(u_a,d_1) & \hat h(s_2,u_b)\hat h(u_b,d_1) \\
  \hat h(s_1,u_a)\hat h(u_a,d_2) & \hat h(s_1,u_b)\hat h(u_b,d_2) \\
\end{vmatrix} \\ 
&  - \hat h(s_2,u_c)\hat h(u_c,d_1)
\begin{vmatrix}
  \hat h(s_1,u_a)\hat h(u_a,d_1) & \hat h(s_1,u_b)\hat h(u_b,d_1) \\
  \hat h(s_1,u_a)\hat h(u_a,d_2) & \hat h(s_1,u_b)\hat h(u_b,d_2) \\
\end{vmatrix} \\ 
&  + \hat h(s_1,u_c)\hat h(u_c,d_2)
\begin{vmatrix}
  \hat h(s_1,u_a)\hat h(u_a,d_1) & \hat h(s_1,u_b)\hat h(u_b,d_1) \\
  \hat h(s_2,u_a)\hat h(u_a,d_1) & \hat h(s_2,u_b)\hat h(u_b,d_1) \\
\end{vmatrix}. 
\end{align*}
The third term in the expansion above can be written as 
\begin{align*}
& \hat h(s_1,u_c)\hat h(u_c,d_2)\hat h(u_a,d_1)\hat h(u_b,d_1)
\begin{vmatrix}
  \hat h(s_1,u_a) & \hat h(s_1,u_b) \\
  \hat h(s_2,u_a) & \hat h(s_2,u_b) \\
\end{vmatrix}, 
\end{align*} }
which is a non-identically zero polynomial since $s_1 \leadsto u_c$, $u_c \in \I(v_p^2)$, $u_a, u_b \in \I(v_p^1)$, and there are two disjoint paths $P_{s_1,u_a}$ and $P_{s_2,u_b}$. Moreover, as we noticed before, one of the variables in $\hat h(u_c,d_2)$ is not shared by any other effective channel gain $\hat h$, and therefore, the term above cannot be canceled by the other terms. This allows us to conclude that $M_1$ is full-rank with probability 1. Now, suppose $\{u_a,u_b\} = \{u_c,u_d\}$. This means that the original network must contain the network shown in Figure \ref{step}. The curvy lines are used to represent paths. 
\begin{figure}[ht]
	\centering
		\includegraphics[height=22mm]{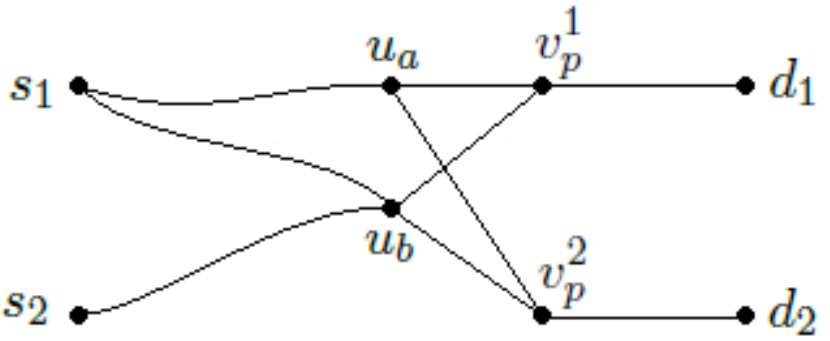}
	\caption{Illustration of a subnetwork contained by the network from \ref{onelayer}, in the case where $\{u_a,u_b\} = \{u_c,u_d\}$}
	\label{step}
\end{figure}
At this point, if $s_2 \leadsto u_a$, then we can remove the nodes in $V_{\ell(v_p^1)-1} \setminus \{u_a,u_b\}$, and we are in the case of \ref{onelayertwo}. If $s_2 \not\leadsto u_a$, then $\hat h(s_2,u_a) = 0$, and by applying Lemma \ref{lem:two} to $v_p^1$, we must have at least one node $u_c' \in \I(v_p^1) \setminus \{u_a,u_b\}$, such that $s_2 \leadsto u_c'$ (since $u_b$ is the other one). Then we choose $\{v_a,v_b,v_c\}=\{u_a,u_b,u_c'\}$. If $\hat h(s_1,u_c') \hat h(u_c',d_2)$ is not identically zero, then the same proof shown above with $u_c'$ instead of $u_c$ will show that $M_1$ is full-rank with probability 1. If we assume that $\hat h(s_1,u_c') \hat h(u_c',d_2)$ is identically zero, then we have
{\small 
\begin{align*} 
& \det M_1 = \\ 
&\begin{vmatrix}
  \hat h(s_1,u_a)\hat h(u_a,d_1) & \hat h(s_1,u_b)\hat h(u_b,d_1) & \hat h(s_1,u_c')\hat h(u_c',d_1) \\
  0 & \hat h(s_2,u_b)\hat h(u_b,d_1) & \hat h(s_2,u_c')\hat h(u_c',d_1) \\
  \hat h(s_1,u_a)\hat h(u_a,d_2) & \hat h(s_1,u_b)\hat h(u_b,d_2) & 0 \\
\end{vmatrix} \\
& = -\hat h(s_1,u_c')\hat h(u_c',d_1) h(s_2,u_b)\hat h(u_b,d_1) \hat h(s_1,u_a)\hat h(u_a,d_2) \\
&+ \hat h(s_2,u_c')\hat h(u_c',d_1) \hat h(s_1,u_b)\hat h(u_b,d_1) \hat h(s_1,u_a)\hat h(u_a,d_2) \\
&- \hat h(s_2,u_c')\hat h(u_c',d_1) \hat h(s_1,u_a)\hat h(u_a,d_1) \hat h(s_1,u_b)\hat h(u_b,d_2) .
\end{align*} }
The last term is non-identically zero since $s_2 \leadsto u_c'$, $u_c' \in \I(v_p^1)$, $s_1 \leadsto u_a$, $u_a \in \I(v_p^1)$, $s_1 \leadsto u_b$ and $u_b \in \I(v_p^2)$. Moreover, as we noticed before, one of the variables in $\hat h(u_a,d_1)$ is not shared by any other effective channel gain $\hat h$, and therefore the last term above cannot be cancelled by the other two terms. Thus, we conclude that $M_1$ is invertible with probability 1. 

From the symmetry between $M_1$ and $M_2$ (they simply have $(s_1,d_1)$ and $(s_2,d_2)$ exchanged), the exact same steps can be used to show that either we can find the nodes $\{v_d,v_e,v_f\} \subset V_{\ell(v_p^1)-1}$ such that $M_2$ is full-rank with probability 1, or we can remove nodes from $V_{\ell(v_p^1)-1}$ so that we are in the case of \ref{onelayertwo}. This concludes the achievability proof of $\dE = 2$ in the cases where we have two disjoint paths with manageable interference.

\end{subsubsection}

\vspace{2mm}

Next, we proceed to providing the achievability scheme for (B\pr), in which case we have a subnetwork with no two disjoint paths, and no node $v$ as described in (A).

\end{subsection}

\begin{subsection}{The butterfly and the grail} \label{notwo}

We start by inferring important properties of the structure of the network, if it does not fall into case (A). We will show that such a network must contain one of the subnetworks in Figure \ref{nets1}. The subnetwork in Figure \ref{nets1}a simply contains two disjoint paths $\p11$ and $\p22$. Next, we formally characterize the other two.

\vspace{2mm}

\begin{definition} \label{butdef} The network $\N$ is a Butterfly network if it contains two nodes $u_0$ and $u_1$ connected by a path $P_{u_0,u_1}$ (if $u_0 = u_1$, then we assume the path consists of a single node), 
two disjoint paths $\p12$ and $\p21$ which do not contain any node from $P_{u_0,u_1}$, and two paths $\p11$ and $\p22$ such that $\p11 \cap \p22 =  P_{u_0,u_1}$. An example is shown in Figure \ref{but1}. 

\begin{figure}[ht]
	\centering
		\includegraphics[height=28mm]{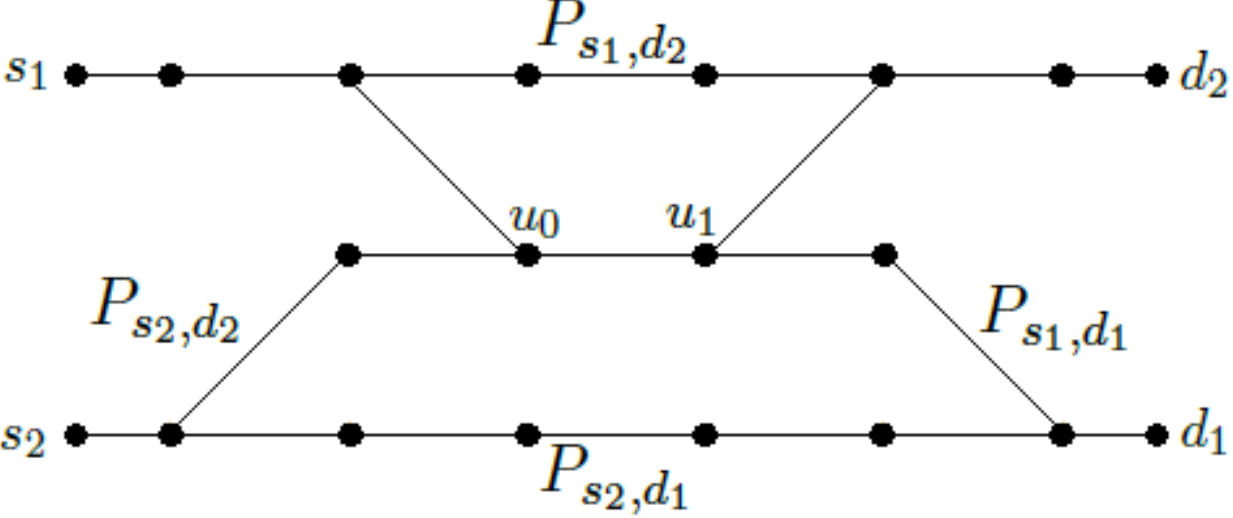}
	\caption{Illustration of a network that contains a Butterfly subnetwork.}
	\label{but1}
\end{figure}

\end{definition}

\begin{definition} \label{graildef} The network $\N$ is a Grail network if it contains two disjoint paths $\p12$ and $\p21$ and nodes  $w_a \in \p12$ and $w_b \in \p21$ such that $s_2 \leadsto w_a$, $w_a \leadsto w_b$, and $w_b \leadsto d_2$. An example is shown in Figure \ref{grail}.

\begin{figure}[ht]
	\centering
		\includegraphics[height=20mm]{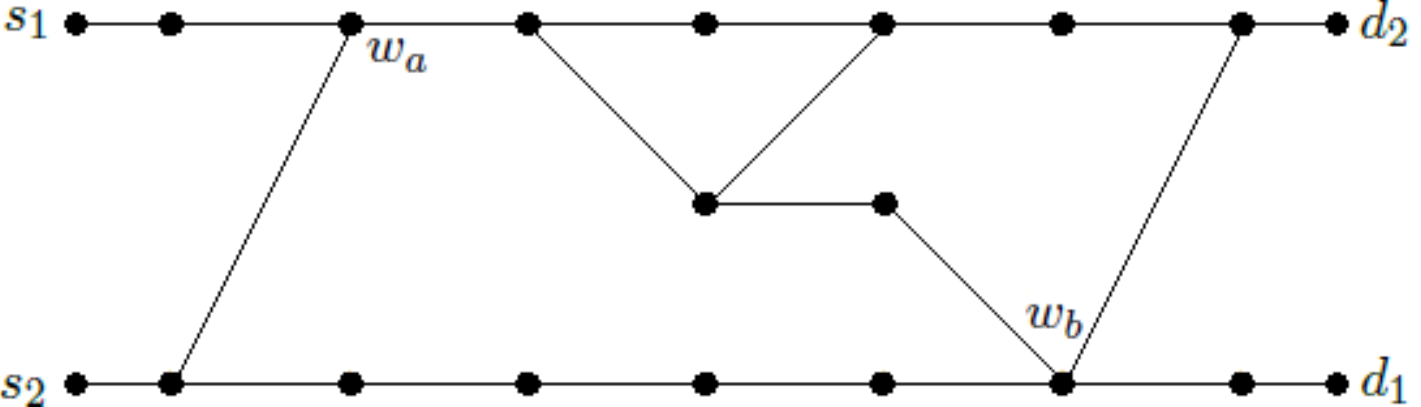}
	\caption{Illustration of a network that contains a Grail subnetwork.}
	\label{grail}
\end{figure}

\end{definition}

Then we can state the following Claim.

\begin{claim} \label{structs}
The absence of a node $v$ whose removal disconnects $d_i$ from both sources and $s_{\bar i}$ from both destinations, for $i=1$ or $i=2$, implies that $\N$ must contain $(i)$ two disjoint paths $\p11$ and $\p22$, $(ii)$ a butterfly subnetwork, or $(iii)$ a grail subnetwork.
\end{claim}

\noindent \emph{Sketch of proof: } We start by building an extended network $\N'$, by transforming each layer of our original network into two copies of itself, and connecting each node to its copy. Then we notice that the absence of a node $v$ whose removal disconnects $d_i$ from both sources and $s_{\bar i}$ from both destinations in the original network, for $i=1$ or $i=2$, implies the absence of an edge $e$ whose removal disconnects $d_i$ from both sources and $s_{\bar i}$ from both destinations, $i=1$ or $i=2$, in the extended network. Therefore, the result obtained in \cite{ShenviDey,Cai2unicast} guarantees that $\N'$ either contains two edge-disjoint paths, a butterfly or a grail. Since any two edge-disjoint paths in $\N'$ are also vertex-disjoint, we conclude that our original network must contain two vertex-disjoint paths, a butterfly or a grail. A more detailed proof can be found in Appendix \ref{proofclaimwireline}.

\vspace{2mm}



Next, we assume that all nodes that do not belong to the subnetwork satisfying the conditions in $(B')$ are removed. Since the resulting network does not contain two disjoint paths, but does not fall in case (A), we conclude from Claim \ref{structs} that we may either have a butterfly network or a grail network. We provide achievability schemes for each case separately.

\begin{subsubsection}{Butterfly network}\label{but_section}

%

We assume we have a subnetwork as described in Definition \ref{butdef} and that any node which does not belong to $\p11$, $\p22$, $\p12$ or $\p21$ is removed from the network. Moreover, we will assume that, if there are several choices for $u_0$ and $u_1$, we choose them so that $u_1$ is as close as possible to the destinations (i.e., we maximize $\ell(u_1)$).

Similar to what we did in the case of two disjoint paths with manageable interference, we will identify a key layer and build a condensed network. Then we will show that by using amplify-and-forward in the nodes in the intermediate key layer, we can make the end-to-end transfer matrix diagonal with non-zero diagonal entries. As our key layer, we will use  $V_{\ell(u_1)}$. Notice that we are guaranteed to have three nodes in $V_{\ell(u_1)}$ (since any extra node would have been removed). The condensed network is shown in Figure \ref{condbut}.

\begin{figure}[ht]
	\centering
		\includegraphics[height=25mm]{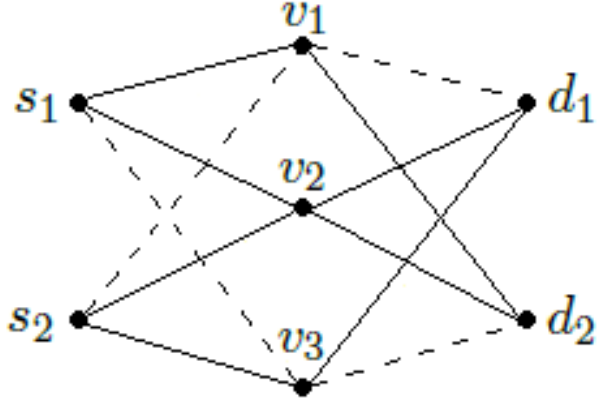}
	\caption{Illustration of the condensed network of a Butterfly network.}
	\label{condbut}
\end{figure}

We let the three nodes in $V_{\ell(u_1)}$ be called $v_1, v_2$ and $v_3$ as shown in Figure \ref{condbut} (notice that $v_2 = u_1$), and associate scaling factors $x_1, x_2$ and $x_3$ to them. We will follow the same steps that we used in \ref{onelayer}, except that now our intermediate layer has exactly three nodes. Thus, we will show that either we can remove one of the nodes in $V_{\ell(u_1)}$ so that the resulting condensed network falls in case \ref{onelayertwo} (i.e., a $2 \times 2 \times 2$ interference channel), or the matrices
\begin{align} 
&M_1 = \left[ \begin{array}{cc}
  \hat h(s_1,v_1)\hat h(v_1,d_1) & \hat h(s_1,v_2)\hat h(v_2,d_1) \\ 
  \hat h(s_2,v_1)\hat h(v_1,d_1) & \hat h(s_2,v_2)\hat h(v_2,d_1) \\ 
  \hat h(s_1,v_1)\hat h(v_1,d_2) & \hat h(s_1,v_2)\hat h(v_2,d_2) \\ 
\end{array} \right. \non
& \quad \quad \quad\quad \quad \left. \begin{array}{c}
  \hat h(s_1,v_3)\hat h(v_3,d_1) \\
  \hat h(s_2,v_3)\hat h(v_3,d_1) \\
  \hat h(s_1,v_3)\hat h(v_3,d_2) \\
\end{array} \right]
\text{ and } \nonumber \\
& M_2 = \left[\begin{array}{cc}
  \hat h(s_2,v_1)\hat h(v_1,d_2) & \hat h(s_2,v_2)\hat h(v_2,d_2) \\ 
  \hat h(s_1,v_1)\hat h(v_1,d_2) & \hat h(s_1,v_2)\hat h(v_2,d_2) \\ 
  \hat h(s_2,v_1)\hat h(v_1,d_1) & \hat h(s_2,v_2)\hat h(v_2,d_1) \\ 
\end{array} \right. \non
& \quad \quad \quad\quad \quad \left. \begin{array}{c}
  \hat h(s_2,v_3)\hat h(v_3,d_2) \\
  \hat h(s_1,v_3)\hat h(v_3,d_2) \\
  \hat h(s_2,v_3)\hat h(v_3,d_1) \\
\end{array} \right] \nonumber
\end{align}
are full-rank with probability 1. In the latter case, the same steps as in \ref{onelayer} guarantee that we can find $x_1, x_2, x_3$ such that the end-to-end transfer matrix is diagonal with non-zero diagonal entries. An important property about the Butterfly structure is that for any two nodes $v_a, v_b \in V_{\ell(u_1)}$, there exists two disjoint paths between $\{s_1,s_2\}$ and $\{v_a,v_b\}$ and two disjoint paths between $\{v_a,v_b\}$ and $\{d_1,d_2\}$. Therefore, we see that if $\hat h(s_2,v_1)\hat h(v_1,d_1)$ is a non-identically zero polynomial in the channel gains, we can remove $v_3$ and we are in \ref{onelayertwo}. Similarly if $\hat h(s_1,v_3)\hat h(v_3,d_2)$ is non-identically zero, we can remove $v_1$ and we are in \ref{onelayertwo}. Therefore, we may assume that either $\hat h(s_1,v_3)$ or $\hat h(v_3,d_2)$ is zero, and either $\hat h(s_2,v_1)$ or $\hat h(v_1,d_1)$ is zero. To show that $M_1$ is full-rank with probability 1, we first consider the case when $\hat h(v_3,d_2) = 0$. We notice that the fact that $\hat h(v_3,d_2) = 0$ and our assumption that $u_1$ was chosen as close as possible to the destinations guarantee that there is no path starting on a node in $P_{v_3,d_1} \cup P_{v_2,d_1} \setminus \{v_2\}$ and ending in $d_2$. Thus, we see that the first channel gain in the $P_{v_2,d_1}$ path only appears as a variable in $\hat h(v_2,d_1)$, and no other $\hat h$. Then we notice that
{\small 
\begin{align*} 
&\det M_1 = \\ &\begin{vmatrix}
  \hat h(s_1,v_1)\hat h(v_1,d_1) & \hat h(s_1,v_2)\hat h(v_2,d_1) & \hat h(s_1,v_3)\hat h(v_3,d_1) \\
  0 & \hat h(s_2,v_2)\hat h(v_2,d_1) & \hat h(s_2,v_3)\hat h(v_3,d_1) \\
  \hat h(s_1,v_1)\hat h(v_1,d_2) & \hat h(s_1,v_2)\hat h(v_2,d_2) & 0 \\
\end{vmatrix} \\
& = - \hat h(s_1,v_1)\hat h(v_1,d_1)\hat h(s_2,v_3)\hat h(v_3,d_1)\hat h(s_1,v_2)\hat h(v_2,d_2) \\
& + \hat h(s_1,v_1)\hat h(v_1,d_2)\hat h(v_2,d_1)\hat h(v_3,d_1)
\begin{vmatrix}
  \hat h(s_1,v_2) & \hat h(s_1,v_3) \\
  \hat h(s_2,v_2) & \hat h(s_2,v_3) \\
\end{vmatrix}.
\end{align*} }
The last term is a non-identically zero polynomial, since $s_1 \leadsto v_1$, $v_1 \leadsto d_2$, $v_2 \leadsto d_1$, $v_3 \leadsto d_1$ and there are two disjoint paths $P_{s_1,v_2}$ and $P_{s_2,v_3}$. Thus, since $\hat h(v_2,d_1)$ contains a variable which cannot be cancelled by the other term, we conclude that $\det M_1$ is non-identically zero, and $M_1$ is full-rank with probability 1. If instead we assume that $\hat h(v_3,d_2)$ is not identically zero, then $\hat h(s_1,v_3) = 0$, and we have that 
{\small
\begin{align*} 
&\det M_1 = \\ &\begin{vmatrix}
  \hat h(s_1,v_1)\hat h(v_1,d_1) & \hat h(s_1,v_2)\hat h(v_2,d_1) & 0 \\
  0 & \hat h(s_2,v_2)\hat h(v_2,d_1) & \hat h(s_2,v_3)\hat h(v_3,d_1) \\
  \hat h(s_1,v_1)\hat h(v_1,d_2) & \hat h(s_1,v_2)\hat h(v_2,d_2) & 0 \\
\end{vmatrix} \\
& = - \hat h(s_2,v_3)\hat h(v_3,d_1)\hat h(s_1,v_1)\hat h(s_1,v_2) \\
& \quad \quad \quad\quad \quad \begin{vmatrix}
  \hat h(v_1,d_1) & \hat h(v_2,d_1) \\
  \hat h(v_1,d_2) & \hat h(v_2,d_2) \\
\end{vmatrix},
\end{align*}}
which is not identically zero, since $s_2 \leadsto v_3$, $v_3 \leadsto d_1$, $s_1 \leadsto v_1$, $s_1 \leadsto v_2$ and there are two disjoint paths $P_{v_1,d_2}$ and $P_{v_2,d_1}$. Therefore, we conclude that $M_1$ is full-rank with probability 1. From the symmetry between $M_1$ and $M_2$, we conclude that the same steps (but considering $\hat h(s_2,v_1)$ or $\hat h(v_1,d_1)$ to be zero) will show that $M_2$ is full-rank with probability 1.

\end{subsubsection}

\begin{subsubsection}{Grail network}\label{grail_section}



We assume that we have a minimal subnetwork which still satisfies Definition \ref{graildef}, i.e., all the unnecessary nodes are removed. As key layers, we will use $V_{\ell(w_a)}$ and $V_{\ell(w_b)}$. Notice that if we assume that the subnetwork is chosen to be minimal, each of these layers must contain exactly two nodes. Therefore, our condensed network will be as shown in Figure \ref{condgrail}.
\begin{figure}[ht]
	\centering
		\includegraphics[height=25mm]{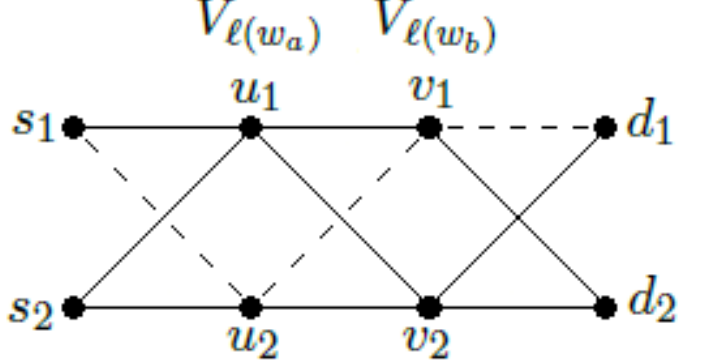}
	\caption{Illustration of the condensed network of a Grail network.}
	\label{condgrail}
\end{figure}
We will let the nodes in $V_{\ell(w_a)}$ be called $u_1$ and $u_2$, and the nodes in $V_{\ell(w_b)}$ be called $v_1$ and $v_2$, as shown in Figure \ref{condgrail}. Next we will show that either we can suppress one of the two intermediate key layers (by assuming their nodes are just forwarding their received signals) and obtain a network as in \ref{onelayertwo}, or we can choose scaling factors $y_1$, $y_2$, $x_1$ and $x_2$ (respectively for $u_1, u_2, v_1$ and $v_2$) so that the end-to-end transfer matrix is diagonal with non-zero diagonal entries. We notice that if $\hat h(s_1,u_2)$ is not identically zero, then the existence of two disjoint paths $\p12$ and $\p21$ guarantees that if we suppress $V_{\ell(w_b)}$ from the condensed network, we obtain the network in \ref{onelayertwo}. Similarly, if $\hat h(v_1,d_1)$ is not identically zero, we can suppress $V_{\ell(w_a)}$ from the condensed network, and we are again in the case of \ref{onelayertwo}. Therefore, we will assume that $\hat h(s_1,u_2)=\hat h(v_1,d_1)=0$, and we will show that there is a choice of $y_1$, $y_2$, $x_1$ and $x_2$ so that the end-to-end transfer matrix is diagonal with non-zero diagonal entries. In order to do that we first consider the transfer matrix between $V_1$ and $V_{\ell(w_b)}$, which is given by
\begin{align}
& F = \left[ \begin{array}{c}
  \hat h(s_1,u_1) \hat h(u_1,v_1) y_1 \\ 
  \hat h(s_1,u_1) \hat h(u_1,v_2) y_1 \\ 
\end{array} \right.  \non
& \quad \quad  \left. \begin{array}{c}
  \hat h(s_2,u_1) \hat h(u_1,v_1) y_1 + \hat h(s_2,u_2) \hat h(u_2,v_1) y_2 \\
  \hat h(s_2,u_1) \hat h(u_1,v_2) y_1 + \hat h(s_2,u_2) \hat h(u_2,v_2) y_2 \\
\end{array}\right]. 
\label{transfer6}
\end{align}
Then we notice that if we let 
\[ M = \begin{bmatrix}
  \hat h(s_2,u_1) \hat h(u_1,v_1) & \hat h(s_2,u_2) \hat h(u_2,v_1) \\
  \hat h(s_2,u_1) \hat h(u_1,v_2) & \hat h(s_2,u_2) \hat h(u_2,v_2) \\
\end{bmatrix},
\] we have
\begin{align*}
\det M & = \begin{vmatrix}
  \hat h(s_2,u_1) \hat h(u_1,v_1) & \hat h(s_2,u_2) \hat h(u_2,v_1) \\
  \hat h(s_2,u_1) \hat h(u_1,v_2) & \hat h(s_2,u_2) \hat h(u_2,v_2) \\
\end{vmatrix} \\ & = \hat h(s_2,u_1)\hat h(s_2,u_2)
\begin{vmatrix}
  \hat h(u_1,v_1) &  \hat h(u_2,v_1) \\
  \hat h(u_1,v_2) &  \hat h(u_2,v_2) \\
\end{vmatrix},
\end{align*}
which is a non-identically zero polynomial on the channel gains, since $s_2 \leadsto u_1$, $s_2 \leadsto u_2$ and there are two disjoint paths $P_{u_1,v_1}$ and $P_{u_2,v_2}$. Thus $M$ is invertible with probability 1. Since we also have that $\hat h(s_2,u_1) \hat h(u_1,v_2) \ne 0$ and $\hat h(s_2,u_2) \hat h(u_2,v_2) \ne 0$ with probability 1, we are guaranteed that if we choose $y_1 \ne 0$ and $y_2 \ne 0$ such that $F_{2,2} = \hat h(s_2,u_1) \hat h(u_1,v_2) y_1 + \hat h(s_2,u_2) \hat h(u_2,v_2) y_2 = 0$, then $F_{1,1} \ne 0$, $F_{1,2} \ne 0$ and $F_{2,1} \ne 0$. Notice that, if $F_{1,2} = \hat h(s_2,u_1) \hat h(u_1,v_1) y_1 + \hat h(s_2,u_2) \hat h(u_2,v_1) y_2$ were zero, we would contradict the fact that the system $M {\bf y} = {\bf 0}$ only has ${\bf y} = {\bf 0}$ as a solution. Therefore, we have that the end-to-end transfer matrix can be expressed as
\begin{align*}
&\begin{bmatrix}
  0 & \hat h(v_2,d_1) \\
  \hat h(v_1,d_2) & \hat h(v_2,d_2) \\
\end{bmatrix} 
\begin{bmatrix}
  x_1 &  0 \\
  0 &  x_2 \\
\end{bmatrix} 
\begin{bmatrix}
  \alpha & \beta \\
  \gamma &  0 \\
\end{bmatrix} \\ & \quad \quad = 
\begin{bmatrix}
  \hat h(v_2,d_1) \gamma x_2 & 0 \\
  \hat h(v_1,d_2) \alpha x_1 + \hat h(v_2,d_2) \gamma x_2 &  \hat h(v_1,d_2) \beta x_1 \\
\end{bmatrix},
\end{align*}
where $\alpha \ne 0$, $\beta \ne 0$ and $\gamma \ne 0$. Therefore, since $\hat h(v_2,d_1)$, $\hat h(v_1,d_2)$ and $\hat h(v_2,d_2)$ are all non-zero with probability 1, we can choose $x_1$ and $x_2$ non-zero to make the end-to-end transfer matrix diagonal with non-zero diagonal entries. This concludes the achievability proof for the case in which we have a grail subnetwork and thus we conclude all cases in which $\dE = 2$ is achievable.

\end{subsubsection}

\end{subsection}

\end{section}

\begin{section}{Networks with $3/2$ degrees-of-freedom} \label{32deg}

In this section, we prove that if our network $\N$ does not fall into cases (A), (A\pr), (B) and (B\pr), then we have $\dE = \frac32$. We start by defining two main categories of networks which belong to (C). If $\N$ does not contain a node $v$ whose removal disconnects $d_i$ from both sources and $s_{\bar i}$ from both terminals, for $i \in \{1,2\}$ (i.e., $\N$ is not in (A)), then, from our discussion in \ref{notwo}, we know that we must have one of the three structures in Figure \ref{nets1}. Moreover, if the network does not contain such a node $v$ and does not contain two disjoint paths $\p11$ and $\p22$, then we are in (B\pr). Therefore, all networks in (C) contain two disjoint paths $\p11$ and $\p22$, but do not contain any pair of disjoint paths $\p11'$ and $\p22'$ with manageable interference, or else we would be in case (B).

We will assume that we have two disjoint paths $\p11$ and $\p22$ and we will first show that we can assume that our network $\N$ falls into one of two cases:
\begin{enumerate}[C1.]
\item $n_1(G) \geq 2$, $n_1^D=1$, $n_2(G)=1$ and $n_2^D=0$. \label{sym}
\item $n_1(G) = n_1^D = 1$ \label{nonsym} 
\end{enumerate}

We see this as follows. Since the interference on $\p11$ and $\p22$ is not manageable, we have that either $n_1(G) = 1$ or $n_2(G) = 1$. Moreover, we must also have either $n_1^D = 1$ or $n_2^D = 1$, because otherwise we can let $S = \p11 \cup \p22$ and $n_1(G[S])=n_1^D$ and $n_2(G[S])=n_2^D$. So we assume WLOG that $n_1^D = 1$. Then, if $n_1(G) = 1$, we are in case C\ref{nonsym}. Thus, we assume $n_1(G) \geq 2$, and we must have $n_2(G) = 1$. If $n_2^D = 1$, we are again in case C\ref{nonsym} by exchanging the names of $(s_1,d_1)$ and $(s_2,d_2)$. Otherwise, if $n_2^D = 0$, we are in case C\ref{sym} (notice that $n_2^D \leq n_2(G)$).

We will provide an achievability and a converse for $\dE = \frac32$ in each case. 

\begin{subsection}{Achievability for case C\ref{sym}} \label{ach_sym}

We will start by considering case C\ref{sym}. Notice that we must have a node $v_1 \notin \p11 \cup \p22$ such that $v_1 \interf \p11$ and thus we have a path $P_{s_2,v_1}$ that is disjoint from $\p11$. We let $v_m$ be the last node in $\p22 \cap P_{s_2,v_1}$, and we have the path $P_{v_m,v_1} = P_{s_2,v_1}[v_m,v_1]$. Next we consider letting $S^* = \p11 \cup \p22 \cup P_{v_m,v_1}$. This guarantees that $n_1(G[S^*]) \geq 2$. Since $\p11$ and $\p22$ do not have manageable interference, we must have $n_2(G[S^*]) = 1$. Moreover, since $n_2^D = 0$, we conclude that we must have a node $v_2 \in P_{v_m,v_1} \setminus \{v_m\}$ such that $v_2 \interf \p22$, and we must have a path $P_{s_1,v_2} \subset S^*$. It can then be seen that our network is as shown in Figure \ref{Csym} up to a change in the position of the edge $(v_3,v_4)$. The curvy lines and the dashed lines indicate paths (which may consist of a single edge or multiple edges). Notice that we may also have $v_1 = v_2$.

\begin{figure}[ht]
\begin{center}
\includegraphics[height=23mm]{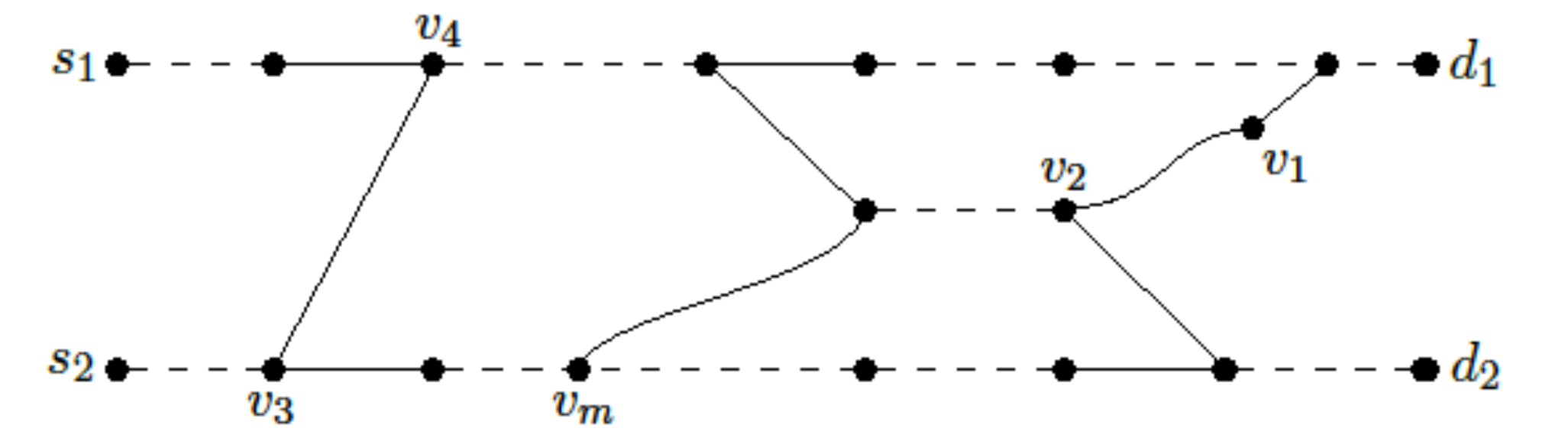}
\caption{Illustration of the network in case C\ref{sym}.} \label{Csym}
\end{center} 
\end{figure}

In order to achieve $\dE = \frac32$, we will describe a scheme in which we use two different modes of operation for the network. During each mode of operation, only a subset of the nodes will be transmitting, while the others will stay silent. During the first mode of operation, one special node will store its received signals. 
Then, in the second mode of operation, it will forward the stored signals. 
We will consider two subcases, according to the position of edge $(v_3,v_4)$ with respect to $v_2$. 

\begin{subsubsection}{$\ell(v_3) < \ell(v_2)$} \label{buf}

In this case, our ``special node'' will be the node from $\p22$ in $V_{\ell(v_2)}$.
In the first mode of operation, it will function as a virtual destination $d_2'$. Node $d_2'$ and any node $u \in \p22$ such that $\ell(u) \geq \ell(d_2')$ will stay silent during Mode 1. Then we notice that the two disjoint paths $\p11$ and $P_{s_2,d_2'}$ have manageable interference. This must be the case, since $n_2(G,\p22) = 1$, and this unique interference is caused by $v_2$ on a node $u \in \p22$ such that $\ell(u) > \ell(d_2')$, and thus $n_2(G,P_{s_2,d_2'})=0$. Moreover, since $\ell(v_3) < \ell(d_2')$ and $\ell(v_m) < \ell(d_2')$, we have $n_1(G,\p11) \geq 2$. 

Therefore, by using the amplify-and-forward scheme described in \ref{allup}, it is possible to guarantee that the transfer matrix between $(s_1,s_2)$ and $(d_1,d_2')$ is diagonal with non-zero diagonal entries. Notice that, even though $d_1$ and $d_2'$ are not in the same layer, one could create a virtual path between $d_2'$ and a virtual node $d_2'' \in V_r$ which does not receive nor cause any interference. Then we can use the scheme from \ref{allup} to guarantee that the transfer matrix between $(s_1,s_2)$ and $(d_1,d_2'')$ is diagonal with non-zero diagonal entries. Then it is easy to see that the same would hold for the transfer matrix between $(s_1,s_2)$ and $(d_1,d_2')$. During Mode 1, $d_2' $ will store its received signals. 

The second mode of operation should last for the same number of time steps as the first one. In Mode 2, $d_2'$ will become a virtual source $s_2'$. Then, we remove all the nodes from the network except those in the paths $\p11$ and $P_{s_2',d_2}$. Now we clearly have two disjoint paths with no interference. Therefore, it is clear that we can have the transfer matrix between $(s_1,s_2')$ and $(d_1,d_2)$ be diagonal with non-zero diagonal entries. Thus, by letting node $d_2' = s_2'$ forward each of the signals received during Mode 1 in Mode 2, it is clear that, over the two modes, we create three parallel AWGN channels, two of them between $s_1$ and $d_1$ and one of them between $s_2$ and $d_2$. Therefore, we achieve $\dE = \frac32$. A visual representation of the scheme is shown in Figure \ref{buffig}. 

\begin{figure}[ht]
\begin{center}
\includegraphics[height=60mm]{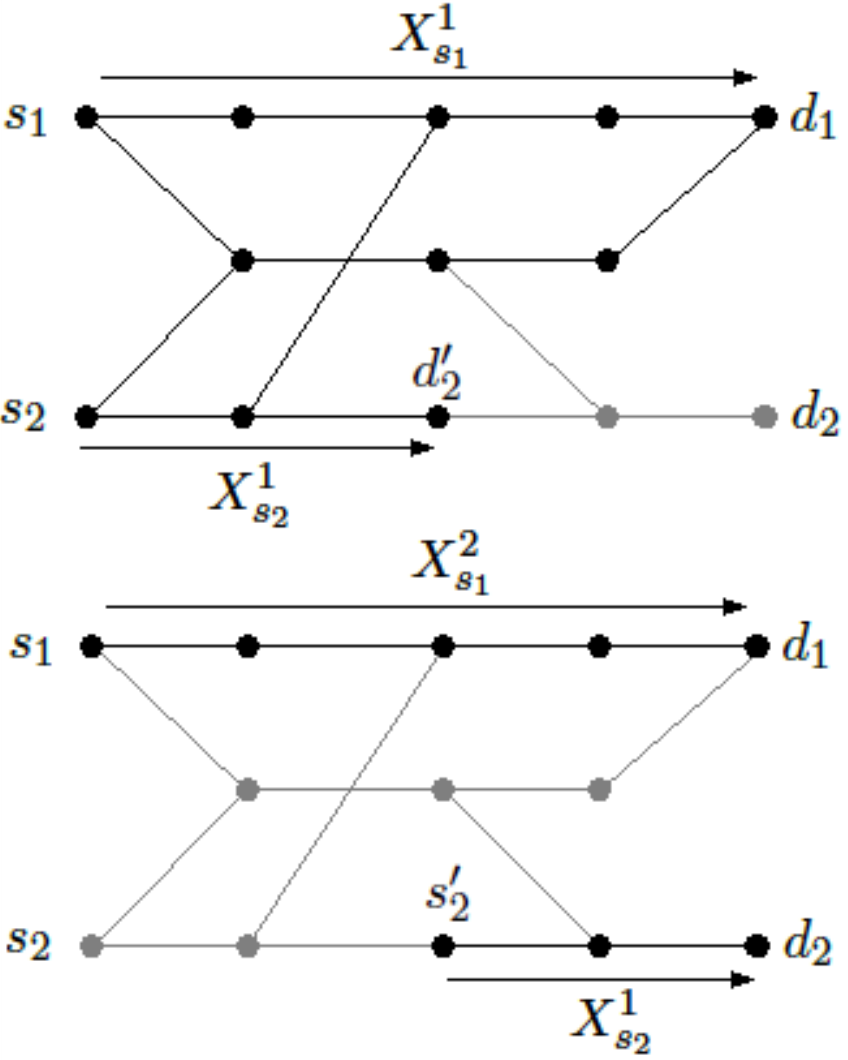}
\caption{Depiction of Mode 1 and Mode 2 for the achievability scheme in case C\ref{sym} if $\ell(v_3) < \ell(v_2)$.} \label{buffig}
\end{center} 
\end{figure}

\end{subsubsection}

\begin{subsubsection}{$\ell(v_3) \geq \ell(v_2)$} \label{buf2}

In this case, in the first mode of operation, we let the node from $\p11$ in $V_{\ell(v_2)}$ be a virtual destination $d_1' $. Then we clearly have two disjoint paths $P_{s_1,d_1'}$ and $\p22$. Any node $v \notin P_{s_1,d_1' } \cup \p22$ will stay silent during Mode 1. Since we assumed that $\ell(v_3) \geq \ell(v_2)$, we cannot have any direct interferences between $P_{s_1,d_1'}$ and $\p22$, or else we would contradict the fact that $n_1^D(\p11,\p22) = 1$ and $n_2^D(\p11,\p22) = 0$. Therefore, during Mode 1, we can have the transfer matrix between $(s_1,s_2)$ and $(d_1' ,d_2)$ be diagonal with non-zero diagonal entries. During Mode 1, $d_1' $ will store its received signals.

The second mode of operation should last for the same number of time steps as the first one. During Mode 2, $d_1'$ will become a virtual source $s_1' $. Any node $v \in \p11$ such that $\ell(v) < \ell(s_1')$ will stay silent. Notice that the paths $P_{s_1',d_1}$ and $\p22$ have manageable interference. Therefore, by assuming the existence of a virtual node $s_1'' \in V_1$ which is connected to $s_1'$ through a virtual path that does not receive nor cause any interferences, we can use the linear scheme from \ref{allup} to guarantee that the transfer matrix from $(s_1',s_2)$ to $(d_1,d_2)$ is diagonal with non-zero diagonal entries. Thus, by letting node $d_1' = s_1'$ forward each of the signals received during Mode 1 in Mode 2, it is clear that, over the two modes, we create three parallel AWGN channels, two of them between $s_2$ and $d_2$ and one of them between $s_1$ and $d_1$. Therefore, we achieve $\dE = \frac32$. A visual representation of the scheme is shown in Figure \ref{buffig2}. 

\begin{figure}[ht]
\begin{center}
\includegraphics[height=60mm]{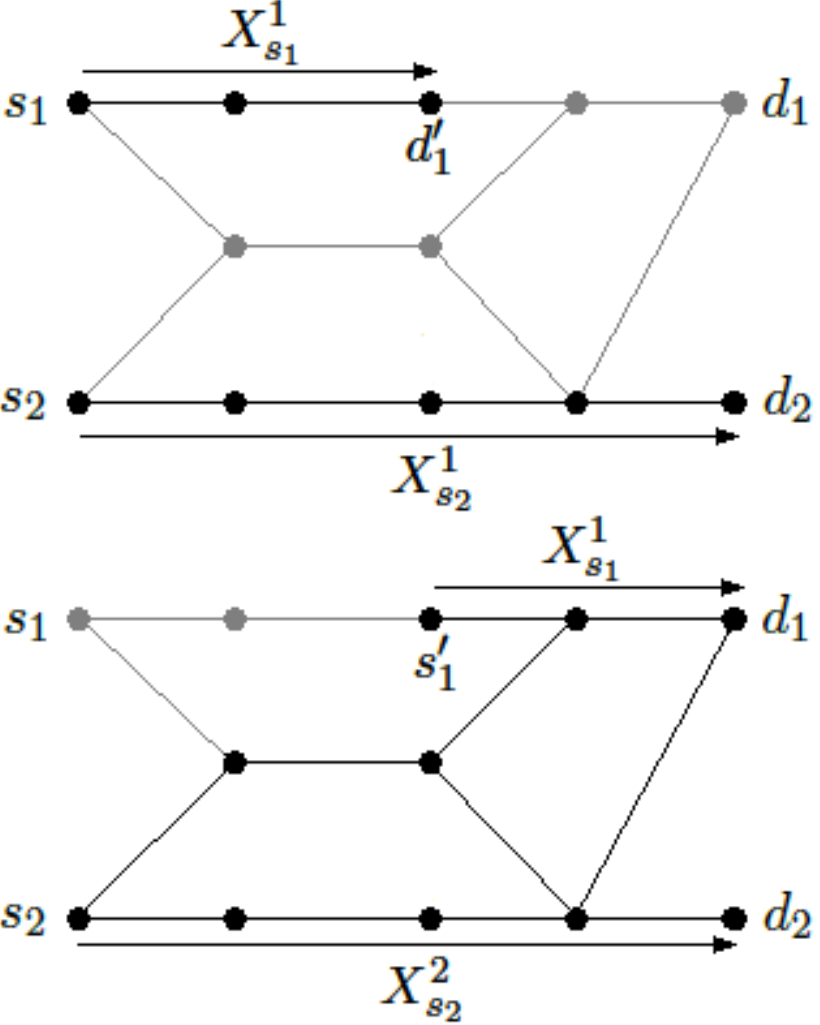}
\caption{Depiction of Mode 1 and Mode 2 for the achievability scheme in case C\ref{sym} if $\ell(v_3) \geq \ell(v_2)$.} \label{buffig2}
\end{center} 
\end{figure}

\end{subsubsection}


\end{subsection}

\begin{subsection}{Converse for case C\ref{sym}}  
\label{conv_sym}

In this section, we will show that if a network falls in C\ref{sym}, but does not contain two disjoint paths with manageable interference, then $\dE \leq \frac32$. We will start by naming some extra nodes which will be important to us, as shown in Figure \ref{fig:zalef1}. We will let $v_0$ be the node on $\p22$ such that $(v_2,v_0) \in E$. From our discussion in \ref{ach_sym}, we know that we have a path $P_{s_1,v_2}$, which must be entirely contained in $S^* = \p11 \cup \p22 \cup P_{v_m,v_1}$. Thus, we let $v_5$ be the last node in $\p11 \cap P_{s_1,v_2}$, and we let $v_6$ be its consecutive node on $P_{s_1,v_2}$ (which must be part of $P_{v_m,v_1}$ as well).


\begin{figure}[htp]
\begin{center}
\includegraphics[height=23mm]{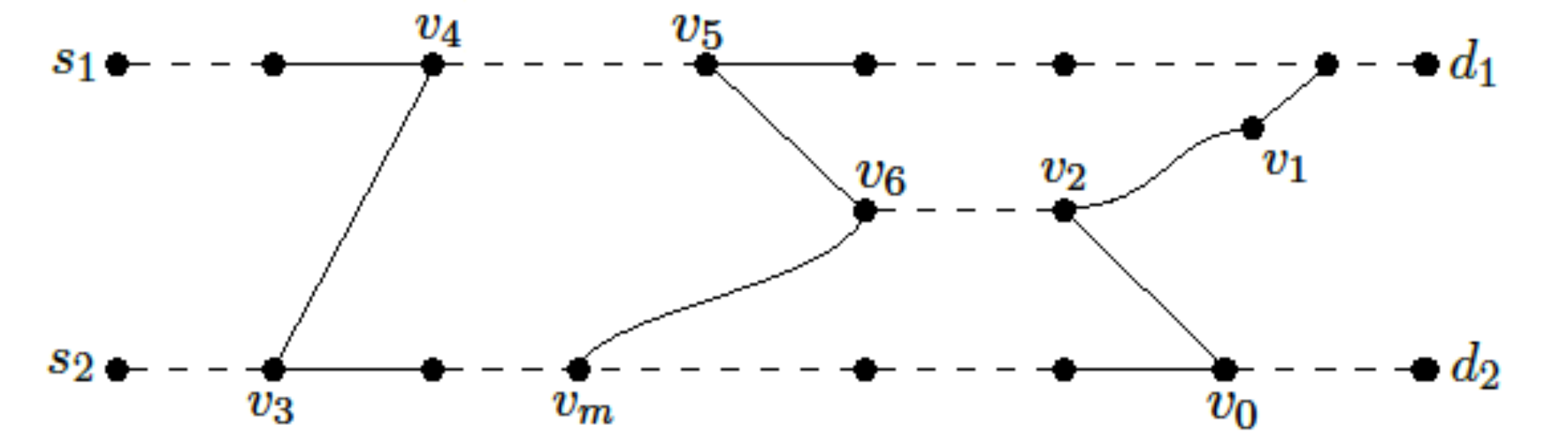}
\caption{Illustration of a network in case C\ref{sym}.} \label{fig:zalef1}
\end{center} 
\end{figure}





The assumption that there are no two disjoint paths with manageable interference allows us to infer some important connectivity properties about networks in case C\ref{sym}, illustrated in Figure \ref{fig:zalef1}. 
Next, we state and prove these properties.

\begin{enumerate}[P1.]


\item All paths from $s_1$ to $d_2$ contain nodes $v_2$ and $v_0$. \label{pv2v0}

It is easy to see that if we have a path $\p12$ not containing $\{v_2,v_0\}$, then we must have a node $v_a \in \p12$ such that $v_a \interf \p22$, and thus we would have $n_2(G,\p22) \geq 2$, which is a contradiction. 


\item All paths from $s_1$ to $d_2$ contain nodes $v_5$ and $v_6$. \label{pv5v6}

First consider the path $\q22 = \p22[s_2,v_m] \concat P_{v_m,v_1}[v_m,v_2] \concat (v_2,v_0) \concat \p22[v_0,d_2]$. Clearly, $\q22 \cap \p11 = \emptyset$ and $v_5 \dinterf \q22$. If we have a path $\p12$ not containing $\{v_5,v_6\}$ we conclude that $n_2(G,\q22) \geq 2$. But since $n_1(G,\p11) \geq 2$ we contradict the fact that there are no two disjoint paths with manageable interference.


\item All paths from $s_2$ to $d_1$ contain $\{v_6,v_2\}$ or $\{v_3,v_4\}$. \label{ps2d1}

Suppose there is a path $\p21$ not containing $\{v_6,v_2\}$ nor $\{v_3,v_4\}$. Then we let $S = \p11 \cup \p22 \cup \p21$ and we have $n_1(G[S],\p11)\geq 2$. But since P\ref{pv2v0} and P\ref{pv5v6} imply that any path from $s_1$ to $d_2$ must contain $\{v_6,v_2\}$, and $\{v_6,v_2\}\not\subset S$, we must have $n_2(G[S],\p22) = 0$, contradicting the fact that there are no two disjoint paths with manageable interference.

%


\item The removal of $v_0$ disconnects $d_2$ from both sources. \label{pv0}

From P\ref{pv2v0}, the removal of $v_0$ disconnects $d_2$ from $s_1$. So suppose the removal of $v_0$ does not disconnect $d_2$ from $s_2$ and we have a path $\q22$ not containing $v_0$. We know that $\q22$ must be disjoint from $\p11$, since otherwise we would contradict the fact that the removal of $v_0$ disconnects $d_2$ from $s_1$ (P\ref{pv2v0}). Moreover, if we let $S = \p11 \cup \q22$, since $v_0 \notin S$, we must have $n_2(G[S],\q22) = 0$. If $n_1(G[S],\p11) \ne 1$, we contradict the assumption of no two disjoint paths with manageable interference. However, if $n_1(G[S],\p11) = 1$, we must have a direct interference from $\q22$ on $\p11$, and we will have $n_1(G[V \setminus \{v_0\}],\p11) \geq 2$ and $n_2(G[V \setminus \{v_0\}],\q22) = 0$, and we again reach a contradiction.



\item The removal of $v_5$ disconnects $s_1$ from both destinations. \label{pv5} 
From P\ref{pv5v6}, the removal of $v_5$ disconnects $s_1$ from $d_2$. So we suppose the removal of $v_5$ does not disconnect $s_1$ from $d_1$ and we have a path $\q11$ not containing $v_5$. The path $\q11$ must be disjoint from $\p22$, or else we would contradict the fact that the removal of $v_5$ disconnects $s_1$ from $d_2$ (P\ref{pv5v6}). So first we let $S = \q11 \cup \p22$, and, since $v_5 \notin S$, we have $n_2(G[S],\p22) = 0$. If we have $n_1(G[S],\q11) \ne 1$, we contradict the assumption of no two disjoint paths with manageable interference. However, if $n_1(G[S],\q11) = 1$, we must have a direct interference from $\p22$ on $\q11$, and we will have $n_1(G[V \setminus \{v_5\}],\q11) \geq 2$ and $n_2(G[V \setminus \{v_5\}],\p22) = 0$, and we again reach a contradiction.



\item The removal of $v_2$ and $v_3$ disconnects $d_2$ from both sources. \label{pv2v3}

From P\ref{pv2v0}, the removal of $v_2$ disconnects $d_2$ from $s_1$. So suppose the removal of $v_2$ and $v_3$ does not disconnect $d_2$ from $s_2$ and we have a path $\q22$ not containing $v_2$ nor $v_3$. We know that $\q22$ is disjoint form $\p11$, or else we would contradict the fact that the removal of $v_2$ disconnects $s_1$ from $d_2$ (P\ref{pv2v0}). Then, we set $S = \p11 \cup \q22$. Since $v_2,v_3 \notin S$, from P\ref{pv2v0}, we must have $n_2(G[S],\q22) = 0$, and from P\ref{ps2d1}, we must have $n_1(G[S],\p11) = 0$. But this contradicts our assumption of no two disjoint paths with manageable interference.


\item The removal of $v_2$ and $v_4$ disconnects $d_1$ from both sources. \label{pv2v4}

From P\ref{ps2d1}, the removal of $v_2$ and $v_4$ disconnects $d_1$ from $s_2$. Thus, we assume that we have a path $\q11$ not containing $v_2$ nor $v_4$. The path $\q11$ must be disjoint of $\p22$, or else we contradict P\ref{ps2d1}. Thus we set $S = \q11 \cup \p22$. Since $v_2,v_4 \notin S$, from P\ref{pv2v0}, we must have $n_2(G[S],\p22) = 0$, and from P\ref{ps2d1}, we must have $n_1(G[S],\q11) = 0$. But this contradicts our assumption of no two disjoint paths with manageable interference.

 


\item All paths from $s_1$ or $s_2$ to $v_2$ contain $v_6$. \label{pv6}

This follows easily from P\ref{pv2v0}, P\ref{pv5v6} and P\ref{ps2d1}, since $v_2 \leadsto d_1$ and $v_2 \leadsto d_2$. 
\end{enumerate}

These properties allow us to infer the information inequalities that will build the converse proof. 
For these derivations, we will let $A \defi \{ v \in V \st s_2 \not\leadsto v \}$ and $B \defi \{ v \in V \st s_1 \not\leadsto v \}$, and let $W_1$ and $W_2$ be independent random variables corresponding to a uniform choice over the messages on sources $s_1$ and $s_2$ respectively.
Before we formally derive the inequalities, we will describe some of the intuition that leads to them, for a specific network example, shown in Figure \ref{zalefex}.
\begin{figure}[htp]
\begin{center}
\includegraphics[height=23mm]{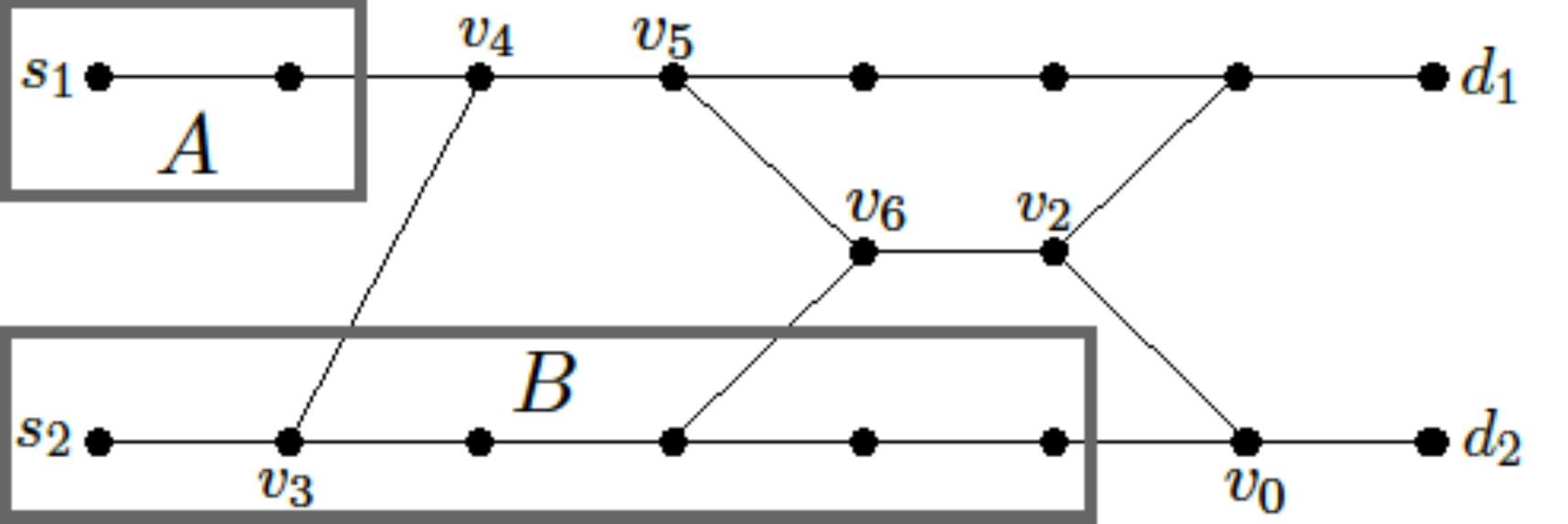}
\caption{Example of a network in case C\ref{sym}.} \label{zalefex}
\end{center} 
\end{figure}

We will consider, for a given $n$, the quantities
\[ \alpha \defi \liminf_{P \to \infty} \frac{I(Y_4^n, X_B^n;\tilde X_2^n)}{\frac n2 \log P}
\text{ and } \beta \defi \liminf_{P \to \infty} \frac{I(X_5^n; Y_{6}^n|\tilde X_B^n)}{\frac n2 \log P} .\]
It is easy to see that $0 \leq \alpha,\beta \leq 1$. 
Intuitively, since all the information from the sources must go through either $v_4$ or the nodes in $B$ to reach $v_2$,
$\alpha$ can be thought of as the number of useful degrees-of-freedom (i.e., carrying information about the sources) transmitted by $v_2$.
Similarly, $\beta$ can be thought of as the number of degrees-of-freedom transmitted by $v_5$, but only counting the degrees-of-freedom with information about message $W_1$ (since we condition on $\tilde X_B^n$).
Based on these quantities we will state three inequalities related to the degrees-of-freedom that can be achieved, and for each one we will provide an intuitive explanation.
The formal proof is omitted, but it follows from the information inequalities we will derive later based on properties P1-P8.
In the sense of Definition \ref{defn_region}, we let $D_i$ be the degrees-of-freedom assigned to $(s_i,d_i)$, for $i=1,2$. 
%
%
First, we have
\begin{equation}
D_1 \leq \beta, \label{conv1}
\end{equation}
since all information from $W_1$ must flow through $v_5$.
Next, we claim that both $W_1$ and $W_2$ can be decoded from $Y_4^n$ and $\tilde X_2^n$, and thus
\begin{equation}
D_1 + D_2 \leq 1 + \alpha. \label{conv2}
\end{equation}
To see this, we first notice that, since the removal of $v_4$ and $v_2$ disconnects $d_1$ from $\{s_1,s_2\}$,
from $Y_4^n$ and $\tilde X_2^n$, $W_1$ can be decoded.
Then, $W_1$ can be used to approximately obtain $X_A^n$ (since the nodes in $A$ cannot be influenced by $W_2$), and, by removing its contribution from $Y_4^n$, we can obtain a noisy version of the transmit signal from $v_3$.
But since all the information about $W_2$ must flow through $v_3$, this allows one to use $Y_4^n$ and $\tilde X_2^n$ to decode $W_2$ as well.
%
For the third inequality we claim that, from $Y_0^n$, we can decode $W_2$ completely and $(\alpha + \beta -1)$ degrees-of-freedom of $W_1$, and thus
\begin{equation}
 D_2 + (\alpha+\beta-1) \leq 1. \label{conv3}
\end{equation}
To see this, we first notice that, since the removal of $v_0$ disconnects $d_2$ from $\{s_1,s_2\}$, from $Y_0^n$, we can decode $W_2$, and thus obtain $X_B^n$ approximately. 
By removing its contribution from $Y_0^n$, we obtain a noisy version of the transmit signal from node $v_2$, which allows us to decode the $\alpha$ degrees-of-freedom transmitted by it.
Now we ask ourselves how many of the $\alpha$ degrees-of-freedom transmitted by $v_2$ must be carrying information about $W_1$.
To answer this question, we notice that all the degrees-of-freedom transmitted by $v_2$ must have come through node $v_6$.
Since node $v_6$ receives $\beta$ degrees-of-freedom with information about $W_1$ from $v_5$, at most $1-\beta$ of its degrees-of-freedom can be not about $W_1$.
Thus, any number of degrees-of-freedom above $1-\beta$ that $v_2$ transmits, i.e., $\alpha - (1-\beta) = \alpha + \beta -1$, must contain information about $W_1$.
%
Finally, by adding inequalities (\ref{conv1}), (\ref{conv2}) and (\ref{conv3}), we obtain $2(D_1 + D_2) \leq 3$, and therefore $\dE \leq 3/2$. 

Next, we formally derive information inequalities that can be used to show that $\dE \leq 3/2$ for all networks in case C\ref{sym}.
The intuition is similar to that of inequalities (\ref{conv1}), (\ref{conv2}) and (\ref{conv3}), but the inequalities are somewhat different since they need to hold for any network in case C\ref{sym}.
First we have \rescnt
\begin{align} 
n R_2 & 
\leq I(W_2;Y_{d_2}^n)+n\epsilon_n \leqnum I(\tilde X_B^n;Y_0^n) + n\epsilon_n \non 
& = I(X_2^n,\tilde X_B^n;Y_0^n)-I(X_2^n;Y_0^n|\tilde X_B^n) + n\epsilon_n \nonumber \\
& \leqnum \frac n 2\log P + n\cons - I(X_2^n;Y_0^n|\tilde X_B^n) + n\epsilon_n,\label{c1eq1}
\end{align} 
where \rescnt
\cnt follows from the Markov chain $\fMC{W_2}{\tilde X_B^n}{Y_0^n}{Y_{d_2}^n}$, which is implied by P\ref{pv0} and the fact that $s_2 \in B$; 
\cnt follows from the fact that $I(X_2^n,\tilde X_B^n;Y_0^n)$ can be upper bounded by $h(Y_0^n)-h(N_{2,0}^n)$ by following the steps in (\ref{deriv}), where $\curcons$ is a positive constant, independent of $P$, for $P$ sufficiently large. We also have that  
\rescnt
\begin{align} 
n& R_1  \leq I(W_1;Y_{d_1}^n)+n\epsilon_n \leqnum I(W_1;\tilde X_5^n,\tilde X_B^n)+n\epsilon_n \non
& \eqnum I(W_1;\tilde X_5^n|\tilde X_B^n)+n\epsilon_n \leqnum I(X_5^n;\tilde X_5^n|\tilde X_B^n)+n\epsilon_n \non
& \leq I(X_5^n;Y_6^n|\tilde X_B^n) + I(X_5^n;\tilde X_5^n|\tilde X_B^n,Y_6^n) + n\epsilon_n  \nonumber \\
& \eqnum I(X_5^n;Y_6^n|\tilde X_B^n) + n\cons + n\epsilon_n, \label{c1eq2}
\end{align} 
 \rescnt
where 
\cnt follows because P\ref{pv5} and the fact that $s_2 \in B$ imply that the removal of $v_5$ and $B$ disconnects $d_1$ from both sources and thus $\MC{W_1}{(\tilde X_5^n,\tilde X_B^n)}{Y_{d_1}^n}$; 
\cnt follows from the fact that $\tilde X_B$ is independent of $W_1$; 
\cnt follows from the fact that, given $\tilde X_B^n$, we have $\MC{W_1}{X_5^n}{\tilde X_5^n}$; 
\cnt follows from Lemma \ref{const}, since P\ref{pv5v6} implies that $\I(v_6)\setminus \{v_5\} \subset B$.
To obtain the next inequalities, we consider two cases, according to the position of $v_4$ and $v_5$.

\vspace{2mm}

\noindent I) $\ell(v_4) \leq \ell(v_5)$: In this case, we have
{
\begin{align} \rescnt
n & R_2 \leq I(W_2;Y_{d_2}^n) + n\epsilon_n \leqnum I(X_{s_2}^n;\tilde X_2^n,\tilde X_3^n) + n\epsilon_n \nonumber \\ 
& \leqnum I(X_{s_2}^n;\tilde X_2^n,\tilde X_3^n| \tilde X_A^n) + n\epsilon_n \non
& \leq I(X_{s_2}^n;\tilde X_2^n,\tilde X_3^n,Y_4^n| \tilde X_A^n) + n\epsilon_n \nonumber \\
& = I(X_{s_2}^n;\tilde X_{3}^n,Y_4^n| \tilde X_A^n)  
+ I(X_{s_2}^n;\tilde X_2^n| \tilde X_A^n, \tilde X_3^n,Y_4^n) + n\epsilon_n \nonumber \\ 
& \leq I(X_{s_2}^n;Y_4^n| \tilde X_A^n)+I(X_{s_2}^n;\tilde X_{3}^n| \tilde X_A^n,Y_4^n) \non
& \quad \quad \quad + I(X_{s_2}^n,\tilde X_3^n;\tilde X_2^n| \tilde X_A^n, Y_4^n) + n\epsilon_n  \nonumber \\
& \leqnum I(X_{s_2}^n;Y_4^n | \tilde X_A^n) +n\cons  \non
& \quad \quad \quad +I(X_{s_2}^n,\tilde X_3^n;\tilde X_2^n| \tilde X_A^n, Y_4^n)+ n\epsilon_n  \nonumber \\
& \leqnum I(X_B^n;Y_4^n | \tilde X_A^n) \non 
& \quad \quad \quad +I(X_{s_2}^n,\tilde X_3^n;\tilde X_2^n| \tilde X_A^n, Y_4^n) +n\curcons+n\epsilon_n  \nonumber \\
& \leq I(X_B^n;Y_4^n | \tilde X_A^n)+I(X_B^n;\tilde X_2^n| \tilde X_A^n, Y_4^n) \non
& \quad \quad \quad +I(X_{s_2}^n,\tilde X_3^n;\tilde X_2^n| \tilde X_A^n, Y_4^n,X_B^n)+n\curcons+n\epsilon_n  \nonumber \\
& \leqnum I(X_B^n;Y_4^n | \tilde X_A^n) 
+I(X_B^n;\tilde X_2^n| \tilde X_A^n, Y_4^n) +n\curcons+n\epsilon_n  \nonumber \\
& \leq I(X_B^n;Y_4^n,\tilde X_2^n| \tilde X_A^n) +n\curcons + n\epsilon_n, \label{c1eq3} 
\end{align}}
where \rescnt
\cnt follows because P\ref{pv2v3} implies the Markov chain $\fMC{W_2}{X_{s_2}^n}{(\tilde X_2^n,\tilde X_3^n)}{Y_{d_2}^n}$; 
\cnt follows from the fact that $\tilde X_A^n$ is independent of $X_{s_2}^n$; 
\cnt follows by applying Lemma \ref{const} to the second term, since $\ell(v_4)\leq\ell(v_5)$ implies that $\I(v_4)\setminus \{v_3\} \subset A$, or else we contradict P\ref{ps2d1}; 
\cnt follows from the fact that $s_2 \in B$; 
and \cnt follows because we have 
$\MC{(X_{s_2}^n,\tilde X_3^n)}{(\tilde X_A^n, Y_4^n,X_B^n)}{\tilde X_2^n}$, since the removal of $A$, $v_4$ and $B$ disconnects $s_2$ and $\Os(v_3)$ from $v_2$. This can be seen as follows. From P\ref{pv6}, all paths from $s_2$ or $v_3$ to $v_2$ must contain a node in $\I(v_6)$. From P\ref{pv5v6}, we know that $\I(v_6)\setminus \{v_5\} \subset B$. From P\ref{ps2d1}, we know that any path from $v_3$ or $s_2$ to $v_5$ must contain $v_4$. Finally, since $\ell(v_4) < \ell(v_6)$, we have that $v_3 \notin \I(v_6)$, and, therefore, any path from $s_2$ or $\Os(v_3)$ to $v_2$ must either contain $v_4$ or a node in $B$. Notice that we had to consider $\Os(v_3)$ instead of simply $v_3$, because we have $\tilde X_3^n$, and not $X_3^n$. Next, we have that 
\rescnt
\begin{align} 
n &R_1  \leq I(W_1;Y_{d_1}^n)+n\epsilon_n \leqnum I(W_1;Y_4^n,\tilde X_2^n) + n\epsilon_n \nonumber \\ 
& \leqnum I(\tilde X_A^n ;Y_4^n,\tilde X_2^n) + n\epsilon_n \non 
& = I(\tilde X_A^n , X_B^n ;Y_4^n,\tilde X_2^n) - I( X_B^n ;Y_4^n,\tilde X_2^n|\tilde X_A^n ) + n\epsilon_n \nonumber \\
&= I(\tilde X_A^n , X_B^n ;Y_4^n)+I(\tilde X_A^n , X_B^n ;\tilde X_2^n|Y_4^n) \non 
& \quad \quad \quad - I( X_B^n ;Y_4^n,\tilde X_2^n|\tilde X_A^n ) + n\epsilon_n \nonumber \\
& \leqnum \frac n2\log P + n\cons +I(\tilde X_A^n , X_B^n, Y_4^n;\tilde X_2^n) \non
& \quad \quad \quad - I( X_B^n ;Y_4^n,\tilde X_2^n|\tilde X_A^n ) + n\epsilon_n, \nonumber
\end{align}
where  \rescnt
\cnt follows because P\ref{pv2v4} implies the Markov chain $\MC{W_1}{(Y_4^n,\tilde X_2^n)}{Y_{d_1}^n}$; 
\cnt follows since $s_1 \in A$; 
\cnt follows from the fact that $I(\tilde X_A^n , X_B^n ;Y_4^n)$ can be upper bounded by $h(Y_4^n)-h(N_{3,4}^n)$ by following the steps in (\ref{deriv}), where $\curcons$ is a positive constant, independent of $P$, for $P$ sufficiently large. The second term in the inequality above can be bounded as 
\begin{align} \rescnt
I(\tilde X_A^n &, X_B^n, Y_4^n;\tilde X_2^n) \leqnum I(\tilde X_A^n ,\tilde X_B^n, Y_4^n;\tilde X_2^n) \nonumber \\
& = I(\tilde X_B^n;\tilde X_2^n) + I(\tilde X_A^n , Y_4^n;\tilde X_2^n|\tilde X_B^n) \nonumber \\
& \leqnum I(\tilde X_B^n;Y_6^n) + I(\tilde X_A^n , Y_4^n;\tilde X_2^n|\tilde X_B^n) \non 
&\leqnum I(\tilde X_B^n;Y_6^n) + I(X_2^n;\tilde X_2^n|\tilde X_B^n) \nonumber \\
& \leq I(X_5^n,\tilde X_B^n;Y_6^n) - I(X_5^n;Y_6^n|\tilde X_B^n) \non
& \quad \quad \quad + I(X_2^n; Y_0^n |\tilde X_B^n) + I(X_2^n;\tilde X_2^n |\tilde X_B^n, Y_0^n) \nonumber \\
& \leqnum I(X_5^n,\tilde X_B^n;Y_6^n) - I(X_5^n;Y_6^n|\tilde X_B^n) \non 
& \quad \quad \quad+ I(X_2^n; Y_0^n |\tilde X_B^n) + n \cons \nonumber \\
& \leqnum \frac n2\log P - I(X_5^n;Y_6^n|\tilde X_B^n) \non
& \quad \quad \quad+ I(X_2^n; Y_0^n |\tilde X_B^n) + n(\curcons + \cons) \label{endcase1}
\end{align}
where \rescnt
\cnt follows because of the Markov chain $\MC{(\tilde X_A^n,X_B^n,Y_4^n)}{(\tilde X_A^n,\tilde X_B^n,Y_4^n)}{\tilde X_2^n}$; 
\cnt follows because P\ref{pv6} implies $\MC{\tilde X_B^n}{Y_6^n}{\tilde X_2^n}$; 
\cnt follows since, given $X_B^n$, we have $\MC{(\tilde X_A^n,Y_4^n)}{X_2^n}{\tilde X_2^n}$; 
\cnt follows by applying Lemma \ref{const} to $I(X_2^n;\tilde X_2^n|\tilde X_B^n, Y_0^n )$, since $\I(v_0)\setminus \{v_2\} \subset B$, or else we contradict P\ref{pv2v0}; 
\cnt follows from the fact that $I(X_5^n,\tilde X_B^n;Y_6^n)$ can be upper bounded by $h(Y_6^n)-h(N_{5,6}^n)$ by following the steps in (\ref{deriv}), where $\curcons$ is a positive constant, independent of $P$, for $P$ sufficiently large. Thus, we obtain
\begin{align}
n  &R_1 \leq n \log P - I(X_5^n;Y_6^n|\tilde X_B^n) + I(X_2^n; Y_0^n |\tilde X_B^n) \non
& - I( X_B^n ;Y_4^n,\tilde X_2^n|\tilde X_A^n ) + n (K_6 + K_7 + K_8) + n\epsilon_n. \label{c1eq4}
\end{align}

\vspace{2mm}

\noindent II) $\ell(v_4) > \ell(v_5)$: We will obtain similar inequalities to the ones in case I. We will define $C \defi \I(v_4) \setminus \{v_2,v_3\}$ and $D \defi \Os(v_3) \setminus \{v_6\}$. Then, we will let $Y_{C,4}^n = \sum_{v_c \in C} \tilde X_{c,4}^n$. We also let $\tilde X_{3,D}^n = \{\tilde X_{3,v_j}^n : v_j \in D\}$. Notice that $\tilde X_{3,D}^n = \tilde X_3^n$ if $v_6 \notin \Os(v_3)$. Then we have \rescnt
\begin{align} 
& n R_2  \leq I(W_2;Y_{d_2}^n) + n\epsilon_n \leqnum I(X_{s_2}^n;\tilde X_2^n,\tilde X_{3,D}^n) + n\epsilon_n \nonumber \\ 
& \leqnum I(X_{s_2}^n;\tilde X_2^n,\tilde X_{3,D}^n| \tilde X_A^n) + n\epsilon_n \non
& = I(X_{s_2}^n;\tilde X_2^n| \tilde X_A^n) + I(X_{s_2}^n;\tilde X_{3,D}^n| \tilde X_A^n, \tilde X_2^n) + n\epsilon_n \nonumber \\
& \leq  I(X_{s_2}^n;\tilde X_2^n| \tilde X_A^n) \non 
& \quad \quad \quad + I(X_{s_2}^n;\tilde X_{3,D}^n,\tilde X_{3,4}^n| \tilde X_A^n, \tilde X_2^n) + n\epsilon_n \nonumber \\
& \leqnum I(X_{B}^n;\tilde X_2^n| \tilde X_A^n) + I(X_{s_2}^n;\tilde X_{3,4}^n| \tilde X_A^n, \tilde X_2^n) \non
& \quad \quad \quad + I(X_{s_2}^n;\tilde X_{3,D}^n| \tilde X_A^n, \tilde X_2^n,\tilde X_{3,4}^n) + n\epsilon_n \nonumber \\
& \leqnum I(X_B^n;\tilde X_2^n| \tilde X_A^n) + I(X_{s_2}^n;\tilde X_{3,4}^n| \tilde X_A^n, \tilde X_2^n) + n \cons + n\epsilon_n \nonumber \\
& \leqnum I(X_B^n;\tilde X_2^n| \tilde X_A^n) + I(X_B^n;\tilde X_{3,4}^n| \tilde X_A^n, \tilde X_2^n) + n \curcons + n\epsilon_n \nonumber \\
& \leqnum I(X_B^n;\tilde X_2^n| \tilde X_A^n) \non 
& \quad \quad \quad + I(X_B^n;\tilde X_{3,4}^n| \tilde X_A^n, \tilde X_2^n, \tilde X_C^n) + n \curcons + n\epsilon_n \label{c2eq3}
\end{align}
where \rescnt
\cnt follows because P\ref{pv2v3} implies that the removal of $\Os(v_3)$ and $v_2$ disconnects $d_2$ from both sources. Then, since P\ref{pv2v0} implies that all paths from $v_6$ to $d_2$ contain $v_2$, we know that the removal of $D$ and $v_2$ also disconnects $d_2$ from both sources, and we have the Markov chain $\fMC{W_2}{X_{s_2}^n}{(\tilde X_2^n,\tilde X_{3,D}^n)}{Y_{d_2}^n}$; 
\cnt follows from the fact that $\tilde X_A^n$ is independent of $X_{s_2}^n$; 
\cnt follows since $s_2 \in B$; 
\cnt follows by applying Lemma \ref{const} to $I(X_{s_2}^n;\tilde X_{3,D}^n| \tilde X_A^n, \tilde X_2^n,\tilde X_{3,4}^n)$, since, in case II, if $u \in D \setminus \{v_4\}$, then $u \not\leadsto v_2$, or else we contradict P\ref{pv6}; 
\cnt follows since $s_2 \in B$; 
and \cnt follows from the fact that, given $\tilde X_2^n$ and $\tilde X_A^n$, $\tilde X_C^n$ is independent of $X_B^n$. This is true because P\ref{ps2d1} implies that any path from a node in $B$ to a node in $C$ must contain $v_2$, and, thus, the removal of $A$ and $v_2$ disconnects $C$ from $B$ and both sources. Notice that $(vi)$ is only non-trivial in the cases where $C \not\subset A$ (when $\ell(v_4) > \ell(v_1)+1$). 
Next, we have that \rescnt
\begin{align} 
n R_1 & \leq I(W_1;Y_{d_1}^n)+n\epsilon_n \leqnum I(W_1;\tilde X_{3,4}^n+ Y_{C,4}^n,\tilde X_2^n) + n\epsilon_n \nonumber \\ 
& \leqnum I(\tilde X_A^n ;\tilde X_{3,4}^n+ Y_{C,4}^n,\tilde X_2^n) + n\epsilon_n \non 
& = I(\tilde X_A^n ;\tilde X_2^n)+I(\tilde X_A^n ;\tilde X_{3,4}^n+ Y_{C,4}^n|\tilde X_2^n) + n\epsilon_n \nonumber \\
& \leq I(\tilde X_A^n ,\tilde X_C^n;\tilde X_{3,4}^n + Y_{C,4}^n|\tilde X_2^n)+ I(\tilde X_A^n ;\tilde X_2^n) + n\epsilon_n \nonumber \\
& = I(\tilde X_A^n,\tilde X_C^n , X_B^n;\tilde X_{3,4}^n + Y_{C,4}^n|\tilde X_2^n) \non 
& \quad \quad \quad \quad- I( X_B^n;\tilde X_{3,4}^n+\tilde Y_{C,4}^n|\tilde X_2^n,\tilde X_A^n ,\tilde X_C^n)  \nonumber \\
& \quad \quad \quad \quad + I(\tilde X_A^n , X_B^n;\tilde X_2^n) - I(X_B^n;\tilde X_2^n|\tilde X_A^n ) + n\epsilon_n \nonumber \\
& \leq I(\tilde X_A^n,\tilde X_C^n , X_B^n, \tilde X_2^n;\tilde X_{3,4}^n+\tilde Y_{C,4}^n) \non
& \quad\quad\quad\quad - I( X_B^n;\tilde X_{3,4}^n|\tilde X_2^n,\tilde X_A^n ,\tilde X_C^n)  \nonumber \\
& \quad \quad \quad \quad + I(\tilde X_A^n , X_B^n;\tilde X_2^n) - I(X_B^n;\tilde X_2^n|\tilde X_A^n ) + n\epsilon_n \nonumber \\
& \leqnum \frac n2\log P + n \cons + I(\tilde X_A^n , X_B^n;\tilde X_2^n) \nonumber \\
& - I( X_B^n;\tilde X_{3,4}^n|\tilde X_2^n,\tilde X_A^n ,\tilde X_C^n) - I(X_B^n;\tilde X_2^n|\tilde X_A^n ) + n\epsilon_n, \nonumber
\end{align}
where \rescnt
\cnt follows because P\ref{pv2v4} implies the Markov chain $\MC{W_1}{(Y_4^n,\tilde X_2^n)}{Y_{d_1}^n}$, and $(Y_4^n,\tilde X_2^n)$ can be constructed from $(\tilde X_{3,4}+ Y_{C,4}^n,\tilde X_2)$ (notice that it may be the case that $Y_4^n = \tilde X_{3,4}+\tilde Y_{C,4}^n+\tilde X_{2,4}$, if $v_2 \in \I(v_4)$); 
\cnt follows since $s_1 \in A$; 
\cnt follows from the fact that $I(\tilde X_A^n,\tilde X_C^n , X_B^n, \tilde X_2^n;\tilde X_{3,4}+Y_{C,4}^n)$ can be upper bounded by $h(\tilde X_{3,4}+Y_{C,4}^n)-h(N_{3,4}^n)$ by following the steps in (\ref{deriv}), where $\curcons$ is a positive constant, independent of $P$, for $P$ sufficiently large. The second term in the inequality above can be bounded as 
\rescnt
\begin{align} 
I(\tilde X_A^n &, X_B^n;\tilde X_2^n) \leq I(\tilde X_A^n, \tilde X_B^n;\tilde X_2^n) \non 
& = I(\tilde X_B^n;\tilde X_2^n) + I(\tilde X_A^n ;\tilde X_2^n|\tilde X_B^n) \nonumber \\
& \leq I(\tilde X_B^n;Y_6^n) + I(X_2^n;\tilde X_2^n|\tilde X_B^n) \nonumber \\
& \leq I(X_5^n,\tilde X_B^n;Y_6^n) - I(X_5^n;Y_6^n|\tilde X_B^n) \non 
& \quad \quad \quad + I(X_2^n;Y_0^n |\tilde X_B^n) + n \cons \nonumber \\
& \leq \frac n2\log P - I(X_5^n;Y_6^n|\tilde X_B^n) + I(X_2^n; Y_0^n |\tilde X_B^n) \non
& \quad \quad \quad  + n (\curcons + \cons) \nonumber
\end{align}
where the inequalities are justified as in (\ref{endcase1}). Therefore, we obtain
\begin{align}
n R_1 & \leq n \log P - I(X_5^n;Y_6^n|\tilde X_B^n) + I(X_2^n; Y_0^n |\tilde X_B^n) \non 
& - I( X_B^n;\tilde X_{3,4}^n|\tilde X_2^n,\tilde X_A^n ,\tilde X_C^n) \nonumber \\ 
&  - I(X_B^n;\tilde X_2^n|\tilde X_A^n )  + n (K_{10}+K_{11}+K_{12}) + n\epsilon_n. \label{c2eq4}
\end{align}
Finally, by adding equations (\ref{c1eq1}), (\ref{c1eq2}), (\ref{c1eq3}) and (\ref{c1eq4}) for case I, and (\ref{c1eq1}), (\ref{c1eq2}), (\ref{c2eq3}) and (\ref{c2eq4}) for case II, we obtain
\begin{align*}
& 2n ( R_1 +  R_2) \leq \frac{3n}2 \log P + 6 n K_{\max} + n\epsilon_n \\
\Rightarrow & \frac{R_1+R_2}{\frac12\log P} \leq \frac32+ \frac{6 K_{\max}+\epsilon_n}{ \log P},
\end{align*} 
where $K_{\max} = \max_j K_j$. Thus, as we let $n\rightarrow \infty$ and then $P \rightarrow \infty$, we obtain
\[ \dE \leq \frac32. \]

\end{subsection}

We now proceed to considering C\ref{nonsym}. We will show that if our network $\N$ does not fall in cases (A), (A\pr), (B), and (B\pr), then $\dE = \frac32$.

\begin{subsection}{Achievability for case C\ref{nonsym}} \label{nonsymsec}

In this section, we will show that if we are in C\ref{nonsym} and no edge as in (A\pr) exists, then we can also achieve $\frac32$ degrees-of-freedom.
%
%
We start by proving properties about the connectivity of our network, if we are in C\ref{nonsym}. Notice that, if for some choice of two disjoint paths $\p11'$ and $\p22'$ we are in C\ref{sym}, our previous result shows that $\dE = \frac32$. Therefore, we may assume that for no choice of two disjoint paths we are in C\ref{sym}. 
So we suppose we have two disjoint paths $\p11$ and $\p22$, but no two disjoint paths with manageable interference. In addition, we assume that we do not have an edge as in (A\pr). Since we are in C\ref{nonsym}, we have that $n_1(G,\p11) = n_1^D(\p22,\p11) = 1$ and we let $(v_2,v_1) \in E$ be the unique edge such that $v_2 \in  \p22$ and $v_1 \in \p11$. 

\begin{enumerate}[P1.]

\item All paths from $s_2$ to $d_1$ contain $v_2$ and $v_1$. \label{qs2d1}

If we have a path $\p21$ not containing $\{v_2,v_1\}$, then we must have $n_1(G,\p11) \geq 2$, thus contradicting the fact that we are in C\ref{nonsym}.

\item There exists a path $\q ii$ such that $v_i \notin \q ii$, and $\q ii \cap \p{\bar i}{\bar i} = \emptyset$, for $i =1$ or 2.   \label{qpath}

Since we have no edge as in (A\pr), we may assume that either the removal of $v_1$ does not disconnect $d_1$ from both sources, or the removal of $v_2$ does not disconnect $s_2$ from both destinations. However, from P\ref{qs2d1}, the removal of $v_1$ or $v_2$ disconnects $s_2$ from $d_1$. Therefore, we must have a path $\q ii$ such that $v_i \notin \q ii$, for $i =1$ or 2. Moreover, if $\q ii$ is not disjoint of $\p{\bar i}{\bar i}$, we would contradict P\ref{qs2d1}, since there would be a path $\p21 \subset \q ii \cup \p{\bar i}{\bar i}$ and $v_i \notin \q ii \cup \p{\bar i}{\bar i}$. 

\item If $i=1$, we have $n_1^D(\p22,\q11) = 0$ and $n_2^D(\q11,\p22) = 1$, and if $i = 2$, we have $n_1^D(\q22,\p11) = 0$ and $n_2^D(\p11,\q22) = 1$. \label{q01}

Since $v_i \notin \q ii \cup \p{\bar i}{\bar i}$, we must have $n_1^D(\p22,\q11) = 0$ (if $i=1$) or $n_1^D(\q22,\p11) = 0$ (if $i=2$), or else we would have a path from $s_2$ to $d_1$ not containing $v_i$, and we would contradict P\ref{qs2d1}. Then, since $\q ii$ and $\p{\bar i}{\bar i}$ do not have manageable interference, we must have $n_2^D(\q11,\p22) = 1$ (if $i=1$) and $n_2^D(\p11,\q22) = 1$ (if $i=2$).

\end{enumerate}

\noindent Since $n_2^D(\q11,\p22) = 1$ (if $i=1$) or $n_2^D(\p11,\q22) = 1$ (if $i=2$), we can assume we have an edge $(v_3,v_4) \in E$ such that $v_{2+ i} \in \q ii$ and $v_{2+\bar i} \in \p{\bar i}{\bar i}$. Then we have the following properties.

\begin{enumerate}[P1.] \setcounter{enumi}{3}

\item All paths from $s_1$ to $d_2$ contain $v_3$ and $v_4$. \label{qv3v4}

Suppose we have a path $\p12$ such that $\{v_3,v_4\} \not\subset \p12$. This implies that $n_2(G,\p22) \geq 2$ if $i=1$ and $n_2(G,\q22) \geq 2$, if $i=2$. From P\ref{q01} we have that $n_1^D(\p22,\q11) = 0$ (if $i=1$) or $n_1^D(\q22,\p11) = 0$ (if $i=2$). Therefore, paths $\q ii$ and $\p{\bar i}{\bar i}$ may not fall in C\ref{nonsym} (not even by exchanging $(s_1,d_1)$ and $(s_2,d_2)$), and must fall in C\ref{sym}. Since we know that, for networks in C\ref{sym}, $\dE = \frac32$, we may disregard such cases.  

\item There exists a path $\z ii$ such that $v_{2+ i} \notin \z ii$, and $\z ii \cap \p{\bar i}{\bar i} = \emptyset$.  \label{zpath}

Since we are not in (A\pr), either the removal of $v_4$ does not disconnect $d_2$ from both sources, or the removal of $v_3$ does not disconnect $s_1$ from both destinations. From P\ref{qv3v4}, we know that the removal of $v_3$ or $v_4$ disconnects $s_1$ from $d_2$. Thus we must either have a path $\z ii$ such that $v_{2+ i} \notin \z ii$ or a path $\z{\bar i}{\bar i}$ such that $v_{2+ \bar i} \notin \z{\bar i}{\bar i}$. If we have a path $\z{\bar i}{\bar i}$ such that $v_{2+ \bar i} \notin \z{\bar i}{\bar i}$, then $\z{\bar i}{\bar i}$ may not intersect $\q ii$, since that would imply the existence of a path from $s_1$ to $d_2$ not containing $\{v_3,v_4\}$ and we would contradict P\ref{qv3v4}. Moreover, we must have $n_1^D(\z22,\q11) = n_2^D(\q11,\z22) = 0$ (if $i=1$) or $n_1^D(\q22,\z11) = n_2^D(\z11,\q22) = 0$ (if $i=2$). Otherwise, since $v_i, v_{2 + \bar i} \notin \q ii \cup \z{\bar i}{\bar i}$, we would contradict either P\ref{qs2d1} or P\ref{qv3v4}. But this means that $\q ii$ and $\z{\bar i}{\bar i}$ have manageable interference, which is a contradiction. Therefore, we have a path $\z ii$ such that $v_{2+ i} \notin \z ii$. The fact that $\z ii \cap \p{\bar i}{\bar i} = \emptyset$ follows since otherwise we would have a path from $s_1$ to $d_2$ not containing $v_{2+i}$.

\item If $i=1$, we have $n_1^D(\p22,\z11) = 1$ and $n_2^D(\z11,\p22) = 0$, and if $i = 2$, we have $n_1^D(\z22,\p11) = 1$ and $n_2^D(\p11,\z22) = 0$. \label{q10}

Since $v_{2+i} \notin \z ii \cup \p{\bar i}{\bar i}$, we must have $n_2^D(\z11,\p22) = 0$ (if $i=1$) or $n_2^D(\p11,\z22) = 0$ (if $i=2$), or else we would have a path from $s_1$ to $d_2$ not containing $v_{2+i}$, and we would contradict P\ref{qv3v4}. Then, since $\z ii$ and $\p{\bar i}{\bar i}$ do not have manageable interference, we must have $n_1^D(\p22,\z11) = 1$ (if $i=1$) and $n_1^D(\z22,\p11) = 1$ (if $i=2$).

\end{enumerate}

\noindent Since $n_1^D(\p22,\z11) = 1$ (if $i=1$) and $n_1^D(\z22,\p11) = 1$ (if $i=2$), we can assume we have an edge $(v_6,v_5) \in E$ such that $v_{4+ i} \in \z ii$ and $v_{4+\bar i} \in \p{\bar i}{\bar i}$. However, we claim that we must have $v_6 = v_2$ and $v_5 = v_1$. If $v_i \in \z ii$ this is obvious because $v_{\bar i} \in \p{\bar i}{\bar i}$. If $v_i \notin \z ii$, then, if $(v_6,v_5) \ne (v_2,v_1)$, we would have a path from $s_2$ to $d_1$ not containing $v_i$, thus contradicting P\ref{qs2d1}. 

Next, we notice that we can assume WLOG that $i=1$. If $i=2$, we can first switch the names of $(s_1,d_1)$ and $(s_2,d_2)$. Then we also switch the names of $\z11$ and $\q11$, and of $(v_2,v_1)$ and $(v_3,v_4)$, and we obtain the case where $i=1$. Thus, from now on we assume $i = 1$. 
\begin{figure*}[ht]
\begin{center}
\includegraphics[height=30mm]{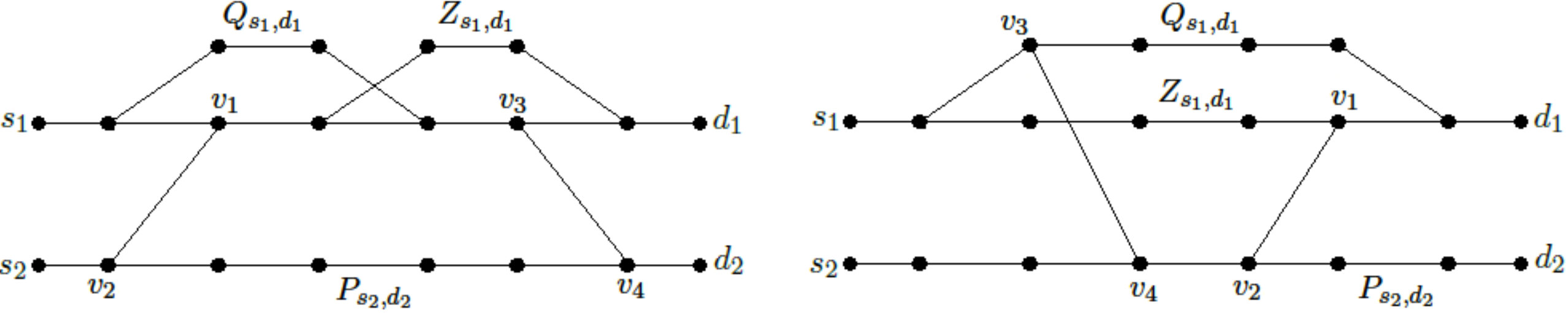}
\caption{Examples of networks in case C\ref{nonsym}.} \label{netscnonsym}
\end{center} 
\end{figure*}
We will build our achievability scheme based on the paths $\z11$, $\q11$ and $\p22$, an edge $(v_2,v_1)$ such that $v_2 \in \p22$ and $v_1 \in \z11$ but $v_1 \notin \q11$, and an edge $(v_3,v_4)$ such that $v_4 \in \p22$ and $v_3 \in \q11$ but $v_3 \notin \z11$. Two examples of networks in C\ref{nonsym} that satisfy P1-P6 for $i=1$ are shown in Figure \ref{netscnonsym}.

We will now consider two cases and provide a scheme to achieve $\frac32$ degrees-of-freedom in each case. Our schemes will once more be based on using two modes of operation and having nodes store the received signals during the first mode of operation and use them during the second mode of operation.

\begin{subsubsection}{$\ell(v_3) \geq \ell(v_1)$}
\label{buf3}

In Mode 1, we let the node from $\p22$ in $V_{\ell(v_1)}$ be a virtual destination $d_2'$. Any node $v \in \p22$ such that $\ell(v) \geq \ell(d_2')$ will stay silent during Mode 1. Then we notice that the two disjoint paths $\q11$ and $P_{s_2,d_2'}$ have no direct edge between them and thus have manageable interference. Therefore, it is possible to guarantee that the transfer matrix between $(s_1,s_2)$ and $(d_1,d_2')$ is diagonal with non-zero diagonal entries. During Mode 1, $d_2' $ will store its received signals. 

The second mode of operation should last for the same number of time steps as the first one. In Mode 2, $d_2'$ will become a virtual source $s_2'$. Then, we remove all the nodes from the network except those in the paths $\z11$ and $P_{s_2',d_2}$. We again have two disjoint paths with no direct interference. Therefore, we can have the transfer matrix between $(s_1,s_2')$ and $(d_1,d_2)$ be diagonal with non-zero diagonal entries. Thus, by letting node $d_2' = s_2'$ forward each of the signals received during Mode 1 in Mode 2, it is clear that, over the two modes, we create three parallel AWGN channels, two of them between $s_1$ and $d_1$ and one of them between $s_2$ and $d_2$. Therefore, we achieve $\dE = \frac32$. A visual representation of the scheme is shown in Figure \ref{buffig3}. 

\begin{figure}[ht]
\begin{center}
\includegraphics[height=80mm]{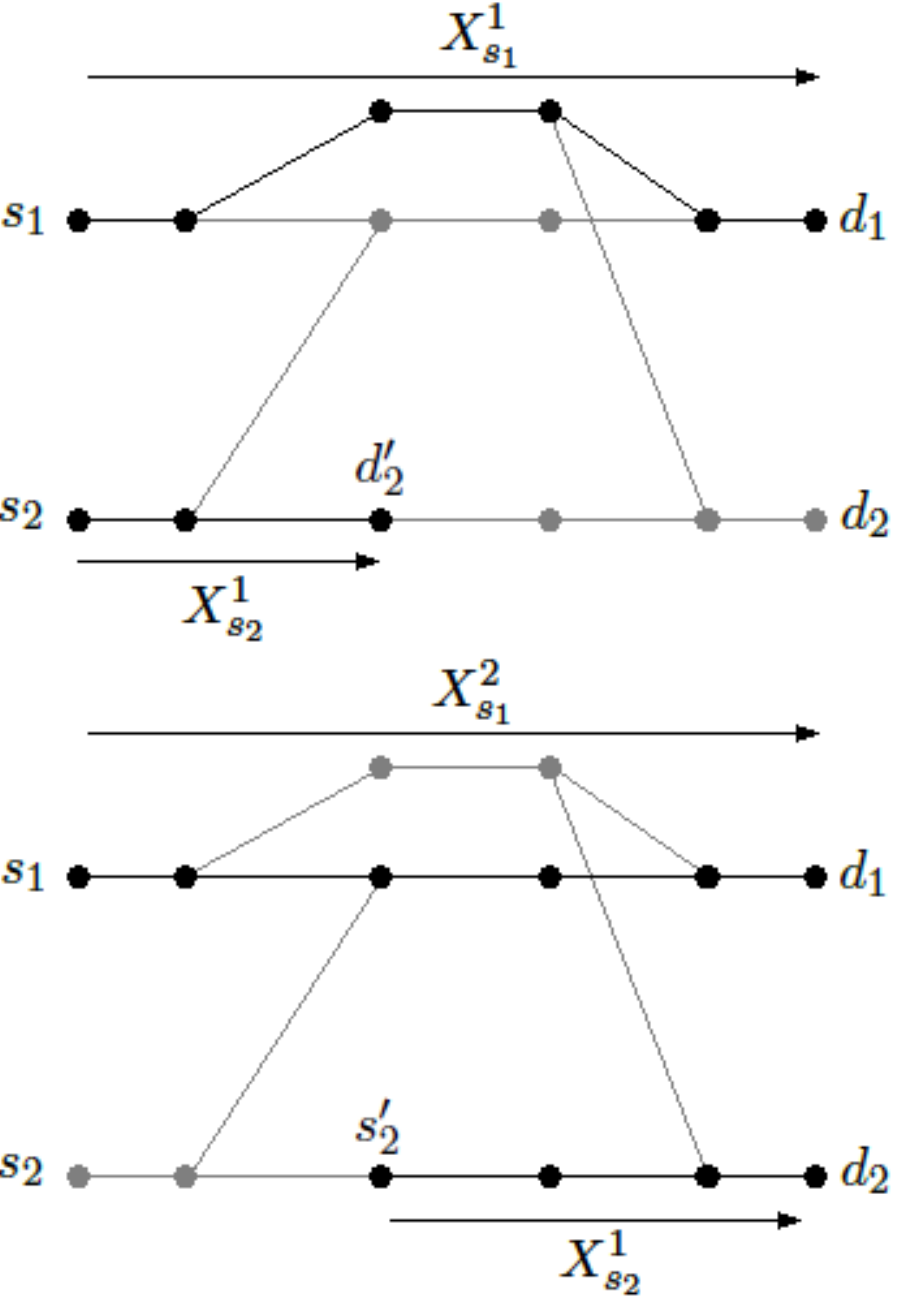}
\caption{Depiction of Mode 1 and Mode 2 for the achievability scheme in case C\ref{nonsym} if $\ell(v_3) \geq \ell(v_1)$.} \label{buffig3}
\end{center} 
\end{figure}

\end{subsubsection}

\begin{subsubsection}{$\ell(v_3) < \ell(v_1)$}
\label{buf4}

In Mode 1, we let $v_1$ be a virtual destination $d_2'$. Then we consider the path $P_{s_2,d_2'} = \p22[s_2,v_2]\concat (v_2,v_1)$. Then we notice that $\q11$ and $P_{s_2,d_2'}$ are disjoint paths. Moreover, we claim that if $v_1 = d_2'$ stays silent, $\q11$ and $P_{s_2,d_2'}$ have manageable interference. We must have $n_1(G,\q11) = 0$, since otherwise we would have a path from $s_2$ to $d_1$ not containing $v_1$, and we would contradict P\ref{qs2d1}. If $\ell(v_4) < \ell(v_1)$, then $\ell(v_4) \leq \ell(v_2)$ and the edge $(v_3,v_4)$ will guarantee that $n_2^D(\q11,P_{s_2,d_2'}) \geq 1$. Moreover, since we have a path $Z_{s_1,d_2'} = \z11[s_1,v_1]$ not containing $v_3$, we must have $n_2(G,P_{s_2,d_2'}) \geq 2$. If $\ell(v_4) = \ell(v_1)$, then $(v_3,v_4)$ will not cause a direct interference from $\q11$ to $P_{s_2,d_2'}$. Then, if we have $n_2^D(\q11,P_{s_2,d_2'}) = 0$, $\q11$ and $\p22$ have manageable interference. If  $n_2^D(\q11,P_{s_2,d_2'}) = 1$, the direct interference must be due to an edge $(v_3,v_1)$ so that $v_3 \dinterf P_{s_2,d_2'}$. Otherwise, that would contradict the fact that $n_2^D(\q11,\p22) = 1$. Therefore, the fact that we have a path $Z_{s_1,d_2'}$ not containing $v_3$ guarantees that $n_2(G,P_{s_2,d_2'}) \geq 2$. We conclude that, in any case, $\q11$ and $P_{s_2,d_2'}$ have manageable interference. Therefore, during Mode 1, it is possible to use an amplify-and-forward scheme which guarantees that the transfer matrix between $(s_1,s_2)$ and $(d_1,d_2')$ is diagonal with non-zero diagonal entries. During Mode 1, $d_2' $ will store its received signals.

The second mode of operation should last for the same number of time steps as the first one. We will remove all nodes except those in $\z11$ and $\p22$.  In Mode 2, $s_2$ will transmit the same signals it transmitted during Mode 1, while $s_1$ will transmit new signals. The only interference between the two paths happens through the edge $(v_2,v_1)$. However, node $v_1$ received, during Mode 1, scaled versions of the transmitted signals at $s_2$. Therefore, by using the signals received during Mode 1, $v_1$ is able to remove the interference due to $s_2$ from its received signal during Mode 2. Hence we can guarantee that the transfer matrix between $(s_1,s_2)$ and $(d_1,d_2)$ during Mode 2 is diagonal with non-zero diagonal entries.
Over the two modes, we again create three parallel AWGN channels, two of them between $s_1$ and $d_1$ and one of them between $s_2$ and $d_2$. Therefore, we achieve $\dE = \frac32$. A visual representation of the scheme is shown in Figure \ref{buffig4}. 

\begin{figure}[ht]
\begin{center}
\includegraphics[height=70mm]{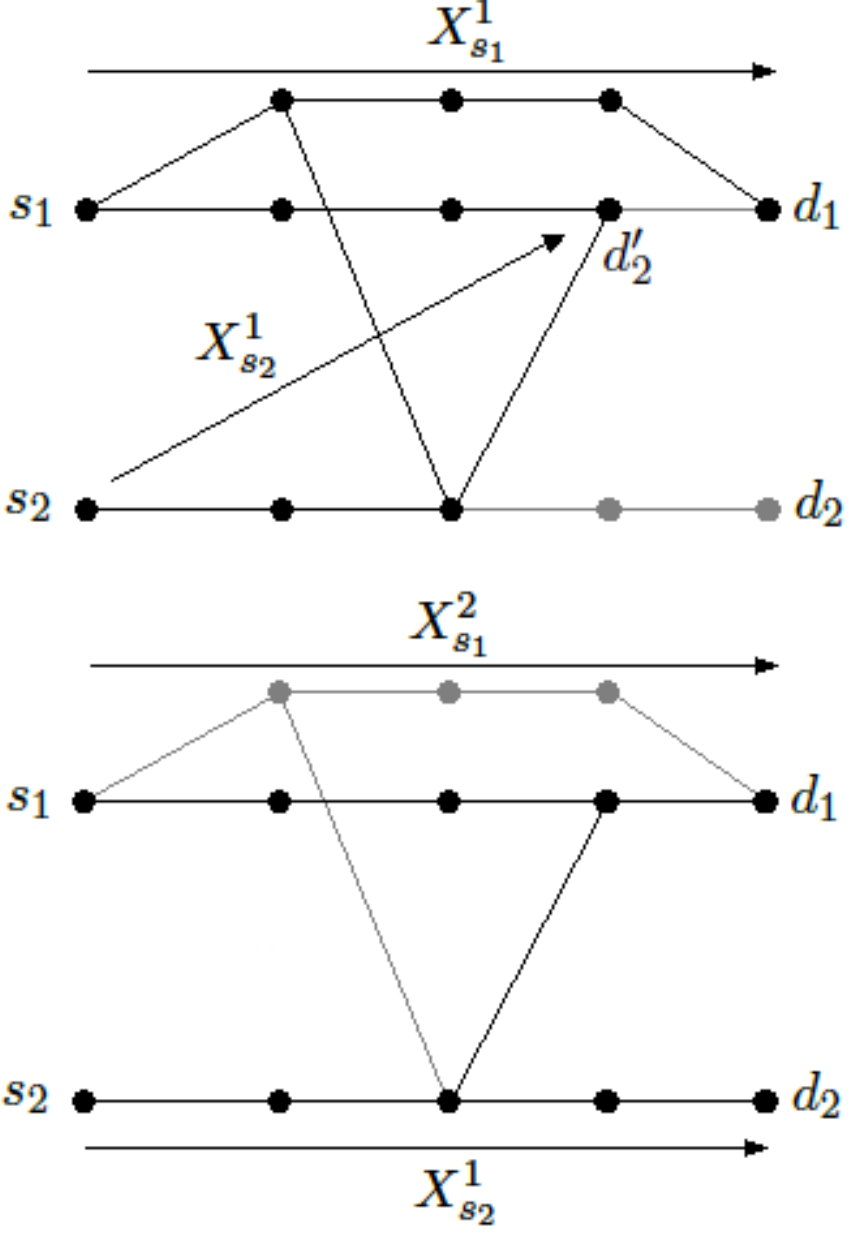}
\caption{Depiction of Mode 1 and Mode 2 for the achievability scheme in case C\ref{nonsym} when $\ell(v_3) < \ell(v_1)$.} \label{buffig4}
\end{center} 
\end{figure}
\end{subsubsection}
\end{subsection}
\begin{subsection}{Converse for case C\ref{nonsym}}

In this section, we will show that if our network falls in C\ref{nonsym}, and does not fall into (A), (A\pr), (B), (B\pr) nor C\ref{sym}, then $\dE \leq \frac32$. We will start by deriving additional connectivity properties, under the assumption that properties P1 to P\ref{q10} are satisfied for $i = 1$. 


\begin{enumerate}[P1.] \setcounter{enumi}{6}

\item The removal of $v_4$ disconnects $d_2$ from both sources \label{qv4}

From P\ref{qv3v4}, we know that the removal of $v_4$ disconnects $d_2$ from $s_1$. 
If the removal of $v_4$ does not disconnect $d_2$ from $s_2$, then we must have a path $\q22$ not containing $v_4$. 
We have that $\q22$ may not intersect $\q11$, since that would imply the existence of a path from $s_1$ to $d_2$ not containing $\{v_3,v_4\}$ and we would contradict P\ref{qv3v4}. Moreover, we must have $n_1^D(\q22,\q11) = n_2^D(\q11,\z22) = 0$. Otherwise, since $v_1, v_4 \notin \q11 \cup \q22$, we would contradict either P\ref{qs2d1} or P\ref{qv3v4}. But this means that $\q11$ and $\q22$ have manageable interference, which is a contradiction.

\item The removal of $v_2$ disconnects $s_2$ from both destinations. \label{qv2}
From P\ref{qs2d1}, we know that the removal of $v_2$ disconnects $s_2$ from $d_1$. 
If the removal of $v_2$ does not disconnect $d_2$ from $s_2$, then we must have a path $\z22$ not containing $v_2$. 
We have that $\z22$ may not intersect $\z11$, or else we would have a path $\p21$ not containing $v_2$, thus contradicting P\ref{qs2d1}. Moreover, we must have $n_1^D(\z22,\z11) = n_2^D(\z11,\z22) = 0$. Otherwise, since $v_2, v_3 \notin \z11 \cup \z22$, we would contradict either P\ref{qs2d1} or P\ref{qv3v4}. Therefore $\z11$ and $\z22$ have manageable interference, which is a contradiction.

\item The removal of $v_1$ and $v_3$ disconnects $d_1$ from both sources \label{qv1v3}

From P\ref{qs2d1}, the removal of $v_1$ disconnects $d_1$ from $s_2$. Thus we assume that the removal of $v_1$ and $v_3$ does not disconnect $d_1$ from $s_1$, and we have a path $\m11$ which does not contain $v_1$ nor $v_3$. Then we have that $\m11 \cap \p22 = \emptyset$, or else we would have a path from $s_2$ to $d_1$ not containing $v_1$, thus contradicting P\ref{qs2d1}. Moreover, P\ref{qs2d1} and the fact that $v_1 \notin \m11 \cup \p22$ imply that $n_1^D(\p22,\m11) = 0$. Likewise, P\ref{qv3v4} and the fact that $v_3 \notin \m11 \cup \p22$ imply that $n_2^D(\m11,\p22) = 0$, and thus $\m11$ and $\p22$ have manageable interference, which is a contradiction.

\item There is no path from $v_1$ to $v_3$ \label{qv1pathv3}

Suppose $v_1 \leadsto v_3$. Then we must have $\ell(v_1) < \ell(v_3)$. We will show that our network must contain a grail subnetwork, and must be in (B\pr). The network on the left of Figure \ref{netscnonsym} is an example. We consider paths $\p21 = \p22[s_2,v_2]\concat (v_2,v_1) \concat \z11[v_1,d_1]$ and $\p12 = \q11[s_1,v_3] \concat (v_3,v_4) \concat \p22[v_4,d_2]$. We claim that $\p12$ and $\p21$ must be disjoint. Since $\ell(v_1) < \ell(v_3)$, we must have $\ell(v_2) < \ell(v_4)$. Thus, $\p22[s_2,v_2]$ and $\p22[v_4,d_2]$ must be disjoint. Therefore, $\p21$ and $\p12$ may only intersect if $\z11[v_1,d_1]$ and $\q11[s_1,v_3]$ intersect. However, since $v_1 \notin \q11[s_1,v_3]$ and $v_3 \notin \z11[v_1,d_1]$, it is easy to see that if $\q11[s_1,v_3] \cap \z11[v_1,d_1] \ne \emptyset$, then there exists a path from $s_1$ to $d_1$ which does not contain $v_1$ nor $v_3$, thus contradicting P\ref{qv1v3}. Therefore, if we let $w_a = v_1 \in \p21$ and $w_b = v_3 \in \p12$, we have $s_1 \leadsto w_a$, $w_a \leadsto w_b$ and $w_b \leadsto d_2$, which satisfies the description of the grail subnetwork, given in \ref{grail_section}, by exchanging $(s_1,d_1)$ and $(s_2,d_2)$. Therefore, if $v_1 \leadsto v_3$, we are in case (B\pr), and we may assume $v_1 \not\leadsto v_3$. 

\end{enumerate}

We may now prove that under properties P1 through P10, $\dE \leq \frac32$. We will derive information inequalities, as we did for C\ref{sym}. Once more, we let $W_1$ and $W_2$ be independent random variables corresponding to a uniform choice over the messages on sources $s_1$ and $s_2$ respectively, and we let $A \defi \{ v \in V \st s_2 \not\leadsto v \}$ and $B \defi \{ v \in V \st s_1 \not\leadsto v \}$. First we have \rescnt
\begin{align} 
nR_2 &  \leq I(W_2;Y_{d_2}^n)+n\epsilon_n \nonumber \\
& \leqnum I(\tilde X_B^n;Y_4^n) + n\epsilon_n = I(\tilde X_B^n,X_3^n;Y_4^n) \non 
& \quad \quad \quad -I(X_3^n;Y_4^n|\tilde X_B^n) + n\epsilon_n \nonumber \\
& \leqnum \frac n2\log P +n \cons - I(X_3^n;Y_4^n|\tilde X_B^n) + n\epsilon_n, \label{c3eq1}
\end{align} 
where \rescnt
\cnt follows from the Markov chain $\fMC{W_2}{\tilde X_B^n}{Y_4^n}{Y_{d_2}^n}$, which is implied by P\ref{qv4} and the fact that $s_2 \in B$; 
\cnt follows from the fact that $I(\tilde X_B^n,X_3^n;Y_4^n)$ can be upper bounded by $h(Y_4^n)-h(N_{3,4}^n)$ by following the steps in (\ref{deriv}), where $\curcons$ is a constant, independent of $P$, for $P$ sufficiently large. 
Next, we have \rescnt
\begin{align}
n R_2 &  \leq I(W_2;Y_{d_2}^n)+n\epsilon_n \nonumber \\
& \leqnum I(W_2;\tilde X_2^n, \tilde X_A^n)+n\epsilon_n \non 
& \eqnum I(W_2;\tilde X_2^n| \tilde X_A^n)+n\epsilon_n \nonumber \\
& \leqnum I(X_2^n;\tilde X_2^n| \tilde X_A^n)+n\epsilon_n \non 
& \leq I(X_2^n;\tilde X_2^n,Y_1^n| \tilde X_A^n)+n\epsilon_n \nonumber \\
& = I(X_2^n;Y_1^n| \tilde X_A^n)+ I(X_2^n;\tilde X_2^n| \tilde X_A^n,Y_1^n) +n\epsilon_n \nonumber \\
& \leqnum I(X_2^n;Y_1^n| \tilde X_A^n)+ n \cons +n\epsilon_n, \label{c3eq2}
\end{align} \rescnt
where 
\cnt follows because from P\ref{qv2}, the removal of $v_2$ disconnects $d_2$ from $s_2$, and therefore, the removal of $v_2$ and $A$ disconnects $d_2$ from both sources, and we have $\MC{W_2}{(\tilde X_2^n,\tilde X_A^n)}{Y_{d_2}^n}$;
\cnt follows since $\tilde X_A^n$ is independent of $W_2$; 
\cnt follows because, given $\tilde X_A^n$, we have $\MC{W_2}{X_2^n}{\tilde X_2^n}$;
\cnt follows by applying Lemma \ref{const} to $I(X_2^n;\tilde X_2^n| \tilde X_A^n,Y_1^n)$, because $\I(v_1)\setminus \{v_2\} \subset A$, or else we would contradict P\ref{qs2d1}. Furthermore, we have \rescnt
\begin{align}
n R_1 &  \leq I(W_1;Y_{d_1}^n)+n\epsilon_n \nonumber \\
& \leqnum I(W_1;\tilde X_3^n, Y_1^n)+n\epsilon_n \non 
& = I(W_1;\tilde X_3^n) + I(W_1; Y_1^n|\tilde X_3^n)+ n\epsilon_n \nonumber \\
& \leqnum I(W_1;\tilde X_3^n|\tilde X_B^n) + I(W_1; Y_1^n|\tilde X_3^n)+ n\epsilon_n \nonumber \\
& \leqnum I(X_3^n;\tilde X_3^n|\tilde X_B^n) + I(W_1; Y_1^n|\tilde X_3^n)+ n\epsilon_n \nonumber \\
& \leqnum I(X_3^n;\tilde X_3^n|\tilde X_B^n) + I(\tilde X_A^n; Y_1^n|\tilde X_3^n)+ n\epsilon_n \nonumber \\
& = I(X_3^n;\tilde X_3^n|\tilde X_B^n) + I(\tilde X_A^n,X_2^n; Y_1^n|\tilde X_3^n) \non 
& \quad \quad - I(X_2^n; Y_1^n|\tilde X_3^n,\tilde X_A^n)+ n\epsilon_n \nonumber \\
& \leq I(X_3^n;\tilde X_3^n|\tilde X_B^n) + I(\tilde X_A^n,X_2^n,\tilde X_3^n; Y_1^n) \non
& \quad \quad - I(X_2^n; Y_1^n|\tilde X_3^n,\tilde X_A^n)+ n\epsilon_n \nonumber \\
& \eqnum I(X_3^n;\tilde X_3^n|\tilde X_B^n) + I(\tilde X_A^n,X_2^n; Y_1^n) \non 
& \quad \quad - I(X_2^n; Y_1^n|\tilde X_A^n)+ n\epsilon_n \nonumber \\
& \leqnum I(X_3^n;\tilde X_3^n|\tilde X_B^n) + \frac n2\log P + n\cons \non
& \quad \quad  - I(X_2^n; Y_1^n|\tilde X_A^n)+ n\epsilon_n \nonumber \\
& \leq I(X_3^n;Y_4^n|\tilde X_B^n)+ I(X_3^n;\tilde X_3^n|\tilde X_B^n,Y_4^n) \non 
& \quad \quad + \frac n2\log P + n \curcons - I(X_2^n; Y_1^n|\tilde X_A^n)+ n\epsilon_n \nonumber \\
& \leqnum I(X_3^n;Y_4^n|\tilde X_B^n)  + \frac n2\log P +  n(\curcons+ \cons) \non 
& \quad \quad - I(X_2^n; Y_1^n|\tilde X_A^n)+ n\epsilon_n, \label{c3eq3}
\end{align}
where \rescnt
\cnt follows from P\ref{qv1v3}, which implies $\MC{W_1}{(\tilde X_3^n, Y_1^n)}{Y_{d_1}^n}$; 
\cnt follows from the fact that $\tilde X_B^n$ is independent of $W_1$; 
\cnt follows from the fact that, given $\tilde X_B^n$, we have $\MC{W_1}{X_3^n}{\tilde X_3^n}$; 
\cnt follows from the fact that $s_1 \in A$; 
\cnt follows because P\ref{qs2d1} and P\ref{qv1pathv3} imply that 
$s_2 \not\leadsto v_3$ and, therefore, $v_3 \in A$; 
\cnt follows from the fact that $I(\tilde X_A^n,X_2^n; Y_1^n)$ can be upper bounded by $h(Y_1^n)-h(N_{2,1}^n)$ by following the steps in (\ref{deriv}), where $K_{15}$ is a constant, independent of $P$, for $P$ sufficiently large; 
and \cnt follows by applying Lemma \ref{const} to $I(X_3^n;\tilde X_3^n|\tilde X_B^n,Y_4^n)$, since $\I(v_4) \setminus \{v_3\} \subset B$, or else we contradict P\ref{qv3v4}. In order to bound the sum degrees-of-freedom, we can use the fact that
\begin{align}
n R_1 &  \leq I(W_1;Y_{d_1}^n)+n\epsilon_n = h(Y_{d_1}^n)-h(Y_{d_1}^n|W_1)+ n\epsilon \non 
& \leq h(Y_{d_1}^n)-h(Y_{d_1}^n|W_1,X_{\I(d_1)}^n)+n\epsilon \nonumber \\
& = h(Y_{d_1}^n)-h(N_{d_1}^n) + n\epsilon \leq \frac n2\log P + n \cons + n\epsilon, \label{c3eq4} 
\end{align}
where the last inequality follows in the same way as (\ref{deriv}). Therefore, we can add inequalities (\ref{c3eq1}), (\ref{c3eq2}), (\ref{c3eq3}) and (\ref{c3eq4}) in order to obtain
\begin{align*}
& 2n ( R_1 +  R_2) \leq \frac{3n}2 \log P + n\sum_{j=13}^{17}K_j + n\epsilon_n \\
\Rightarrow & \frac{R_1+R_2}{\frac12\log P} \leq \frac32+ \frac{\sum_{j=13}^{17}K_j+\epsilon_n}{ \log P}.
\end{align*} 
Thus, if we let $n\rightarrow \infty$ and then $P \rightarrow \infty$, we obtain
$ \dE \leq \frac32$.

\begin{section}{Obtaining the full degrees-of-freedom region} \label{ext}

In this section, we extend the results from Theorem \ref{mainth} and characterize the full degrees-of-freedom region of two-unicast layered Gaussian networks. 
The degrees-of-freedom region (see Definition \ref{defn_region}) can be understood as a high-SNR approximation to the capacity region, scaled down by $\frac12 \log P$. Since the sum degrees-of-freedom is given by
\[ 
\dE = \lim_{P \goesto \infty}\left( \sup_{(R_1,R_2) \in \C(P)}{\frac{R_1 + R_2}{\frac12 \log P}}\right),
\]
we conclude that if $(D_1,D_2) \in \D$, then we must have $D_1 + D_2 \leq \dE$. Thus, the results from Theorem \ref{mainth} provide an outer bound to the degrees-of-freedom region with at least one achievable point. Moreover, by following the steps in (\ref{c3eq4}), it is always possible to bound each individual rate, for $P$ sufficiently large, as
\[ 
R_i \leq \frac12 \log P + K,
\]
where $K$ is independent of $P$, for $i=1,2$. Hence, we conclude that $\D$ is always a subset of 
$
\left\{(D_1,D_2) \in \R^2_+  \st D_1\leq 1, D_2 \leq 1 \right\}.
$
It is straightforward to show that $\D$ is convex.
Therefore, for networks that belong to cases (A) and (A\pr) from Theorem \ref{mainth}, the fact that $\dE = 1$ guarantees that the degrees-of-freedom region is given by 
\[ 
\D = \left\{(D_1,D_2) \in \R^2_+  \st D_1 + D_2 \leq 1 \right\}.
\]
This region is depicted in Figure \ref{reg1}. The degrees-of-freedom region for networks in cases (B) and (B\pr) can also be easily obtained from the result in Theorem \ref{mainth}. Since for all networks in cases (B) and (B\pr), we have $(1,1) \in \D$, we conclude that the degrees-of-freedom region in these cases is given by
\[ 
\D = \left\{(D_1,D_2) \in \R^2_+  \st D_1\leq 1, D_2 \leq 1 \right\},
\]
as depicted in Figure \ref{reg2}.

For networks in case (C), we will once again consider the division into cases C\ref{sym} and C\ref{nonsym}, as described in Section \ref{32deg}. 
If our network $\N$ falls into case C\ref{nonsym}, then, under the assumption that properties P\ref{qpath}, P\ref{q01}, P\ref{zpath} and P\ref{q10} in Section \ref{nonsymsec} are satisfied for $i=1$, we have inequalities (\ref{c3eq1}), (\ref{c3eq2}) and (\ref{c3eq3}). By adding these three inequalities, we obtain
\begin{align*}
& n( R_1 +  2 R_2) \leq n \log P + n\sum_{j=13}^{16}K_j + n\epsilon_n \\
\Rightarrow & \frac{ R_1+ 2R_2}{\frac12\log P} \leq 2+ \frac{\sum_{j=13}^{16}K_j+\epsilon_n}{\frac12 \log P} \\
\Rightarrow & 
\lim_{P \goesto \infty}\left( \sup_{(R_1,R_2) \in \C(P)}{\frac{ R_1 + 2 R_2}{\frac12 \log P}}\right) \leq 2,
\end{align*}
which implies that, if $(D_1,D_2) \in \D$, then $D_1 + 2 D_2 \leq 2$. Since the achievability scheme described in Section \ref{nonsymsec} shows that $(1,1/2) \in \D$, we conclude that the degrees-of-freedom region for networks which belong to case C\ref{nonsym} (but do not belong to cases (A), (A\pr), (B), (B\pr) and C\ref{sym}) and satisfy properties P\ref{qpath}, P\ref{q01}, P\ref{zpath} and P\ref{q10} in Section \ref{nonsymsec} for $i=1$ is given by
\begin{equation}
\D = \left\{(D_1,D_2) \in \R^2_+  \st D_1\leq 1, D_1+ 2 D_2 \leq 2 \right\}. \label{dofnonsym}
\end{equation}
This region is depicted in Figure \ref{reg32a}. If properties P\ref{qpath}, P\ref{q01}, P\ref{zpath} and P\ref{q10} in Section \ref{nonsymsec} are instead satisfied for $i=2$, then it is easy to see that analogous results will hold, and the degrees-of-freedom region is given by
\begin{equation}
\D = \left\{(D_1,D_2) \in \R^2_+  \st D_1\leq 1, 2 D_1+ D_2 \leq 2 \right\}, \label{dofnonsym2}
\end{equation}
as depicted in Figure \ref{reg32a_inv}.
The only networks that we still need to consider are networks which are in case C\ref{sym}. We will show that in these cases the degrees-of-freedom region will be given by 
\begin{equation}
\D = \left\{(D_1,D_2) \in \R^2_+  \st D_1\leq 1, D_2\leq 1, D_1+ D_2 \leq 3/2 \right\}, \label{dofsym}
\end{equation}
as shown in Figure \ref{reg32b}. 
In order to do that we will assume that our network $\N$ contains disjoint paths $\p11$ and $\p22$ such that $n_1(G) \geq 2$, $n_1^D = 1$, $n_2(G) = 1$ and $n_2^D = 0$, and that the nodes from $\N$ are named as described in Section \ref{ach_sym} and depicted in Figure \ref{Csym_again}. We will then consider the condensed network formed by layers $V_1$, $V_{\ell(v_2)}$ and $V_r$, which is shown in Figure \ref{condsym}.
\begin{figure*}[ht]
     \centering
     \subfigure[]{
       \includegraphics[height=28mm]{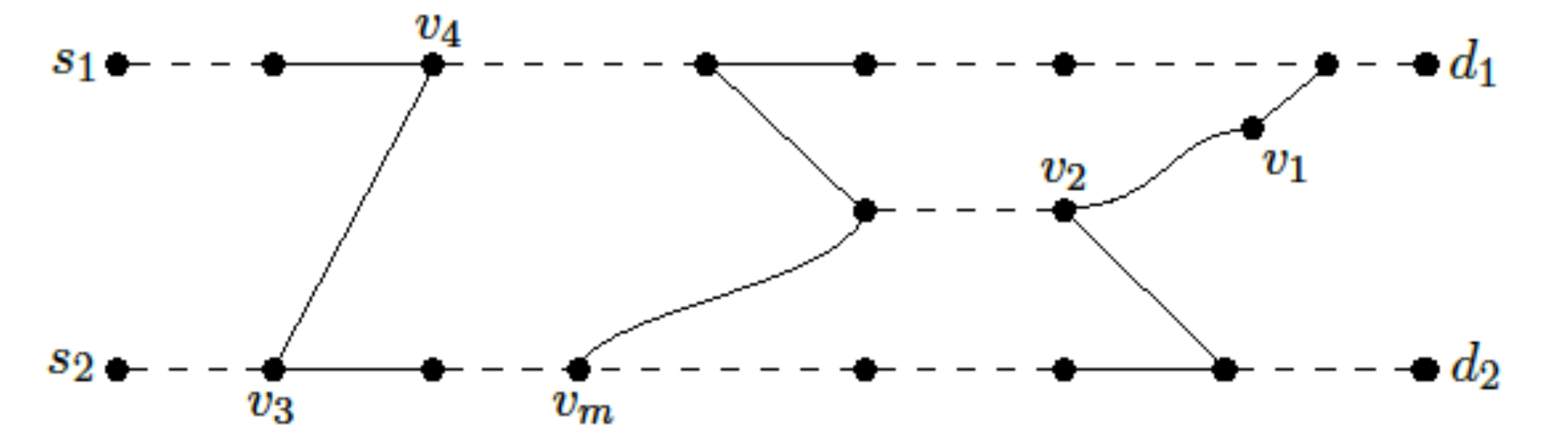} \label{Csym_again} }
       \hspace{2mm}
     \subfigure[]{
       \includegraphics[height=28mm]{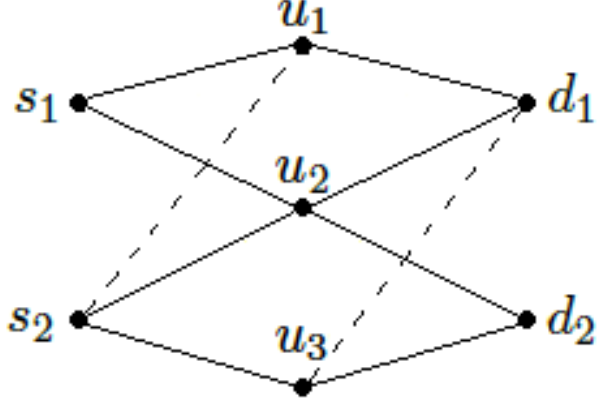} \label{condsym} }
     \caption{(a) Illustration of network from case C\ref{sym}, up to a change in the position of edge $(v_3,v_4)$; (b) Condensed network for networks in case C\ref{sym}.} 
\end{figure*}
Notice that $u_2$ is the same node as $v_2$ in the original network.
Therefore, edges $\hat h(s_1,u_3)$ and $\hat h(u_1,d_2)$ may not exist due to property P\ref{pv2v0} in Section \ref{conv_sym}. Edges $\hat h(s_2,u_1)$ and $\hat h(u_3,d_1)$ may or may not exist, and that will depend on the position of the edge $(v_3,v_4)$ in the original network.  We will show that points $(1,1/2)$ and $(1/2,1)$ are included in $\D$ and, by convexity and the fact that $\dE \leq 3/2$, $\D$ must be as shown in Figure \ref{reg32b}. 

First note that we may assume that exactly one of the edges $\hat h(s_2,u_1)$ and $\hat h(u_3,d_1)$ exists. Otherwise, by removing $u_2$, $\p11$ and $\p22$ have manageable interference, thus $(1,1)$ is in $\D$ and we are in case (B). Hence, we restrict ourselves to two cases: (1) $\hat h(u_3,d_1) \ne 0$; and (2) $\hat h(s_2,u_1) \ne 0$. We consider each one separately.

\begin{subsubsection}{$\hat h(u_3,d_1) \ne 0$}

This network is depicted in Figure \ref{condsym1}. We see that it
falls in the case described in Section \ref{buf2}. The achievability scheme provided for $\dE = 3/2$ shows that $(1/2,1) \in \D$. In order to achieve the point $(1,1/2)$, we need to use a scheme based on real interference alignment, similar to the ones described in \cite{MotahariKhandani} and \cite{xx}.

At source $s_1$, the message $W_1$ will be split into two submessages $W_1^{(1)}$ and $W_1^{(2)}$, while $s_2$ will have a single message $W_2$. Each of these messages will be encoded using a single codebook with codewords of length $n$, which is obtained by uniform i.i.d.~sampling of the set
\begin{equation}
\U = \Z \cap \left[ -\gamma P^{\frac{1-\ep}{2(2+\ep)}} ,\gamma P^{\frac{1-\ep}{2(2+\ep)}} \right], \label{codeset}
\end{equation}
for a small $\ep > 0$ and a constant $\gamma$.
The rate of this code, i.e., the number of codewords, will be determined later.
We will let $a_j[1]$, $a_j[2],...,a_j[n]$ be the $n$ symbols of the codeword associated to message $W_1^{(j)}$, $j=1,2$, and $b[1]$, $b[2],...,b[n]$ be the $n$ symbols of the codeword associated to message $W_2$. At time $t \in \{1,...,n\}$, source $s_1$ will transmit 
\[X_{s_1}[t] = G(a_1[t]+T a_2[t]),\] 
where $T$ is an irrational number, and $G= \beta P^{\frac{1+2\ep}{2(2+\ep)}}$ is chosen to satisfy the power constraint 
for a constant $\beta$ to be determined. Source $s_2$ will transmit 
\[X_{s_2}[t] = G \frac{\hat h(s_1,u_2)}{\hat h(s_2,u_2)} b[t].\] 
The maximum power of a transmit signal from $s_1$ is upper-bounded by
\begin{align*}
 \beta^2 P^{\frac{1+2\ep}{2+\ep}} (1+T^2) \gamma^2 P^{\frac{1-\ep}{2+\ep}} & = \beta^2 (1+T^2) \gamma^2 P,
\end{align*}
and the maximum power of a transmit signal from $s_2$ is upper-bounded by
\begin{align*}
 \beta^2 P^{\frac{1+2\ep}{2+\ep}} \frac{\hat h(s_1,u_2)^2}{\hat h(s_2,u_2)^2} \gamma^2 P^{\frac{1-\ep}{2+\ep}} & = \beta^2 \frac{\hat h(s_1,u_2)^2}{\hat h(s_2,u_2)^2}  \gamma^2 P.
\end{align*}
Thus, for any choice of $T$ and $\gamma$, parameter $\beta$ can be chosen so that the maximum transmit power at the sources is less than $P$. 
Next we write the received signals at $u_1$, $u_2$ and $u_3$. We will drop the time $t$ from the notation for simplicity.
\[ Y_{u_1} = G \hat h(s_1,u_1) (a_1+T a_2) + N_{u_1}
\]
\[ Y_{u_2} = G \hat h(s_1,u_2) (a_1 + b +T a_2) + N_{u_2}
\]
\[ Y_{u_3} = \frac{G \hat h(s_2,u_3) \hat h(s_1,u_2)}{\hat h(s_2,u_2)} b + N_{u_3}
\]
Nodes $u_1$ and $u_3$ will simply perform amplify-and-forward. More precisely, their transmit signals will be given by 
\[ X_{u_1} = \alpha Y_{u_1} = \alpha G \hat h(s_1,u_1) (a_1+T a_2) + \alpha N_{u_1},
\]
\begin{align*} X_{u_3} & = -\alpha \frac{\hat h(s_2,u_2)}{\hat h(s_2,u_3) \hat h(s_1,u_2)} \frac{\hat h(s_1,u_1)\hat h(u_1,d_1)}{\hat h(u_3,d_1)} Y_{u_3} \non 
& = - \alpha G \frac{\hat h(s_1,u_1)\hat h(u_1,d_1)}{\hat h(u_3,d_1)} b 
+ \alpha K_1 N_{u_3},
\end{align*}
where $\alpha$ is a constant and $K_1$ is a function of the channel gains. We will choose $\alpha$ so that the power constraint at each relay is satisfied. For example, consider node $u_1$. The noise term at its received signal, $N_{u_1}$, is a linear combination of noises in the original networks, whose coefficients are products of channel gains. Therefore, we may assume that $E[N_{u_1}^2]$ is a constant $\sigma_{u_1}^2$. Thus, the transmitted power at $u_1$ is 
\begin{align}
E[\alpha^2 & Y_{u_1}^2]  \non
& = \alpha^2 \beta^2 P^{\frac{1+2\ep}{2+\ep}} \hat h^2(s_1,u_1)(1+T^2)\gamma^2 P^{\frac{1-\ep}{2+\ep}} + \alpha^2\sigma_{u_1}^2 \nonumber \\
& = \alpha^2 \beta^2 \hat h^2(s_1,u_1)(1+T^2)\gamma^2 P + \alpha^2\sigma_{u_1}^2.
\end{align}
It is now easy to see that $\alpha$ can be chosen independently of $P$ to make sure that $E[\alpha^2 Y_{u_1}^2] \leq P$,  for $P$ sufficiently large.
The received signal at node $u_2$ can be seen as a noisy observation of a point in the set 
\[
\U_{u_2} = G \hat h(s_1,u_2) \left\{ x_1 + T x_2 \st x_1 \in \U + \U, x_2 \in \U \right\},
\]
for $x_1 = a_1 + b$ and $x_2 = a_2$, where $\U + \U = \{u_1 + u_2 \st u_1,u_2 \in \U \} \subset \Z \cap \left[ -2\gamma P^{\frac{1-\ep}{2(2+\ep)}} , 2\gamma P^{\frac{1-\ep}{2(2+\ep)}} \right]$.
As explained in \cite{MotahariKhandani}, the fact that $T$ is irrational guarantees that there is a one-to-one map from the points in $\U_{u_2}$ to the points $(x_1,x_2) \in (\U+\U) \times \U$. Moreover, from Theorem 1 of \cite{MotahariKhandani} and subsequent remarks, we conclude that, for almost all choices of $T$, the minimum distance between two points in $\U_{u_2}$ satisfies
\[
d_{\min } > G \hat h(s_1,u_2) \frac{\kappa}{(\max_{x \in \U} x)^{1+\ep}},
\]
for some constant $\kappa$. Thus we have
\[
d_{\min } >  \hat h(s_1,u_2) \frac{\kappa \beta P^{\frac{1+2\ep}{2(2+\ep)}} }{ \gamma^{1+\ep} P^{\frac{(1-\ep)(1+\ep)}{2(2+\ep)}} }
= \frac{\hat h(s_1,u_2) \kappa \beta}{\gamma^{1+\ep}} P^{\ep/2}. 
\]
Node $u_2$ will map its received signal to the nearest point in $\U_{u_2}$, and then use the fact that this point uniquely determines $(a_1 + b)$ and $a_2$ to decode these two integers. We will refer to the output of this procedure as $\hat a_1 + \hat b$ and $\hat a_2$. If the variance of $N_{u_2}$ is given by $\sigma^2_{u_2}$, then the probability of a wrong decoding at $u_2$ is given by
\begin{align*}
\Pr [\hat a_1 & + \hat b \ne a_1+b,\hat a_2 \ne a_2 ] \leq 2 \, Q\left(\frac{d_{\min }}{2 \sigma_{u_2} } \right) \non
& < \exp\left({-\frac{d_{\min }^2}{8 \sigma^2_{u_2}}}\right) = \exp\left(-\delta P^\ep\right),
\end{align*}
where $\delta$ is a positive constant, independent of $P$. The transmit signal at $u_2$ will then be
\[ X_{u_2} = \alpha G \frac{\hat h(s_1,u_1)\hat h(u_1,d_1)}{\hat h(u_2,d_1)} (\hat a_1 + \hat b).
\]
We choose $\alpha$ independently of $P$, so that the power constraints at $u_1$, $u_2$ and $u_3$ are simultaneously satisfied, for $P$ sufficiently large. The received signal at the destination $d_1$ is given by
\begin{align*} Y_{d_1} &= \hat h(u_1,d_1) X_{u_1} + \hat h(u_2,d_1) X_{u_2} \non 
& \quad \quad +\hat h(u_3,d_1) X_{u_3} + N_{d_1} \\
&=\alpha G \hat h(s_1,u_1) \hat h(u_1,d_1) \left(a_1+T a_2+\hat a_1 + \hat b - b\right) \non 
& \quad \quad  + N_{d_1}^{\text{eff}},
\end{align*}
where $N_{d_1}^{\text{eff}} = \alpha \hat h(u_1,d_1) N_{u_1} + \alpha \hat h(u_2,d_1) K_1 N_{u_3} + N_{d_1}$. 
The received signal at $d_2$ is given by
 \begin{align*} Y_{d_2} &= \hat h(u_2,d_2) X_{u_2} + \hat h(u_3,d_2) X_{u_3} + N_{d_2} \\
&=\alpha G \frac{\hat h(u_2,d_2)\hat h(s_1,u_1)\hat h(u_1,d_1)}{\hat h(u_2,d_1)} (\hat a_1 + \hat b) \non 
& \quad - \alpha \frac{\hat h(u_3,d_2) \hat h(s_1,u_1)\hat h(u_1,d_1)}{\hat h(u_3,d_1)} b 
+ N_{d_2}^{\text{eff}},
\end{align*}
where $N_{d_2}^{\text{eff}} = \alpha \hat h(u_3,d_2) K_1 N_{u_3} + N_{d_2}$. Notice that with probability at least $1- \exp(-\delta P^\ep)$, $Y_{d_1}$ and $Y_{d_2}$ are given by
\[ 
Y_{d_1} = \alpha G \hat h(s_1,u_1) \hat h(u_1,d_1) (2 a_1+T a_2) + N_{d_1}^{\text{eff}},
\]
\begin{align*}
Y_{d_2} & =\alpha G \hat h(s_1,u_1)\hat h(u_1,d_1) \\ 
& \left( \frac{\hat h(u_2,d_2)}{\hat h(u_2,d_1)} (a_1 + b) - \frac{\hat h(u_3,d_2)}{\hat h(u_3,d_1)} b \right) 
+ N_{d_2}^{\text{eff}}.
\end{align*}
The destinations will first perform a hard-decoding, similar to the one performed by $u_2$. If we assume that the decoding at node $u_2$ was correct, the signal received by $d_1$ is a noisy version of a point in the set
\[
\U_{d_1} = \alpha G \hat h(s_1,u_1) \hat h(u_1,d_1) \left\{ 2 x_1 + T x_2 \st x_1,x_2 \in \U \right\},
\]
for $x_1 = a_1$ and $x_2 = a_2$.
Thus, by using the same argument used previously, it can be shown that $d_1$ can decode $a_1$ and $a_2$ with probability of error smaller than $\exp(-\delta_2 P^\ep)$, for some positive constant $\delta_2$. Assuming that the decoding at node $u_2$ was correct, the signal received by $d_2$ is a noisy version of a point in the set
\begin{align*}
\U_{d_2} & =  \alpha G \hat h(s_1,u_1)\hat h(u_1,d_1) \frac{\hat h(u_2,d_2)}{\hat h(u_2,d_1)} \\ 
& \left\{ x_1 + T_2 x_2 \st x_1 \in \U + \U, x_2 \in \U \right\},
\end{align*}
for $x_1 = a_1 + b$ and $x_2 = b$,
where 
$T_2 = - \frac{\hat h(u_2,d_1)\hat h(u_3,d_2)}{\hat h(u_2,d_2)\hat h(u_3,d_1)}$. 
Next we notice that $\hat h(u_2,d_1)$, $\hat h(u_3,d_2)$, $\hat h(u_2,d_2)$ and $\hat h(u_3,d_1)$ are each a polynomial on the channel gains $h_e$ of the original network with only coefficients 1. 
Since each $h_e$ is independently drawn from a continuous distribution, we see that if the polynomials $\hat h(u_2,d_1)\hat h(u_3,d_2)$ and $\hat h(u_2,d_2)\hat h(u_3,d_1)$ are not identical, then, with probability 1, $T_2$ is an irrational number. From the description of the original network, given in Section \ref{ach_sym}, we see that $u_2 = v_2$ is on a path $P_{v_m,v_1}$ such that $P_{v_m,v_1} \cap \p22 = \{v_m\}$, $v_m \ne v_2$ and $v_1 \interf \p11$. Therefore, there must exist two disjoint paths $P_{u_2,d_1}$ and $P_{u_3,d_2}$. This implies that the determinant
\begin{align*}
& \begin{vmatrix}
  \hat h(u_2,d_1) & \hat h(u_3,d_1) \\
  \hat h(u_2,d_2) & \hat h(u_3,d_2) \\
\end{vmatrix} \\
& = \hat h(u_2,d_1)\hat h(u_3,d_2) - \hat h(u_3,d_1)\hat h(u_2,d_2)
\end{align*}
is non-zero with probability 1. This implies that $\hat h(u_2,d_1)\hat h(u_3,d_2)$ and $\hat h(u_3,d_1)\hat h(u_2,d_2)$ are not identical polynomials. Hence, we conclude that, with probability 1, $T_2$ is an irrational number and $d_2$ can decode $b$ (and also $a_1 + b$, and thus $a_1$, even though $d_2$ does not require the message encoded by the $a_1$'s) with probability at least $1- \exp(-\delta_3 P^\ep)$, for some $\delta_3 > 0$. 

By applying these hard-decoders, destination $d_1$ obtains the estimates $\hat a_1[t]$ and $\hat a_2[t]$, and destination $d_2$ obtains the estimates $\hat b[t]$, for $t=1,...,n$. Then they can apply typicality-based decoders in order to decode the messages $W_1^{(1)}$, $W_1^{(2)}$ and $W_2$.

\begin{figure}[t]
     \centering
     \subfigure[]{
       \includegraphics[height=25mm]{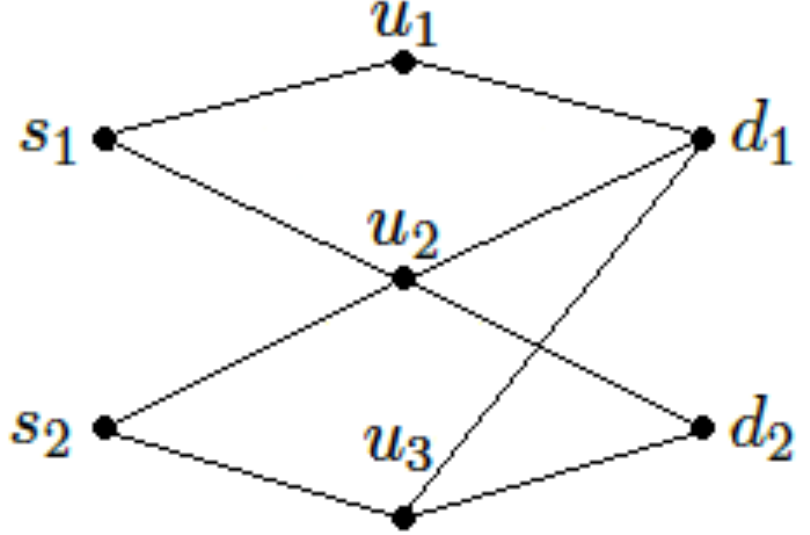} \label{condsym1}}
    \hspace{0mm}
    \subfigure[]{
       \includegraphics[height=25mm]{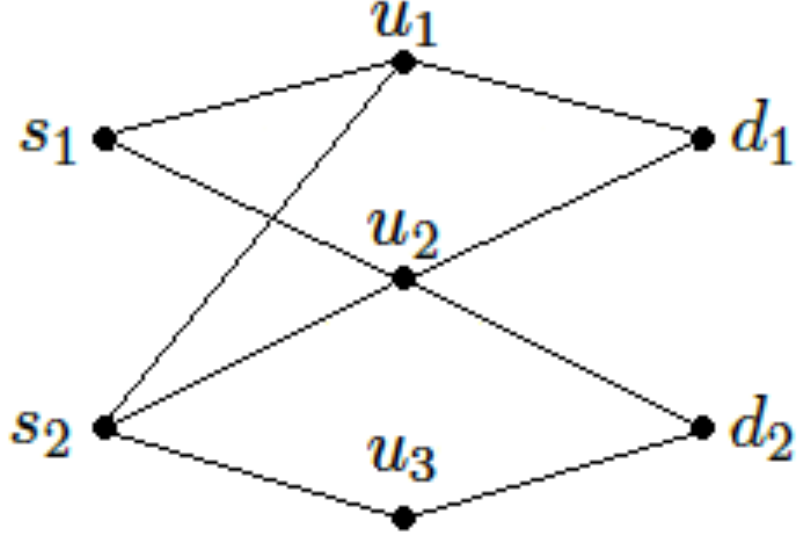} \label{condsym2}}
     \caption{(a) Condensed network in case (1); (b) Condensed network in case (2).}
\end{figure}

We now determine the rate of the codebook which is used to encode $W_1^{(1)}$, $W_1^{(2)}$ and $W_2$. We notice that for each of the messages $W_1^{(1)}$, $W_1^{(2)}$ and $W_2$, we effectively have a point-to-point discrete channel with input and output alphabet $\U$. Even though we do not calculate the actual transition probabilities, we know that the error probability is upper-bounded as
\begin{align*}
P_e & \leq 1 - (1- \exp(-\delta P^\ep)) \\ 
& \quad \quad (1- \exp(-\delta_2 P^\ep))(1- \exp(-\delta_3 P^\ep))\\
& \leq 1 - (1- \exp(-\delta_{\min } P^\ep))^3 \leq 4 \exp(-\delta_{\min } P^\ep),
\end{align*}
where $\delta_{\min } = \min (\delta,\delta_2,\delta_3)$. This allows us to lower bound the mutual information between the input $U$ and the output $\hat U$ of this channel, for a uniform distribution over the input alphabet. Using Fano's inequality, we have
\begin{align*}
I(U;\hat U) & \geq H(U) - H(U|\hat U) \\
& \geq \log|\U| - (1+P_e \log|\U|) \\
& = (1-P_e)\log|\U| - 1 \\
& \geq (1-4 \exp(-\delta_{\min } P^\ep))\left(\frac{1-\ep}{2+\ep} \frac{\log P}2 + 1\right) - 1 \\
& \geq (1-4 \exp(-\delta_{\min } P^\ep)) \frac{1-\ep}{2+\ep} \frac{\log P}2 -4.
\end{align*}
Therefore, since we constructed our code by taking independent samples uniformly at random from the set $\U$, it can achieve rate $R = (1-4 \exp(-\delta_{\min } P^\ep))\frac{1-\ep}{2+\ep} \frac{\log P}2 - 4$, by having the codebook contain $2^{nR}$ codewords. Thus, each of the messages $W_1^{(1)}$, $W_1^{(2)}$ and $W_2$ possesses
\[
\lim_{P \goesto \infty} \frac{R}{\frac12 \log P} = \frac{1-\ep}{2+\ep}
\]
degrees-of-freedom. Since $\ep$ can be chosen arbitrarily small, we conclude that each of the messages may in fact achieve arbitrarily close to $1/2$ degrees-of-freedom. Therefore, we achieve the point $(1,1/2)$ in the degrees-of-freedom region, and complete the proof in this case.


\end{subsubsection}

\begin{subsubsection}{$\hat h(s_2,u_1) \ne 0$}

As seen in Figure \ref{condsym2}, the network in this case falls in the category of networks described in Section \ref{buf}. The achievability scheme provided for $\dE = 3/2$ shows that $(1,1/2) \in \D$. Once more, we will use real interference alignment to achieve the other extreme point, i.e.,  $(1/2,1)$. We will follow the steps in case (1) closely.

This time, at source $s_2$, message $W_2$ will be split into two submessages $W_2^{(1)}$ and $W_2^{(2)}$, while $s_1$ will have a single message $W_1$. These three messages will be encoded using a single codebook with codewords of length $n$, which are obtained by uniform i.i.d.~sampling of the set $\U$, defined in (\ref{codeset}).
The rate of this code will be determined later.
We let $a[1]$, $a[2],...,a[n]$ be the $n$ symbols of the codeword associated to message $W_1$, and $b_j[1]$, $b_j[2],...,b_j[n]$ be the $n$ symbols of the codeword associated to message $W_2^{(j)}$, $j=1,2$. Sources $s_1$ and $s_2$ will respectively transmit
\[X_{s_1}[t] = G a[t],\]
\[X_{s_2}[t] = G \left( \frac{\hat h(s_1,u_1)}{\hat h(s_2,u_1)} b_1[t] + \frac{\hat h(s_1,u_2)}{\hat h(s_2,u_2)} b_2[t] \right).\]
As in (1), $G$ can be chosen as $G = \beta P^{\frac{1+2\ep}{2(2+\ep)}}$ to satisfy the sources power constraint, for some constant $\beta$.
We again drop $t$ from the notation for simplicity. 
The received signals at $u_1$, $u_2$ and $u_3$ are given by
\[ Y_{u_1} = G \hat h(s_1,u_1) \left(a +b_1 + \frac{\hat h(s_1,u_2)\hat h(s_2,u_1)}{\hat h(s_2,u_2)\hat h(s_1,u_1)} b_2\right) + N_{u_1},
\]
\[ Y_{u_2} = G \hat h(s_1,u_2)\left(\frac{\hat h(s_2,u_2)\hat h(s_1,u_1)}{\hat h(s_1,u_2)\hat h(s_2,u_1)} b_1 + b_2 + a \right) + N_{u_2},
\]
\[ Y_{u_3} = G \hat h(s_2,u_3)\left( \frac{\hat h(s_1,u_1)}{\hat h(s_2,u_1)} b_1 + \frac{\hat h(s_1,u_2)}{\hat h(s_2,u_2)} b_2 \right) + N_{u_3}.
\] 
Nodes $u_1$ and $u_3$ will simply perform amplify-and-forward. More precisely, their transmit signals will be given by 
{\small 
\begin{align*}
&X_{u_1}  = \frac{\alpha \hat h(u_2,d_1)}{\hat h(s_1,u_1)\hat h(u_1,d_1)} Y_{u_1} \\ 
&=\frac{\alpha G \hat h(u_2,d_1)}{\hat h(u_1,d_1)} \left(a +b_1 + \frac{\hat h(s_1,u_2)\hat h(s_2,u_1)}{\hat h(s_2,u_2)\hat h(s_1,u_1)} b_2\right) + \alpha K_1  N_{u_1},
\end{align*}
\begin{align*} & X_{u_3} = \frac{-\alpha \hat h(s_2,u_1) \hat h(u_2,d_2)}{\hat h(s_1,u_1) \hat h(s_2,u_3)\hat h(u_3,d_2)} Y_{u_3} \\
& = \frac{\alpha G \hat h(u_2,d_2)}{\hat h(u_3,d_2)} \left(-b_1 - \frac{\hat h(s_1,u_2)\hat h(s_2,u_1)}{\hat h(s_2,u_2)\hat h(s_1,u_1)} b_2\right)
 +\alpha K_2  N_{u_3},
\end{align*}}
for some constant $\alpha$, where $K_1$ and $K_2$ are functions of the channel gains only.
The received signal at node $u_2$ can be seen as a noisy observation of a point in the set 
\[
\U_{u_2} = G \hat h(s_1,u_2) \left\{ T x_1 + x_2 \st x_1 \in \U, x_2 \in \U  + \U \right\},
\]
for $x_1 = b_1$ and $x_2 = a+b_2$, where $T = \frac{\hat h(s_2,u_2)\hat h(s_1,u_1)}{\hat h(s_1,u_2)\hat h(s_2,u_1)}$. 
Next we notice that $\hat h(u_2,d_1)$, $\hat h(u_3,d_2)$, $\hat h(u_2,d_2)$ and $\hat h(u_3,d_1)$ are each a polynomial on the channel gains $h_e$ of the original network with only coefficients 1. From the description of the original network from Section \ref{ach_sym}, we see that $u_2 = v_2$ is on a path $P_{v_m,v_1}$ such that $P_{v_m,v_1} \cap \p11 = \emptyset$ and $v_m \in \p22$. Therefore, there must exist two disjoint paths $P_{s_1,u_1}$ and $P_{s_2,u_2}$. This implies that the determinant
\begin{align*}
&\begin{vmatrix}
  \hat h(s_1,u_1) & \hat h(s_2,u_1) \\
  \hat h(s_1,u_2) & \hat h(s_2,u_2) \\
\end{vmatrix} \\
& = \hat h(s_2,u_2)\hat h(s_1,u_1) - \hat h(s_1,u_2)\hat h(s_2,u_1)
\end{align*}
is non-zero with probability 1. As we argued before, this implies that, with probability 1 on the value of $T$, after mapping the received signal to the nearest point in $\U_{u_2}$, $u_2$ can decode $b_1$ and $a+b_2$ with probability at least $1-\exp(-\delta_4 P^\ep)$, for some positive constant $\delta_4$. The transmit signal at $u_2$, will then be
\[ X_{u_2} = - \alpha G \hat b_1,
\]
where $\hat b_1$ is the output of the hard-decoding performed by $u_2$.
We again notice that $\alpha$ can be chosen independently of $P$, for $P$ sufficiently large, guaranteeing that the power constraints at $u_1$, $u_2$ and $u_3$ are simultaneously satisfied. The received signal at destination $d_1$ is given by
\begin{align*}&Y_{d_1} = \hat h(u_1,d_1) X_{u_1} + \hat h(u_2,d_1) X_{u_2}+ N_{d_1} \\
&=\alpha G \hat h(u_2,d_1) \hspace{-1mm} \left(\hspace{-0.6mm} a +b_1\hspace{-0.6mm} - \hat b_1+ \frac{\hat h(s_1,u_2)\hat h(s_2,u_1)}{\hat h(s_2,u_2)\hat h(s_1,u_1)} b_2\hspace{-0.6mm}\right)\hspace{-0.6mm} + N_{d_1}^{\text{eff}},
\end{align*}
where $N_{d_1}^{\text{eff}} = \alpha \hat h(u_1,d_1) K_1 N_{u_1} + N_{d_1}$. 
The received signal at $d_2$ is given by
 \begin{align*} & Y_{d_2} = \hat h(u_2,d_2) X_{u_2} + \hat h(u_3,d_2) X_{u_3} + N_{d_2} \\
&=\alpha G \hat h(u_2,d_2) \left(-b_1 - \hat b_1 - \frac{\hat h(s_1,u_2)\hat h(s_2,u_1)}{\hat h(s_2,u_2)\hat h(s_1,u_1)} b_2\right)
+ N_{d_2}^{\text{eff}},
\end{align*}
where $N_{d_2}^{\text{eff}} = \alpha \hat h(u_3,d_2) K_2 N_{u_3} + N_{d_2}$. 
Notice that with probability at least $1- \exp(-\delta_4 P^\ep)$, $Y_{d_1}$ and $Y_{d_2}$ are given by
\[ 
Y_{d_1} = \alpha G \hat h(u_2,d_1) \left(a +\frac{\hat h(s_1,u_2)\hat h(s_2,u_1)}{\hat h(s_2,u_2)\hat h(s_1,u_1)} b_2\right)  + N_{d_1}^{\text{eff}},
\]
\[
Y_{d_2} =\alpha G \hat h(u_2,d_2) \left(-2 b_1 - \frac{\hat h(s_1,u_2)\hat h(s_2,u_1)}{\hat h(s_2,u_2)\hat h(s_1,u_1)} b_2\right)
+ N_{d_2}^{\text{eff}}.
\]
The destinations will first perform a hard-decoding, similar to the one performed by $u_2$. If we assume that the decoding at node $u_2$ was correct, the signal received by $d_1$ is a noisy version of a point in the set
\[
\U_{d_1} = \alpha G  \hat h(u_2,d_1) \left\{ x_1 + T^{-1} x_2 \st x_1,x_2 \in \U \right\},
\]
for $x_1 = a$ and $x_2 = b_2$.
Thus, it can be shown that, with probability 1 over the value of $T^{-1}$,  $d_1$ can decode $a$ (and also $b_2$) with probability of error smaller than $\exp(-\delta_5 P^\ep)$, for some positive constant $\delta_5$. 
Again assuming that the decoding at node $u_2$ was correct, the signal received by $d_2$ is a noisy version of a point in the set
\[
\U_{d_2} =  \alpha G  \hat h(u_2,d_2) \left\{ x_1 + T x_2 \st x_1 \in 2\,\U , x_2 \in \U \right\},
\]
for $x_1 = -2 b_1$ and $x_2 = b_2$. 
With probability 1 over the value of $T$, $d_2$ can decode $b_1$ and $b_2$ with probability at least $\exp(-\delta_6 P^\ep)$, for some positive constant $\delta_6$ (if the decoding at $u_2$ was also correct). 
Therefore, destination $d_1$ obtains $a[t]$ and destination $d_2$ obtains both $b_1[t]$ and $b_2[t]$, for $t=1,...,n$, and, by applying typicality-based decoders, the messages $W_1$, $W_2^{(1)}$ and $W_2^{(2)}$ can be decoded by their intended destinations. 
By following the same steps as in case (1), our codebook can have rate
\[
R = (1-4 \exp(-\delta_{\min } P^\ep))\frac{1-\ep}{2+\ep} \frac{\log P}2 - 4,
\]
where $\delta_{\min } = \min (\delta_4,\delta_5,\delta_6)$. 
Thus, each of the messages carries $\frac{1-\ep}{2+\ep}$ degrees-of-freedom. Since $\ep$ can be chosen arbitrarily small, we conclude that $(1/2,1) \in \D$, and $\D$ is as given in (\ref{dofsym}).
This concludes the derivation of the degrees-of-freedom region of all two-unicast layered Gaussian networks. 
\end{subsubsection}

In order to state the result in a concise way, we will use the notion of disjoint paths with $(s_i,d_i)$-manageable interference (see Definition \ref{semimanag}).
Notice that if $\p11$ and $\p22$ have interference that is both $(s_1,d_1)$-manageable and $(s_2,d_2)$-manageable, they do not necessarily have manageable interference, since the latter requires a single set $S$ for which $n_1(G[S]) \ne 1$ and $n_2(G[S]) \ne 1$. We will describe case C\ref{sym} 
in terms of $(s_i,d_i)$-manageable interference through the following claim.

\begin{claim}
A network $\N$ is in case C\ref{sym} and not in cases (A), (A\pr), (B) and (B\pr) if and only if it has disjoint paths $\p11$ and $\p22$ with interference that is not manageable, but is both $(s_1,d_1)$-manageable and $(s_2,d_2)$-manageable.
\end{claim}

\begin{IEEEproof} By definition, if $\N$ is in case C\ref{sym}, we have WLOG $n_1(G) \geq 2$, $n_1^D=1$, $n_2(G)=1$ and $n_2^D=0$, which implies that $\p11$ and $\p22$ have $(s_1,d_1)$-manageable interference. Moreover, we see from Section \ref{conv_sym} (and Figure \ref{Csym_again}) that $\p11 \cup \p22 \subset V \setminus \{v_2\}$ and, from property P\ref{pv2v0}, we have $n_2(G[V\setminus\{v_2\}]) = 0$, which implies that the interference between $\p11$ and $\p22$ is also $(s_2,d_2)$-manageable. Next we argue that, conversely, if $\p11$ and $\p22$ do not have manageable interference but have interference that is both $(s_1,d_1)$-manageable and $(s_2,d_2)$-manageable, then we must be in case C\ref{sym}. Since $\p11$ and $\p22$ do not have manageable interference, we must have either $n_1^D = 1$ or $n_2^D = 1$. If $n_i^D=1$, for $i=1$ or $2$, then since $\p11$ and $\p22$ have $(s_i,d_i)$-manageable interference, we must have $n_i(G)\geq 2$. Therefore, we cannot have $n_1^D = n_2^D = 1$, or else we would have $n_1(G)\geq 2$ and $n_2(G) \geq 2$. We conclude that the only possible case is $n_i(G) \geq 2$, $n_i^D = 1$, $n_{\bar i}(G) = 1$ and $n_{\bar i}^D = 0$, and we are in case C\ref{sym}.
\end{IEEEproof}

For networks in case C\ref{nonsym}, we can describe whether properties P\ref{qpath}, P\ref{q01}, P\ref{zpath} and P\ref{q10} in Section \ref{nonsymsec} are satisfied for $i=1$ or $i=2$ in terms of $(s_i,d_i)$-manageable interference with the following claim. 

\begin{claim}
A network $\N$ that is in case  C\ref{nonsym} and not in cases (A), (A\pr), (B),  (B\pr) and C\ref{sym} satisfies properties P\ref{qpath}, P\ref{q01}, P\ref{zpath} and P\ref{q10} in Section \ref{nonsymsec} for $i= k$ if and only if it contains paths $\z kk$, $\q kk$ and $\p{\bar{k}}{\bar{k}}$ such that $\q kk$ and $\p{\bar{k}}{\bar{k}}$ are disjoint and have $(s_1,d_1)$-manageable interference, $\z kk$ and $\p{\bar{k}}{\bar{k}}$ are disjoint and have $(s_2,d_2)$-manageable interference.
\end{claim}

\begin{IEEEproof}
If $\N$ satisfies properties P\ref{qpath}, P\ref{q01}, P\ref{zpath} and P\ref{q10} in Section \ref{nonsymsec} for $i= k$, then it is easy to see that it must contain paths $\z kk$, $\q kk$ and $\p{\bar{k}}{\bar{k}}$ as in the statement of the claim. 
Conversely, we want to argue (WLOG) that if we have paths $\q11$ and $\z11$, each disjoint from $\p22$, such that $\z11$ and $\p22$ have $(s_2,d_2)$-manageable (but not $(s_1,d_1)$-manageable) interference and $\q11$ and $\p22$ have $(s_1,d_1)$-manageable (but not $(s_2,d_2)$-manageable) interference, then the network $\N$ must satisfy properties P\ref{qpath}, P\ref{q01}, P\ref{zpath} and P\ref{q10} for $i=1$. Since $\z11$ and $\p22$ have $(s_2,d_2)$-manageable (but not $(s_1,d_1)$-manageable) interference, we must have $n_1(G,\z11) = n_1^D(\p22,\z11)=1$. 
Similarly, we must have $n_2(G,\p22) = n_2^D(\q11,\p22)=1$. We let $(v_2,v_1)$ be the unique edge such that $v_2 \in \p22$ and $v_1 \in \z11$.
Now we notice that we must have $n_1^D(\p22,\q11) = 0$, since otherwise we would have a path $\p21$ which does not contain $\{v_2,v_1\}$, and this would imply that $n_1(G,\z11) \geq 2$, which contradicts the fact that $\z11$ and $\p22$ do not have $(s_1,d_1)$-manageable interference. Similarly, we must have $n_2^D(\z11,\p22) = 0$.
\end{IEEEproof}

The results obtained in this section regarding the complete degrees-of-freedom region are summarized in Theorem \ref{extth}.

\end{section}

%
%

\end{subsection}

\end{section}

\begin{section}{Conclusion}\label{concl} 

We explored the degrees-of-freedom of two-unicast layered Gaussian networks. 
Our result shows that, in terms of degrees-of-freedom, there are essentially three categories of such networks. 
In the first one, the network connectivity creates a bottleneck for the information flow, forcing all the messages to be decodable at a single node. Therefore, we only have one degree-of-freedom. 
In the second, the interference can be completely avoided or neutralized and we achieve two degrees-of-freedom. 
Networks which contain a grail or a butterfly achieve their two degrees-of-freedom by ``borrowing'' their achievability schemes from linear network coding. 
When no such structures exist, we must find two disjoint paths and verify whether it is possible to completely neutralize their interference. 
The notion of manageable interference arises quite naturally in these cases:
we simply want to make sure that no path receives a single interference from the other path. 
This way, a path may either receive no interference at all or receive at least two interferences from the other path, which allows for interference cancellation techniques. 
The remaining networks have exactly $3/2$ sum degrees-of-freedom. 
The intuitive reason is that these networks do not have a single node through which all information must pass restricting the degrees-of-freedom to one, but their interference is not manageable. 
One way to achieve the $3/2$ sum degrees-of-freedom is to perform a sort of ``scheduling'' of the transmissions, in order to be able to avoid or neutralize the interference. 
The buffering involved in this scheduling costs us half a degree-of-freedom. 
This third class of networks can be further subdivided into three subclasses, according to which of the points $(1,1/2)$ and $(1/2,1)$ are included in the degrees-of-freedom region.  


The achievability schemes considered show that only a small fraction of the nodes must in fact be careful in their relaying operations. 
In fact, we achieve the sum degrees-of-freedom by converting any multi-layered network into a condensed network with at most four layers.
Thus, the nodes in at most four layers (the source layer, the destination layer, and two intermediate layers) perform relaying operations that require channel state information.
All the other nodes are simply forwarding their received signals.
Another interesting aspect of the achievability schemes is that, in most cases, a linear scheme suffices to achieve two degrees-of-freedom. 
In some cases, however, we must resort to a more sophisticated scheme such as real interference alignment. 
For example, this is the case of the $2 \times 2 \times 2$ interference channel studied in \cite{xx}, where a simple linear scheme cannot achieve two degrees-of-freedom (unless the channel gains vary with time).
%

Our assumption of a layered network topology serves mainly to simplify the problem.
If non-layered networks are considered, the main issue is that interference may occur not only between signals originated at different sources, but also between signals originated at different times.
Therefore, in order to perform interference cancellation, for example, one needs to make sure that the two cancelling signals correspond to the same time-version of the source signal. 
For non-layered networks, other techniques such as signal delaying and backward decoding must be used to achieve the degrees-of-freedom, and the problem becomes significantly more difficult.
However, the layered network assumption is not artificial.
Since the layered topology simplifies the analysis and the implementation of coding schemes, it is desirable in practice, and it can actually be simulated in practical contexts by having the transmitters on each layer transmit on a different frequency band, which allows us to assume that the links only exist between consecutive layers.
Moreover, a layered structure can also arise from the scheduling of the transmitting nodes in a wireless network with half-duplex nodes.
In this context, each hop would capture which nodes are transmitting and which nodes are receiving at a given time slot, and the same node could appear in multiple layers, since they may be transmitting at multiple time-slots.

The natural extension of this work would be to consider more than two information flows in the network.
Recently, in \cite{WangX2Unicast}, networks with two source-destination pairs where each source has a message to each destination (for a total of four messages) were considered.
Interestingly, it was shown that the sum degrees-of-freedom can also take values $4/3$ and $5/3$, in addition to the values $1$, $3/2$ and $2$ that are possible in the setup considered in this paper.
The next step would thus be to consider more than two source-destination pairs.
The main issue lies in the combinatorial complexity of a larger number of source-destination pairs.
For example, if one were to extend a notion such as manageable interference to more than two source-destination pairs, not only would the number of interferences on a $(s_1,d_1)$-path have to be considered, but also which subset of the other sources contributes to each interference.
Moreover, it is not clear whether the non-linear schemes that were necessary for networks such as the $2 \times 2 \times 2$ interference channel can in fact be easily extended to networks with more than two source-destination pairs.
In particular, the characterization of degrees-of-freedom of the $3 \times 3 \times 3$ interference channel remains an open problem.

%

\end{section}


\appendix \label{appsection}

\subsection{Proof of Lemma \ref{const}} \label{prooflemmaconstant}

{\small 
\begin{align*} \rescnt
&I (X_S^n;\tilde{X}_c^n|Y_b^n,\tilde X_A^n,X_T^n) \non
& = I(X_S^n;\{\tilde X_{c,j}^n : j \text{ s.t. } v_j\in \Os(v_c)\}|Y_b^n,\tilde X_A^n,X_T^n) \\
& \eqnum I(X_S^n;\{\tilde X_{c,j}^n-\tfrac{h_{c,j}}{h_{c,b}}\tilde X_{c,b}^n : j \text{ s.t. } v_j\in \Os(v_c)\}|Y_b^n,\tilde X_A^n,X_T^n) \\
& \eqnum I(X_S^n;\{N_{c,j}^n-\tfrac{h_{c,j}}{h_{c,b}}N_{c,b}^n : j \text{ s.t. } v_j\in D \}|Y_b^n,\tilde X_A^n,X_T^n) \\
& \leq h(\{N_{c,j}^n-\tfrac{h_{c,j}}{h_{c,b}}N_{c,b}^n : j \text{ s.t. } v_j\in D \}) \\
& \quad -h(\{N_{c,j}^n-\tfrac{h_{c,j}}{h_{c,b}}N_{c,b}^n : j \text{ s.t. } v_j\in D \} | Y_b^n,\tilde X_A^n,X_T^n,X_S^n) \\
& \leqnum \frac{n |D|}2 \log(2 \pi e \kappa) \\
& \quad -h(\{N_{c,j}^n-\tfrac{h_{c,j}}{h_{c,b}}N_{c,b}^n : j \text{ s.t. } v_j\in D \} | Y_b^n,\tilde X_A^n,X_T^n,X_S^n) \\
& \leqnum \frac{n|D|}2 \log(2 \pi e \kappa) \\
& \quad - h(\{N_{c,j}^n : j \text{ s.t. } v_j\in D \} | N_{c,b}^n , Y_b^n,\tilde X_A^n,X_T^n,X_S^n) \\
& \eqnum \frac{n|D|}2 \log(2 \pi e \kappa)   - h(\{N_{c,j}^n : j \text{ s.t. } v_j\in D \} ) \\
& = n \left( \frac{|D|}2 \log(2 \pi e \kappa)  - \sum_{j : v_j \in D} \frac12\log\left(\frac{2 \pi e}{|\I(v_j)|} \right) \right),
\end{align*} }
 \rescnt
where \cnt follows from the fact that $Y_b^n - \sum_{v_a \in A} \tilde X_{a,b}^n = \tilde X_{c,b}^n$; \cnt follows since, for $j = b$, $N_{c,j}^n-\tfrac{h_{c,j}}{h_{c,b}}N_{c,b}^n = 0$; 
\cnt follows by letting $\kappa \defi 1 + (\max_{e,f \in E} h_e/h_f)^2$;
\cnt follows because conditioning reduces entropy and thus we can condition on $N_{c,b}^n$; $(iv)$ follows from the fact that, since for $u \in D$ and $w \in T$, $u \not\leadsto w$, $N_{c,u}^n$ is independent of all the random variables conditioned on.




\subsection{Proof of Lemma \ref{lem:paths}}

Consider the nodes in $\I(v_i^p)$. 
Assume, by contradiction, that there are no two paths $P_{s_1,v_p^i}$ and $P_{s_2,v_p^i}$ such that $P_{s_1,v_p^i} \cap P_{s_2,v_p^i} = \{v_p^i\}$. 
Then, we do not have two vertex-disjoint paths starting in $\{s_1,s_2\}$ and ending in $\I(v_i^p)$. 
From Menger's Theorem, there exists a node $v_d$ whose removal disconnects $\{s_1,s_2\}$ from $\I(v_i^p)$, and thus from $v_p^i$. The existence of the path $\p{i}i$ containing $v_p^i$ guarantees that $v_d \in \p{i}i$. Since the removal of $v_p^i$ disconnects $s_{\bar i}$ from $d_i$, and the removal of $v_d$ disconnects $\{s_1,s_2\}$ from $v_p^i$, we conclude that the removal of $v_d$ also disconnects $s_{\bar i}$ from $d_i$. But this is a contradiction to the fact that $v_p^i$ was the first such node.



\subsection{Proof of Claim \ref{structs}} \label{proofclaimwireline}


We let $G = (V,E)$ be the graph of our original network, and we construct an extended network $\N$ with graph $G=(V',E')$ in the following way. We let the layers in $V'$ be $V_1,V_1',V_{2},V_{2}',...,V_{r},V_r'$, where $V_j'$ is a copy of $V_j$, $j=1,...,r$. The edges between $V_j'$ and $V_{j+1}$, for $j = 1,2,...,r-1$, are the same as the edges between $V_j$ and $V_{j+1}$ in $G$. To add the edges between $V_j$ and $V_j'$, for $j = 1,2,...,r$, we simply connect each $v_k \in V_j$ to its copy in $V_j'$. The source-destination pairs of $\N$ are the same as of $\N$. 

Next we claim that if we have an edge $e \in E'$ whose removal from $\N'$ disconnects $d_i$ from both sources and $s_{\bar i}$ from both destinations, $i \in \{1,2\}$, then our original network falls in (A). Suppose we have such an edge $e \in E'$. If $e \in V_j \times V_j'$ for some $j$, then it is easy to see that in the original network, this edge corresponds to a single node in $V_j$ whose removal disconnects $d_i$ from both sources and $s_{\bar i}$ from both destinations, and we must be in (A). Otherwise, if $e \in V_j' \times V_{j+1}$ for some $j$, then the removal of the edge $\tilde e$ in $V_j \times V_j'$ (or $V_{j+1} \times V_{j+1}'$) which is adjacent to $e$ must also disconnect $d_i$ from both sources and $s_{\bar i}$ from both terminals. This is because all paths from any source to any destination which contain the nodes in $e$ must also contain the nodes in $\tilde e$. Therefore, $\tilde e$ can be translated to a node $v$ in $\N$ whose removal disconnects $d_i$ from both sources and $d_i$ from both destinations, and $\N$ falls into case (A).

Therefore, the absence of a node $v$ as described in (A) in our network $\N$ implies that $\N'$ does not contain an edge whose removal disconnects $d_i$ from both sources and $s_i$ from both destinations for some $i \in \{1,2\}$. Thus, we employ a result for double unicast networks, shown in both \cite{ShenviDey} and \cite{Cai2unicast}, which guarantees that the extended network $\N'$ must contain one of the three structures shown in Figure \ref{nets1}: two edge-disjoint paths $\p11$ and $\p22$, a butterfly, or a grail. Moreover, we notice that, in $\N'$, any pair of edge-disjoint paths is also vertex-disjoint, and corresponds to a pair of vertex-disjoint paths in $\N$. Thus, we conclude that if our network $\N$ is not in (A), then it must contain two vertex-disjoint paths $\p11$ and $\p22$, a grail structure or a butterfly structure.

\newpage

\bibliographystyle{unsrt}

\bibliography{refs}

\end{document}